\documentclass[12pt,english]{article}
\pdfoutput=1

\usepackage{caption}

\usepackage{xargs}                      
\usepackage[pdftex,dvipsnames]{xcolor}  
\usepackage[colorinlistoftodos,prependcaption,textsize=tiny]{todonotes}

\newcommandx{\unsure}[2][1=]{\todo[linecolor=red,backgroundcolor=red!25,bordercolor=red,#1]{#2}}
\newcommandx{\change}[2][1=]{\todo[linecolor=blue,backgroundcolor=blue!25,bordercolor=blue,#1]{#2}}
\newcommandx{\info}[2][1=]{\todo[linecolor=OliveGreen,backgroundcolor=OliveGreen!25,bordercolor=OliveGreen,#1]{#2}}
\newcommandx{\improvement}[2][1=]{\todo[linecolor=Plum,backgroundcolor=Plum!25,bordercolor=Plum,#1]{#2}}
\newcommandx{\thiswillnotshow}[2][1=]{\todo[disable,#1]{#2}}

\usepackage[verbose,letterpaper,tmargin=2.54cm,bmargin=2.54cm,lmargin=2.54cm,rmargin=2.54cm]{geometry}
\usepackage{setspace}
\usepackage[displaymath, mathlines]{lineno}

\usepackage[affil-it]{authblk}
\providecommand{\keywords}[1]{\indent \textit{Key words: #1}}

\usepackage{lmodern}

\usepackage{amssymb,amsmath}
\usepackage{ifxetex,ifluatex}
\usepackage{fixltx2e} 
\ifnum 0\ifxetex 1\fi\ifluatex 1\fi=0 
  \usepackage[T1]{fontenc}
  \usepackage[utf8]{inputenc}
\else 
  \ifxetex
    \usepackage{mathspec}
    \usepackage{xltxtra,xunicode}
  \else
    \usepackage{fontspec}
  \fi
  \defaultfontfeatures{Mapping=tex-text,Scale=MatchLowercase}
  
\fi
\IfFileExists{upquote.sty}{\usepackage{upquote}}{}
\IfFileExists{microtype.sty}{\usepackage{microtype}}{}
\usepackage{longtable,booktabs}

\usepackage{graphicx}
\makeatletter
\def\maxwidth{\ifdim\Gin@nat@width>\linewidth\linewidth\else\Gin@nat@width\fi}
\def\maxheight{\ifdim\Gin@nat@height>\textheight\textheight\else\Gin@nat@height\fi}
\makeatother
\setkeys{Gin}{width=\maxwidth,height=\maxheight,keepaspectratio}
\ifxetex
  \usepackage[setpagesize=false, 
              unicode=false, 
              xetex]{hyperref}
\else
  \usepackage[unicode=true]{hyperref}
\fi
\hypersetup{breaklinks=true,
            bookmarks=true,
            pdfauthor={},
            pdftitle={Choosing the observational likelihood in state-space stock assessment models},
            colorlinks=true,
            citecolor=blue,
            urlcolor=blue,
            linkcolor=blue,
            pdfborder={0 0 0}}
\title{Choosing the observational likelihood in state-space stock assessment
models}

\author[]{Christoffer Moesgaard
Albertsen\thanks{Corresponding author. Email: cmoe@aqua.dtu.dk}}
\author[]{Anders Nielsen}
\author[]{Uffe Høgsbro Thygesen}

\affil[]{Technical University of Denmark, National Institute of Aquatic
Resources, Charlottenlund Castle, DK-2920 Charlottenlund, Denmark}

\date{}
\newcommand{\at}{@}
\usepackage{booktabs}

\begin{document}
\maketitle
\begin{abstract}
Data used in stock assessment models result from combinations of
biological, ecological, fishery, and sampling processes. Since different
types of errors propagate through these processes it can be difficult to
identify a particular family of distributions for modelling errors on
observations a priori. By implementing several observational
likelihoods, modelling both numbers- and proportions-at-age, in an age
based state-space stock assessment model, we compare the model fit for
each choice of likelihood along with the implications for spawning stock
biomass and average fishing mortality. We propose using AIC intervals
based on fitting the full observational model for comparing different
observational likelihoods. Using data from four stocks, we show that the
model fit is improved by modelling the correlation of observations
within years. However, the best choice of observational likelihood
differs for different stocks, and the choice is important for the
short-term conclusions drawn from the assessment model; in particular,
the choice can influence total allowable catch advise based on reference
points.

\end{abstract}

\newcommand{\needref}{\improvement{ref}}

\section{Introduction}\label{introduction}

Stock assessment models often use aggregated and uncertain data such as
surveys and landings-at-age which rely on age classification of
effectively few individuals (Aanes and Pennington 2003). Commercial
fishing and scientific surveys sample from populations that vary
according to, for example, season, sex, age and region. From this catch,
samples are weighed and measured to estimate the length distribution,
weight-at-length, and total catch in numbers. Additional sub-samples are
taken to age classify individuals for estimating proportions-at-age;
either directly or through an age-length key. The samples consist of
many individuals from few hauls (Aanes and Pennington 2003), which may
lead to underestimated uncertainties of estimates if ignored. Finally,
all this information is aggregated to numbers-at-age for each year. This
aggregation may be via models including e.g.~spatial location, season,
gear and length effects. Even though the stock population growth process
at this level of aggregation is well described (Each year the fish age
by one year, some die of natural causes, and others die from fishing)
aggregating the different sources of uncertainty makes it difficult to
find the optimal (or true) distributions of the observations a priori.

Age-based stock assessment models can be divided into two classes
depending on the way they utilize the data. Either the data can be
modelled as numbers-at-age or as proportions-at-age along with total
weight or numbers. Most currently used age-based stock assessment models
exclusively consider either numbers- or proportions-at-age and only one
or few observational likelihoods (ICES 2010a). When modelling
numbers-at-age, the normal distribution, parameterized to avoid too much
probability on negative observations, has been used (Gudmundsson 1994,
Fryer 2002) along with the log-normal distribution (Cook 2013, Nielsen
and Berg 2014) and its multivariate extension (Myers and Cadigan 1995).
Although recommended over the log-normal by Cadigan and Myers (2001),
the gamma distribution is infrequently used to model numbers-at-age in
assessment models (ICES 2010a).

The multinomial distribution has been popular when modelling
proportions-at-age (Fournier and Archibald 1982, Methot Jr. and Wetzel
2013, Williams and Shertzer 2015). Based on the age classification
sampling, it is an intuitive choice; however, when using the true number
of data generating samples, the variances of the modelled proportions
are often too small, and the correlation structure too restrictive
(Crone and Sampson 1998, Aanes and Pennington 2003, Francis 2014).
Efforts have been made to increase the variance by estimating an
effective sample size (McAllister and Ianelli 1997, Francis 2011, Hulson
et al. 2011, 2012). Nonetheless, the effective sample size must be
estimated by iterative optimization (McAllister and Ianelli 1997,
Francis 2011, Maunder 2011) since the multinomial distribution is
improper when used for continuous data (Francis 2014). Hence, the
multinomial distribution will not be considered here. To avoid iterative
estimation of the effective sample size, it has been suggested to
replace the multinomial with the Dirichlet distribution (Williams and
Quinn 1998, Francis 2014) in which the variance is only determined by
parameters.

While the Dirichlet distribution is an improvement over the multinomial
distribution, they both have a very restrictive variance-covariance
structure that only allows negative correlations, which may not be
appropriate (Francis 2014). Therefore distributions based on
transformations of multivariate normals, such as the additive logistic
normal (Francis 2014) and the multiplicative logistic normal (Cadigan
2015), have recently been proposed for proportions-at-age in stock
assessment models.

Although several authors have compared different proportions-at-age
models (Maunder 2011, Francis 2014), not much effort has been given to
compare different observational likelihoods for numbers-at-age data
(Cadigan and Myers 2001), and even less has been given to compare
between the proportions- and numbers-at-age. Using the R-package
Template Model Builder (Kristensen et al. 2016), we implement 13
observational likelihoods, including both numbers- and
proportions-at-age models, in an age-based state-space stock assessment
model. Using assessment data from four European stocks, we compare the
model fit for each choice of likelihood along with the implications for
key outputs such as spawning stock biomass (SSB) and average fishing
mortality (\(\bar{F}\)).

\section{Methods}\label{methods}

We implemented age-based state-space stock assessment models (Nielsen
and Berg 2014) with 13 different observational likelihoods
(\autoref{tab:likelihoods}) for four different European stocks. For
simplicity the same observational likelihood was used for both
commercial catch data and survey indices. While the process model was
kept unchanged for each stock, we compared the goodness-of-fit of the
observational likelihoods by AIC. We considered models for
numbers-at-age and proportions-at-age combined with total catch. We
considered seven different distributions for numbers-at-age. When using
data in the form of a total and proportions-at-age we followed the
wide-spread convention of modelling total catch as univariate
log-normal, but considered two alternatives where the total was either
in numbers or biomass. These two alternatives for total catch were
crossed with three alternative distributions for the proportions. The
observational likelihoods implemented cover frequently used
distributions in fisheries stock assessments and close extensions.

\captionsetup{width=\linewidth}
\begin{longtable}[c]{@{}lllllll@{}}
\caption{Overview of the observational models used in the case studies and some properties: if zero observations are allowed; whether the Baranov catch equation determines the mean, median or location; the number of estimated observational parameters per age (\(a\)) and fleet (\(f\)); and whether a correlation parameter is estimated. The models are divided in to model classes: Univariate numbers-at-age (UN\at A), multivariate numbers-at-age (MN\at A), proportions-at-age with log-normal total numbers (P\at AwN), and proportions-at-age with log-normal total weight (P\at AwW).\label{tab:likelihoods}}
\tabularnewline
\toprule
Model & Distribution & Class & Allows 0 & Baranov & Est. par.s & Est.
cor.\tabularnewline
\midrule
\(M_{1}\) & log-Normal & UN\at A & No & Median & 1 \(a\)
\(f\)\footnote{Should be read: One per age per fleet.} &
No\tabularnewline
\(M_{2}\) & Gamma & UN\at A & Some & Mean & 1 \(a\) \(f\) &
No\tabularnewline
\(M_{3}\) & Generalized Gamma & UN\at A & Some & Location & 2 \(a\)
\(f\) & No\tabularnewline
\(M_{4}\) & Normal & UN\at A & Yes & Mean & 1 \(a\) \(f\) &
No\tabularnewline
\(M_{5}\) & Left Truncated Normal & UN\at A & Yes & Location & 1 \(a\)
\(f\) & No\tabularnewline
\(M_{6}\) & log-Student's t & UN\at A & No & Location & 2 \(a\) \(f\) &
No\tabularnewline
\(M_{7}\) & Multivariate log-Normal & MN\at A & No & Median & 1 \(a\)
\(f\)+1
\(f\)\footnote{Should be read: One per age per fleet and one additional per fleet.}
& Yes\tabularnewline
\(M_{8}\) & Additive Logistic Normal & P\at AwN & No & Location & 1
\(a\) \(f\)+1 \(f\) & Yes\tabularnewline
\(M_{9}\) & Multiplicative Logistic Normal & P\at AwN & No & Location &
1 \(a\) \(f\) + 1 \(f\) & Yes\tabularnewline
\(M_{10}\) & Dirichlet & P\at AwN & No & Mean & 1 \(f\) &
No\tabularnewline
\(M_{11}\) & Additive Logisitc Normal & P\at AwW & No & Location & 1
\(a\) \(f\)+1 \(f\) & Yes\tabularnewline
\(M_{12}\) & Multiplicative Logistic Normal & P\at AwW & No & Location &
1 \(a\) \(f\) + 1 \(f\) & Yes\tabularnewline
\(M_{13}\) & Dirichlet & P\at AwW & No & Mean & 1 \(f\) &
No\tabularnewline
\bottomrule
\end{longtable}

\subsection{Process model}\label{process-model}

The processes described in the state-space model involved the true
unobserved numbers-at-age in the stock, and the true unobserved fishing
mortality (See Nielsen and Berg 2014 or Appendix A for details).
Following Nielsen and Berg (2014), the logarithm of the fishing
mortality was assumed to follow a multivariate Gaussian random walk,
where the correlation had an AR(1) structure (model D in Nielsen and
Berg 2014). The true population numbers-at-age were assumed to follow an
exponential decay model where the natural mortality is known. The model
included recruitment to the first age group. The error-terms for the
true numbers-at-age were assumed to follow a log-normal distribution
without correlation. All variance, correlation and stock-recruitment
parameters were estimated. Quantities such as weight-at-age and maturity
were assumed to be known. The process model was related to the
observations through the Baranov catch equation for catch data, and
through an assumption of proportionality to abundance-at-age for
surveys. The proportionality constants were estimated. We denoted the
calculated catch (or survey index) by \(\tilde{C}_{a,f,y}\).

\subsection{Observational models}\label{observational-models}

Our model \(M_1\) was the log-normal distribution with its usual
parameterization. The median was determined by \(\tilde{C}_{a,f,y}\),
while a scale parameter was estimated for each age and fleet. The model
\(M_2\) was the gamma distribution parameterized to have constant
coefficient of variation (Cadigan and Myers 2001). The mean was
determined by \(\tilde{C}_{a,f,y}\), while a coefficient of variation
(CV) was estimated for each age and fleet. The generalized gamma
distribution was included as model \(M_3\) with the parameterization of
Prentice (1974). This parameterization was preferred over the Stacy
(1962) parameterization as it both extends it, and is numerically more
stable when reducing to the log-normal distribution (Prentice 1974,
Farewell and Prentice 1977). The log-location parameter was determined
by \(\log(\tilde{C}_{a,f,y})\) while a shape and scale parameter was
estimated for each age and fleet. The models \(M_4\) and \(M_5\) were
the normal and truncated normal (with left truncation at zero). Both
were parameterized based on the mean determined by \(\tilde{C}_{a,f,y}\)
and separate CV parameters for each age and fleet (which applied to the
un-truncated values for the truncated normal). The Student's
t-distribution on log-scale was our model \(M_6\). The distribution was
parameterized with a log-location parameter determined by
\(\log(\tilde{C}_{a,f,y})\) along with log-scale and
log-degrees-of-freedom parameters estimated separately for each age and
fleet.

Model \(M_7\) was the multivariate log-normal with its usual
parameterization. The marginal medians were determined by
\(\tilde{C}_{a,f,y}\), while a one-parameter AR(1) structure was used
for the correlation between ages on logarithmic scale (Pinheiro and
Bates 2000, Francis 2014). Separate correlation parameters were
estimated for each fleet along with scale parameters estimated for each
age and fleet. Our models \(M_8\) and \(M_9\) were the additive logistic
normal and the multiplicative logistic normal (Aitchison 2003) with
log-normal total numbers. For both models, the location parameters were
determined by the \(\tilde{C}_{a,f,y}\)s while the scale matrices were
parameterized as \(M_7\). For the log-normal distributions, the medians
were determined by \(\sum_a\tilde{C}_{a,f,y}\), while a separate scale
parameter was estimated for each fleet. The Dirichlet distribution with
log-normal total numbers was our model \(M_{10}\). The Dirichlet
distribution was parameterized with concentration parameters
proportional to \((\tilde{C}_{1,f,y},\ldots,\tilde{C}_{A,f,y})^T\). A
proportionality parameter was estimated for each fleet. The log-normal
distributions for total numbers were parameterized as \(M_8\) and
\(M_9\). Finally, the models \(M_{11}\), \(M_{12}\), and \(M_{13}\) were
the additive logistic normal, multiplicative logistic normal, and
Dirichlet with log-normal total weight parameterized as \(M_8\),
\(M_9\), and \(M_{10}\) respectively. All estimated observational
parameters were assumed to be constant over years. The densities and
further details can be seen in Appendix A.

\subsection{Comparing by AIC}\label{comparing-by-aic}

To compare the different observational models, we employed the Akaike
Information Criterion (Akaike 1974). However, the AIC applies to
comparison between specific models, whereas each observational model
represents an entire family of models, differing in assumed
relationships between parameters for different age groups. These
families include ``full models'' where each age group and fleet are
assigned independent parameters, a ``minimal model'' where all age
groups share common parameters, as well as a range of models between
these two extremes. A standard application of the AIC would require that
the optimal model in each family is identified, a task that would
involve estimation of parameters in thousands of models.

To avoid this step, which is computationally very demanding and
tangential to our purpose, we chose to identify an AIC interval which
characterized each model family. This AIC interval gave an upper and a
lower bound on the optimal AIC within that family. The upper bound of
the interval was attained by the AIC for the full model. The lower bound
of the interval was calculated as the AIC that would hypothetically be
obtained by the smallest possible nested sub-model if the negative
log-likelihood would not increase compared to the full model. The
difference between the upper and lower bound is thus twice the
difference in the number of parameters between the full model and the
minimal model.

A model family was considered clearly superior to another if the upper
bound of its AIC interval was below the lower bound of the other model
family's interval (i.e., the other model family had a higher interval).
Clearly inferior model families could be discarded. To compare the
remaining model families, it would be possible to narrow the AIC
intervals through testing within each model family, but for simplicity
we base the comparison on the full model in each family.

Using AIC to compare the models required that the models were defined on
the same data, which was not the case when we compared between
numbers-at-age models and proportions-at-age models. The
proportions-at-age data were, however, a one-to-one transformation of
the numbers-at-age data. Thus, using a standard transformation of
densities we derived the log-likelihood for the numbers-at-age data that
is consistent with our specified distributions based on proportions and
totals (Appendix B). Using the transformed likelihood in the AIC
calculation allowed for valid comparisons of models using numbers-at-age
directly versus those using totals and proportions-at-age. Note that a
similar transformation was required so that models that used total
weight could be compared to models that used total numbers with the
proportions.

\subsection{Case study}\label{case-study}

We implemented the models for four different data sets used for
assessments (\autoref{tab:data}): The Blue Whiting data set was the
basis of the 2014 ICES advice (ICES 2014a) for Subareas I-IX and XIV;
the North-East Arctic Haddock data was used for the 2014 ICES advice
(ICES 2014b) for Subarea IV (North Sea) and Division IIIa West
(Skagerrak); The North Sea Cod data was obtained from the 2012 ICES
advice (ICES 2012) for subarea IV (North Sea) and Divisions VII (Eastern
Channel) and IIIa West (Skagerrak); and the Northern Shelf Haddock data
from the ICES advice for Subarea IV (North Sea) and Division IIIa West
(Skagerrak) in 2012 (ICES 2012). A Beverton-Holt curve was assumed for
the relationship between stock and recruitment for North Sea Cod,
whereas a random walk was assumed for the other stocks. Two of the data
sets had missing data. For the Blue Whiting data set a whole year was
missing for the survey, whereas for the North-East Arctic Haddock data
set, values were missing for at most three ages per year. The
proportions-at-age models above could not easily handle years with
missing age observations. Hence, to give a fair comparison of the
univariate and multivariate models, we treated years with missing age
observations as if it was missing all ages. Process model parameters
were assumed equal between ages in the same way as they were for the
advice.

\begin{table}
\caption{Overview of the data sources used in the case study. Q1-Q4 indicates at which quarter of the year the survey is conducted.}
\label{tab:data}
\begin{tabular}{lcccclc}
\toprule
Fleet & First year & Last year & First age & Last age & Years with missing & Process parameters \tabularnewline
\midrule
\multicolumn{6}{l}{\textbf{Blue Whiting}} & 7 \tabularnewline
Commercial & 1981 & 2013 & 1 & 10 & & \tabularnewline
Survey Q1 & 2004 & 2014 & 3 & 8 & 2010 & \tabularnewline
\multicolumn{6}{l}{\textbf{North-East Arctic Haddock}} & 30 \tabularnewline
Commercial & 1950 & 2013 & 3 & 11 & & \tabularnewline
Survey Q4 & 1991 & 2013 & 3 & 7 & & \tabularnewline
Survey Q1 & 1992 & 2013 & 3 & 7 & 1992-1995 & \tabularnewline
&      &      &   &   & 2000-2002,2004  & \tabularnewline
Survey Q1 & 1992 & 2013 & 3 & 8 & 1994,1995,2001 & \tabularnewline
Survey Q3 & 2004 & 2013 & 3 & 8 & 2005 & \tabularnewline
\multicolumn{6}{l}{\textbf{North Sea Cod}} & 11 \tabularnewline
Commercial & 1950 & 2011 & 1 & 7 & & \tabularnewline
Survey Q1 & 1983 & 2012 & 1 & 5 & & \tabularnewline
\multicolumn{6}{l}{\textbf{Northern Shelf Haddock}} & 33 \tabularnewline
Commercial & 1963 & 2011 & 0 & 8 & & \tabularnewline
Survey Q3 & 1977 & 1991 & 0 & 6 & & \tabularnewline
Survey Q3 & 1992 & 2011 & 0 & 6 & & \tabularnewline
Survey Q3 & 1982 & 1997 & 0 & 6 & & \tabularnewline
Survey Q3 & 1998 & 2011 & 0 & 6 & & \tabularnewline
Survey Q1 & 1982 & 2011 & 0 & 4 & & \tabularnewline
\bottomrule
\end{tabular}

\end{table}

\section{Results}\label{results}

\begin{figure}[htbp]
\centering
\includegraphics{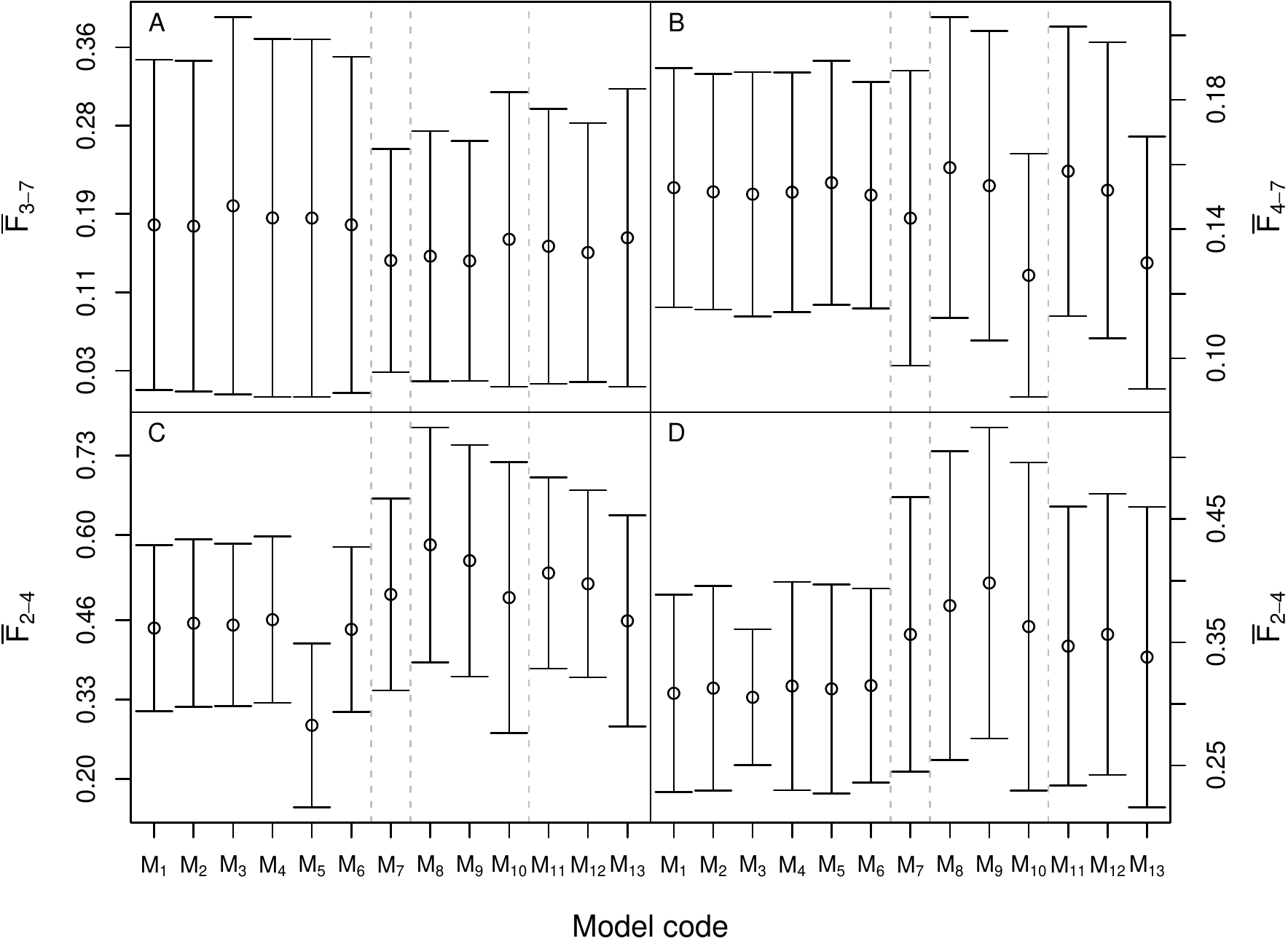}
\caption{Last year fishing mortalities with 95\% confidence intervals
for models \(M_1\) to \(M_{13}\) (Table 1) in the case studies: Blue
Whiting (A), North-East Arctic Haddock (B), North Sea Cod (C), and
Northern Shelf Haddock (D). Vertical dashed grey lines separates the
models in model classes (Table 1). Subscripts to \(\bar{F}\) indicates
the ages the average is over. All ages are weighed equally in the
average.\label{fig:FBARlast}}
\end{figure}

\begin{figure}[htbp]
\centering
\includegraphics{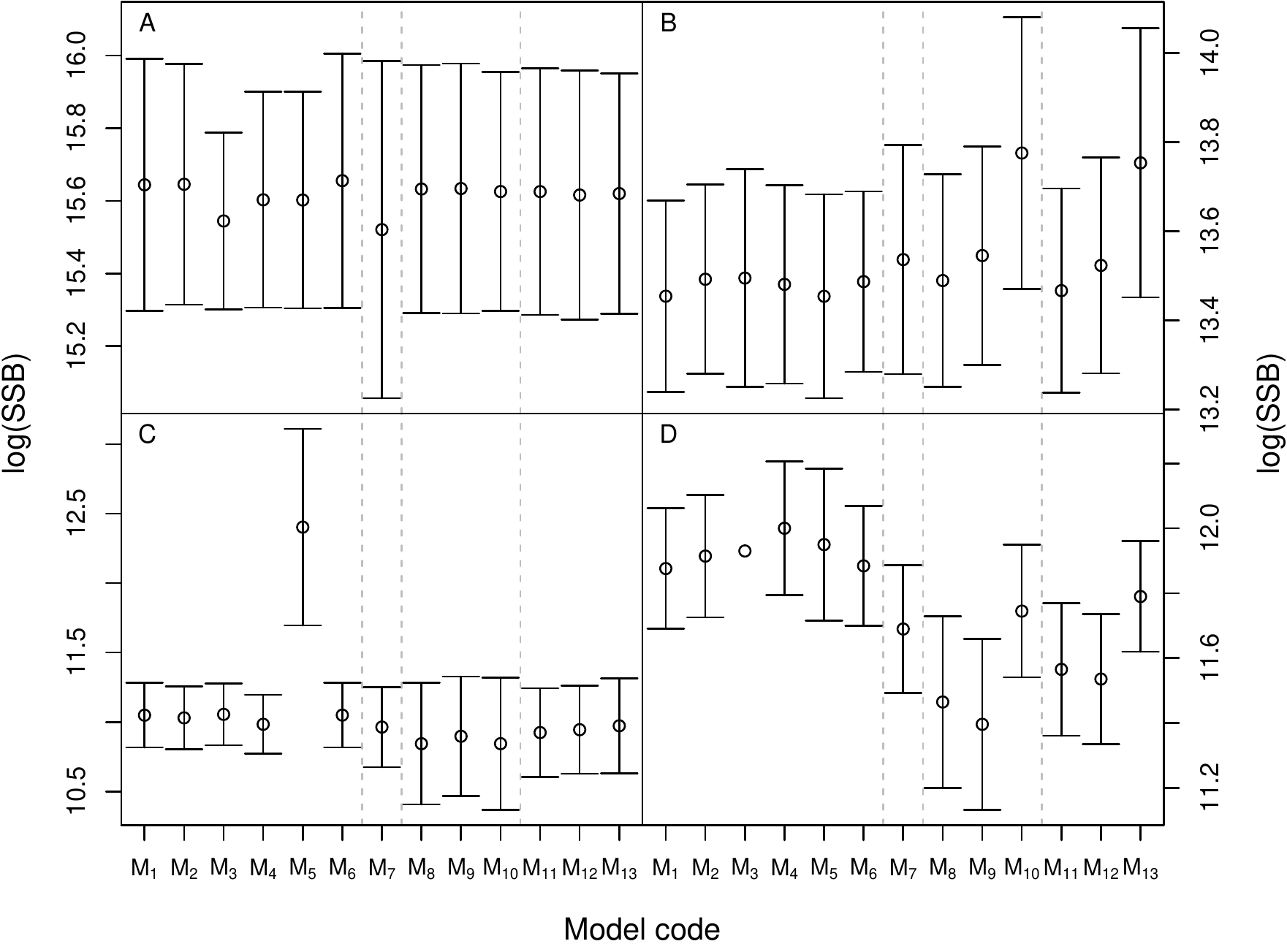}
\caption{Last year spawning stock biomass logarithmic (log(SSB)) for
models \(M_1\) to \(M_{13}\) (Table 1) in the case studies: Blue Whiting
(A), North-East Arctic Haddock (B), North Sea Cod (C), and Northern
Shelf Haddock (D). Vertical dashed grey lines separates the models in
model classes (Table 1).\label{fig:SSBlast}}
\end{figure}

In all four case studies we found that the estimated average fishing
mortality (\autoref{fig:FBARlast}), spawning stock log-biomass
(\autoref{fig:SSBlast}), and their standard errors in the final year
differed between models. In particular, we found that for North Sea Cod,
the highest fishing mortality was 2 times the lowest fishing mortality,
and the widest confidence interval was 1.7 times the narrowest. For
Northern Shelf Haddock, the confidence interval of the estimated final
year spawning stock biomass for model \(M_4\), which had the highest
estimate, did not overlap with the confidence interval for model
\(M_9\), which had the lowest estimate.

\begin{figure}[htbp]
\centering
\includegraphics{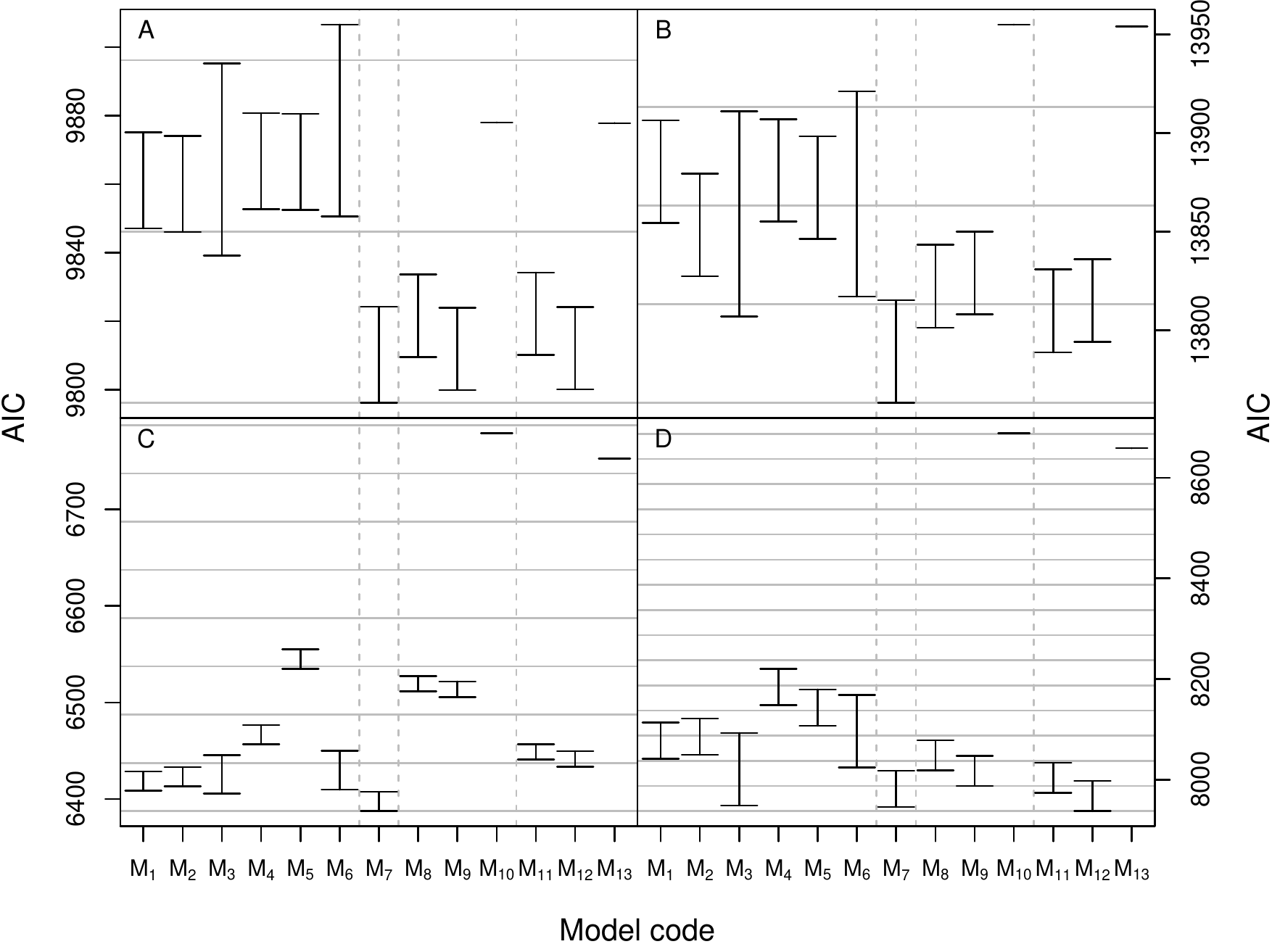}
\caption{AIC intervals for models \(M_1\) to \(M_{13}\) (Table 1) in the
case studies: Blue Whiting (A), North-East Arctic Haddock (B), North Sea
Cod (C), and Northern Shelf Haddock (D). The horizontal grey lines
indicate AIC differences of 50 starting at the lowest lower bound of the
models. Vertical dashed grey lines separates the models in model classes
(Table 1).\label{fig:AIC}}
\end{figure}

We found that the models including correlation parameters obtained
better fit to the data for the full models than models without
correlation parameters within each model class (\autoref{fig:AIC}); the
AIC for the full model (upper bound of the interval) was lower for the
multivariate log-normal than for the univariate numbers-at-age, and
similarly the logistic normals had better model fits than the Dirichlet
distribution, which provided one of the highest AIC intervals of all
models in all case studies. In the North Sea Cod and Northern Shelf
Haddock cases, the lower bounds of the AIC intervals for the Dirichlet
distribution were clearly separated from the upper AIC bounds of all
other models with an AIC difference of more than 190.

For North Sea Cod, the multivariate log-normal achieved the lowest AIC
for the full model. The AIC interval for this distribution (upper bound:
6407.32) only barely overlapped with the intervals for the generalized
gamma (lower bound: 6405.52). Hence among the observational likelihoods
we considered, the multivariate log-normal was the most appropriate for
the North Sea Cod data and this particular process model. The
multivariate log-normal also had the lowest AIC for the full model for
North-East Arctic Haddock, whereas it was the multiplicative logistic
normal with total weight for Northern Shelf Haddock and the additive
logistic normal with total numbers for Blue Whiting. However, in these
cases the AIC intervals of the multivariate log-normal, the additive
logistic normals, and the multiplicative logistic normals all
overlapped. For the two haddock cases, the AIC interval of the
generalized gamma also overlapped with the AIC interval of the model
with the lowest AIC interval upper bound. We further found that the AIC
intervals for proportions-at-age models using total weight overlapped
with the corresponding model using total numbers-at-age, except for the
North Sea Cod where the total weight models had lower AIC intervals. In
addition, the intervals for the additive and multiplicative logistic
normals overlapped for all four data sets.

\begin{figure}[htbp]
\centering
\includegraphics{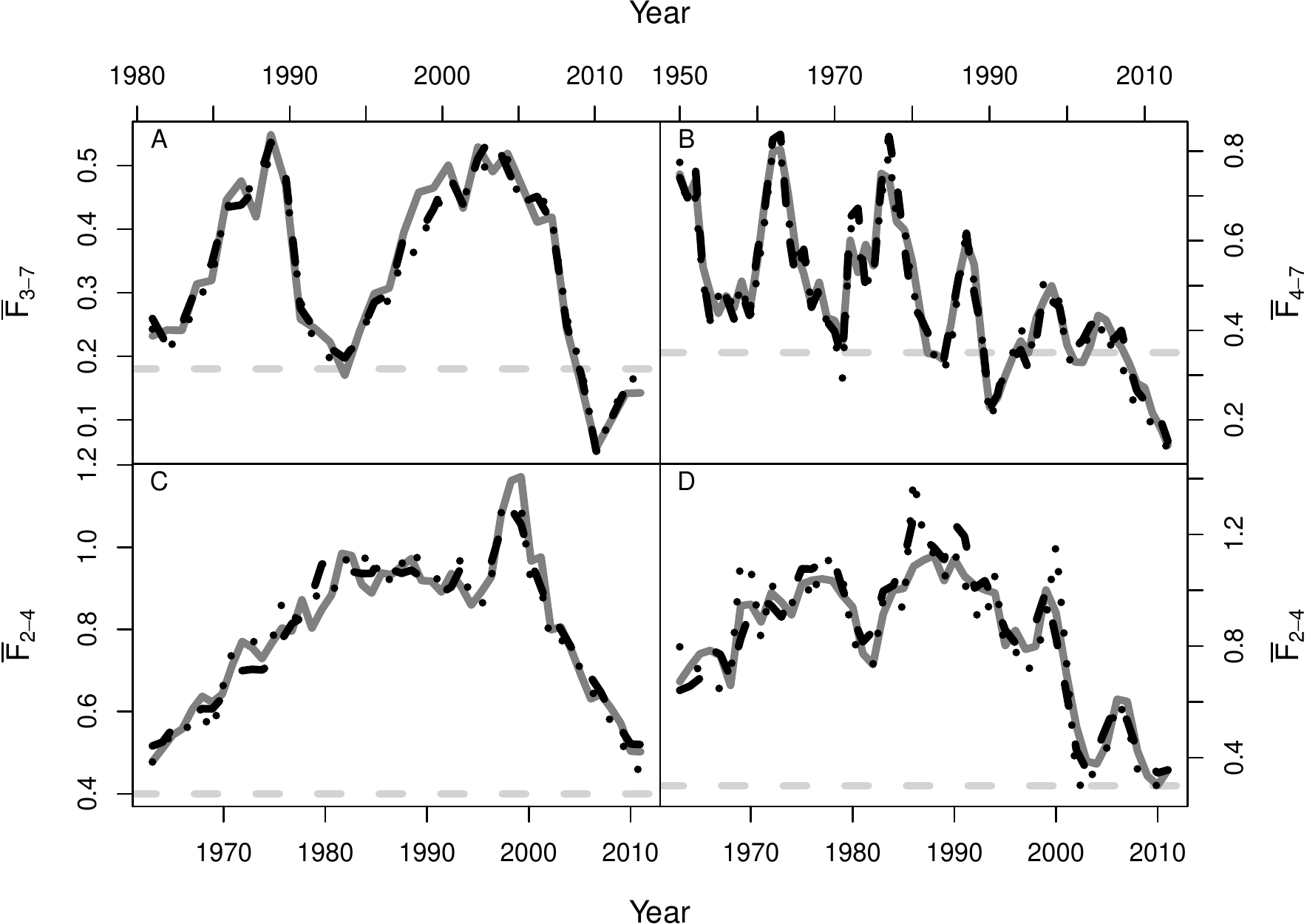}
\caption{Estimated average fishing mortality, \(\bar{F}\), for
Multivariate log-Normal (dark grey line), Multiplicative Logistic Normal
with log-Normal Weight (dashed black line), and Dirichlet with
log-Normal Weight (dotted black line) in the case studies: Blue Whiting
(A), North-East Arctic Haddock (B), North Sea Cod (C), and Northern
Shelf Haddock (D). Horizontal dashed grey lines show the management plan
reference point. Subscripts to \(\bar{F}\) indicates the ages the
average is over. All ages are weighed equally in the
average.\label{fig:fbar}}
\end{figure}

\begin{figure}[htbp]
\centering
\includegraphics{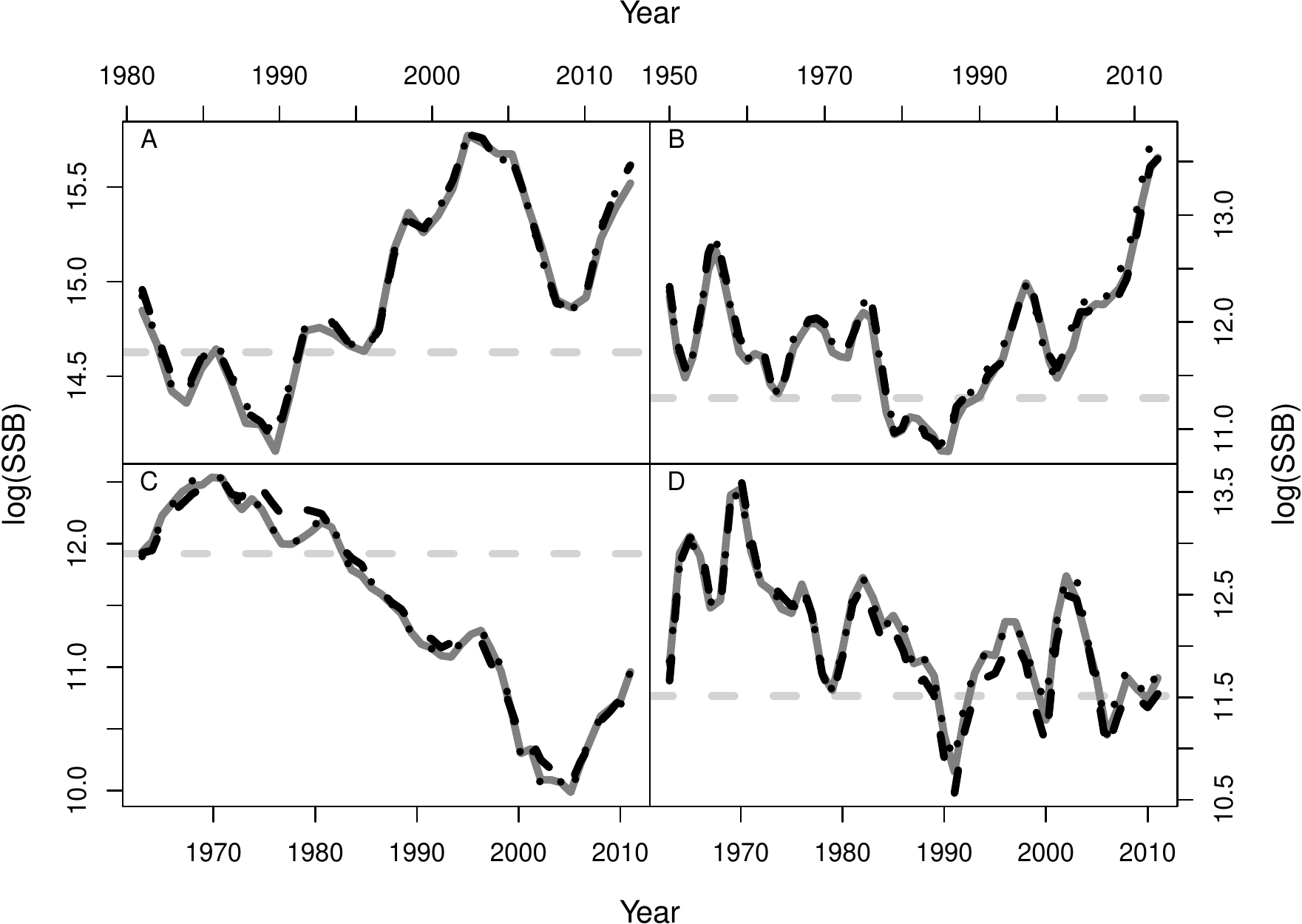}
\caption{Natural logarithm of estimated spawning stock biomass,
\(\log(SSB)\), for Multivariate log-Normal (dark grey line),
Multiplicative Logistic Normal with log-Normal Weight (dashed black
line), and Dirichlet with log-Normal Weight (dotted black line) in the
case studies: Blue Whiting (A), North-East Arctic Haddock (B), North Sea
Cod (C), and Northern Shelf Haddock (D). Horizontal dashed grey lines
show the management plan reference point.\label{fig:ssb}}
\end{figure}

Overall, the trends in estimated fishing mortality (\autoref{fig:fbar})
and spawning stock biomass (\autoref{fig:ssb}) were similar between the
models (See also supplementary material). However, there were noticeable
differences in single years. For Blue Whiting, the estimated fishing
mortality and spawning stock biomass for multivariate log-normal,
multiplicative logistic-normal, and Dirichlet distribution followed each
other closely. The largest difference in average fishing mortality
between the multiplicative logistic-normal and the multivariate
log-normal was 0.07 (16\%), and the difference in spawning stock biomass
was up to 12\%. In the North Sea Cod case, the multivariate log-normal
and the logistic normals had larger differences in fishing mortality and
spawning stock biomass. The largest difference in mortality was 0.12
(11.2\%), while the spawning stock biomass differed as much as 23\%. The
resulting confidence intervals also differed between the models. For
North Sea Cod the standard errors of the estimated average fishing
mortality were up to 76.6\% larger for the Dirichlet model, which had
the highest AIC, compared to the multivariate log-normal, which had the
lowest AIC. Although the trends were similar to the other models, the
spawning stock biomass was estimated to be 3.8 to 14.2 times higher for
the left truncated normal than for the other models
\footnote{Figure S31}. Likewise, the average fishing mortality was
estimated to be lower. For both North Sea Cod
\footnote{Figures S34, S35, S37, S38} and Northern Shelf
Haddock\footnote{Figures S47, S48, S50, S51}, the logistic normals
provided less volatile estimated time series of fishing mortality and
spawning stock biomass than other models. For these models the CVs for
commercial catch were estimated to be higher than for the other data
sets\footnote{Table S5}.

For Blue Whiting, North-East Arctic Haddock, and Northern Shelf Haddock,
estimated CVs were similar for the two logistic
normals\footnote{Table S5}. The CVs were estimated to be between 0.09
and 0.37. For North Sea Cod, the logistic normals with total numbers had
higher CVs (0.32 for commercial catch; 0.27 for survey) than the
logistic normals with total weight (0.15 for commercial catch; 0.20 for
survey). For the multivariate log-normal, the estimated marginal CVs
were estimated to be between 0.01 and 1.86\footnote{Table S1-S4}. The
CVs were typically estimated to be higher for the first and last ages.

\section{Discussion}\label{discussion}

When modelling highly aggregated stock assessment data, the optimal
observational likelihood to use can not be known a priori. We provide an
objective method for limiting the number of candidate models. By fitting
each model with the largest possible number of parameters, upper and
lower bounds for the lowest attainable AIC can be calculated. After
discarding models with AIC intervals that does not intersect with the
lowest interval, the remaining AIC intervals can be narrowed by
combining parameters until a single observational likelihood is left.
Once a final stock assessment model is found, model validation tools
such as residuals and retrospective analysis should be used. AIC was
used to determined the optimal observational likelihood, since the AIC
estimates the Kullback-Liebler divergence between the candidate model
and the true data generating system (Akaike 1974), i.e.~the information
lost by using the candidate model instead of the true data generating
system (Burnham and Anderson 2002). The Kullback-Liebler divergence can
not be used directly because the calculation requires full knowledge of
the true data generating system. Although the AIC was used here, other
criterias allowing multivariate data could be used instead.

Choosing the best possible observational likelihood for the data is
vital for the short-term management and conservation of fish stocks
(\autoref{fig:FBARlast}, \autoref{fig:SSBlast}). In 2012, the ICES
advice for Northern Shelf Haddock (ICES 2012), based on XSA (Shepherd
1999), suggested a 15\% increase in the total allowable catch. The
suggested increase was based on the difference between the estimated
average fishing mortality in the last year and the management reference
point. XSA can be seen as a special case of the univariate log-normal
(ICES 2010b) by fixing parameters in a suitable way. Here we saw that
the multivariate log-normal and multiplicative logistic normal were more
suitable for the Northern Shelf Haddock data than the univariate
log-normal, which in turn provides a better fit than the XSA. The
estimated average fishing mortality from both the multivariate
log-normal and the multiplicative logistic normal are above the
reference point. Hence, using one of these models would have suggested a
19\% decrease of the total allowable catch to get the fishing mortality
at a sustainable level (\autoref{fig:fbar}; panel D). Thus, the choice
of observational likelihood in an assessment model can have a
substantial effect on the advice brought forward to the fisheries
management system.

All models implemented in this study fall in to one of two categories;
either they are formulated for numbers-at-age or proportions-at-age with
total catch. Most assessment tools only consider one of these
categories. We have shown how to compare models between these two ways
of using the stock assessment data, and that models from both categories
can be suitable, depending on the specific stock. Besides the choice
between modelling numbers or proportions there are other differences
between the models in, e.g., tail probabilities and skewness, but there
are also more subtle differences. When we choose between the log-normal
distribution and the gamma distribution, we also choose between whether
the Baranov catch equation should model the median or the mean observed
catch. Some likelihoods can be re-parameterized to link the Baranov
catch equation to either the mean, median or modal observed catch.
Although subtle, this difference is important for prediction and
interpretation of the results. These choices can be compared objectively
by including them in the analysis.

Accounting for the correlations in the data is also important for more
reliable stock assessment models. In all four case studies, the
distributions with correlation parameters, in particular the
multivariate log-normal, performed well. However, the correlation
structure must be flexible enough to mimic the data, unlike the
Dirichlet distribution, which performed poorly, even compared to the
models assuming independence between ages. The Dirichlet distribution
arises as the distribution of proportions of gamma distributed numbers,
where the gamma distributions have equal scale parameters. Stock
assessment data is often believed to have constant CV (Cadigan and Myers
2001), which leads to a parameterization of the gamma distribution where
the scale parameters are not equal. Therefore the covariance structure
of the Dirichlet distribution does not match the appropriate structure
for the gamma distributions. This corresponds with previous findings,
that the additive logistic normal generally is more suitable than the
Dirichlet distribution in describing the correlation structure in stock
assessment data (Francis 2014). For simplicity we restricted ourselves
to the simple AR(1) structure for correlations in this study. The
correlation structure can easily be exchanged for other structures such
as a linear, AR(2), ARMA(\(p\),\(q\)), compound symmetric, or
unstructured covariance matrix (Pinheiro and Bates 2000, Francis 2014),
and some of these may be even more suitable than the AR(1) structure
(Berg and Nielsen 2016).

We only compared frequently used distributions in fisheries stock
assessments and close extensions of them, yet any conceivable
distribution can be used in the framework we presented, as long as they
are all comparable. Correlations between age classes could be introduced
in the univariate models through copulas or multivariate extensions such
as the multivariate t-distribution or multivariate gamma distributions.
Likewise, different ways of handling zero observations, such as zero
inflating the models, could be included. For simplicity, years with
missing data were removed to compare between univariate and multivariate
models. For multivariate numbers-at-age models, missing data may be
handled by finding the marginal distribution of the remaining ages. We
also restricted the study to use the same likelihood for both commercial
catches and surveys. This can be relaxed by including combinations of
models (such as Cadigan 2015), or if the survey index generation is
well-understood, a suitable observational likelihood may be derived a
priori. Further, the analyses were made conditional on the process
models specified in the assessments from which the data was collected. A
similar analysis could be made to choose the most appropriate process
model conditional on the observational model, or the analyses could be
combined. This may influence the specific choice of observational
likelihood, as the observational model can, somewhat, compensate for
misspecification in the process model and vice versa. Finally, these
methods could just as well be used for, e.g., length-based models.

Statistical assessment modelling involves a choice of observational
likelihood, and current practice is often to make this choice
arbitrarily and subjectively. This applies particularly to the choice of
whether the data inputs should be numbers-at-age or proportions-at-age
along with total catch in numbers. Here, we have provided methods for an
objective choice, by correcting the AICs so that they can be compared
between these two families, and we have outlined a computationally
efficient method for choosing between families of distributions, by
bounding the AICs, which avoids elaborate hypothesis testing within each
family, and demonstrated that the best fitting family depends on the
particular case. These results will allow stock assessment modellers to
choose objectively between these representations of uncertainty on
observations, thereby improving model fit and ultimately allowing more
accurate assessments.

\section{Acknowledgements}\label{acknowledgements}

The authors wish to thank Casper W. Berg and three anonymous reviewers
for their valuable comments to improve the presentation of this
manuscript.

\section*{References}\label{references}
\addcontentsline{toc}{section}{References}

\hyperdef{}{ref-aanes2003a}{\label{ref-aanes2003a}}
Aanes, S., and Pennington, M. 2003. On estimating the age composition of
the commercial catch of Northeast Arctic cod from a sample of clusters.
ICES Journal of Marine Science: Journal du Conseil \textbf{60}(2):
297--303.
\href{http://doi.org/10.1016/S1054-3139(03)00008-0}{doi: 10.1016/S1054-3139(03)00008-0}.

\hyperdef{}{ref-aitchison2003a}{\label{ref-aitchison2003a}}
Aitchison, J. 2003. The Statistical Analysis of Compositional Data. The
Blackburn Press, Caldwell, N.J.

\hyperdef{}{ref-akaike1974a}{\label{ref-akaike1974a}}
Akaike, H. 1974. A new look at the statistical model identification.
Automatic Control, IEEE Transactions on \textbf{19}(6): 716--723.
\href{http://doi.org/10.1109/TAC.1974.1100705}{doi: 10.1109/TAC.1974.1100705}.

\hyperdef{}{ref-berg2016a}{\label{ref-berg2016a}}
Berg, C.W., and Nielsen, A. 2016. Accounting for correlated observations
in an age-based state-space stock assessment model. ICES Journal of
Marine Science: Journal du Conseil \textbf{73}(7): 1788--1797.
\href{http://doi.org/10.1093/icesjms/fsw046}{doi: 10.1093/icesjms/fsw046}.

\hyperdef{}{ref-burnham2002a}{\label{ref-burnham2002a}}
Burnham, K.P., and Anderson, D.R. 2002. Model Selection and Multimodel
Inference: A Practical Information-Theoretic Approach. \emph{In} 2nd
editions. Springer.

\hyperdef{}{ref-cadigan2015a}{\label{ref-cadigan2015a}}
Cadigan, N.G. 2015. A state-space stock assessment model for northern
cod, including under-reported catches and variable natural mortality
rates. Canadian Journal of Fisheries and Aquatic Sciences.
\href{http://doi.org/10.1139/cjfas-2015-0047}{doi: 10.1139/cjfas-2015-0047}.
In press.

\hyperdef{}{ref-cadigan2001a}{\label{ref-cadigan2001a}}
Cadigan, N.G., and Myers, R.A. 2001. A comparison of gamma and lognormal
maximum likelihood estimators in a sequential population analysis.
Canadian Journal of Fisheries and Aquatic Sciences \textbf{58}(3):
560--567. \href{http://doi.org/10.1139/f01-003}{doi: 10.1139/f01-003}.

\hyperdef{}{ref-cook2013a}{\label{ref-cook2013a}}
Cook, R. 2013. A fish stock assessment model using survey data when
estimates of catch are unreliable. Fisheries Research \textbf{143}:
1--11.
\href{http://doi.org/http://dx.doi.org/10.1016/j.fishres.2013.01.003}{doi: http://dx.doi.org/10.1016/j.fishres.2013.01.003}.


\hyperdef{}{ref-crone1998a}{\label{ref-crone1998a}}
Crone, P., and Sampson, D. 1998. Evaluation of assumed error structure
in stock assessment models that use sample estimates of age composition.
\emph{In} Fishery Stock Assessment Models. \emph{Edited by} Q. Funk F.
Alaska Sea Grant College Program Report No. AK-SG-98-01, University of
Alaska Fairbanks. pp. 355--370.

\hyperdef{}{ref-farewell1977a}{\label{ref-farewell1977a}}
Farewell, V.T., and Prentice, R.L. 1977. A Study of Distributional Shape
in Life Testing. Technometrics \textbf{19}(1): 69--75.

\hyperdef{}{ref-fournier1982a}{\label{ref-fournier1982a}}
Fournier, D., and Archibald, C.P. 1982. A General Theory for Analyzing
Catch at Age Data. Canadian Journal of Fisheries and Aquatic Sciences
\textbf{39}(8): 1195--1207.
\href{http://doi.org/10.1139/f82-157}{doi: 10.1139/f82-157}.

\hyperdef{}{ref-francis2011a}{\label{ref-francis2011a}}
Francis, R.I.C.C. 2011. Data weighting in statistical fisheries stock
assessment models. Canadian Journal of Fisheries and Aquatic Sciences
\textbf{68}(6): 1124--1138.
\href{http://doi.org/10.1139/f2011-025}{doi: 10.1139/f2011-025}.

\hyperdef{}{ref-francis2014a}{\label{ref-francis2014a}}
Francis, R.I.C.C. 2014. Replacing the multinomial in stock assessment
models: A first step. Fisheries Research \textbf{151}(0): 70--84.
\href{http://doi.org/http://dx.doi.org/10.1016/j.fishres.2013.12.015}{doi: http://dx.doi.org/10.1016/j.fishres.2013.12.015}.

\hyperdef{}{ref-fryer2002a}{\label{ref-fryer2002a}}
Fryer, R. 2002. TSA: is it the way? \emph{In} Report of Working Group on
Methods of Fish Stock Assessment, Dec. 2001. ICES CM 2002/D:01. pp.
86--93.

\hyperdef{}{ref-gudmundsson1994a}{\label{ref-gudmundsson1994a}}
Gudmundsson, G. 1994. Time Series Analysis of Catch-At-Age Observations.
Journal of the Royal Statistical Society. Series C (Applied Statistics)
\textbf{43}(1): pp. 117--126.

\hyperdef{}{ref-hulson2011a}{\label{ref-hulson2011a}}
Hulson, P.-J.F., Hanselman, D.H., and Quinn, T.J. 2011. Effects of
process and observation errors on effective sample size of fishery and
survey age and length composition using variance ratio and likelihood
methods. ICES Journal of Marine Science: Journal du Conseil
\textbf{68}(7): 1548--1557.
\href{http://doi.org/10.1093/icesjms/fsr102}{doi: 10.1093/icesjms/fsr102}.

\hyperdef{}{ref-hulson2012a}{\label{ref-hulson2012a}}
Hulson, P.-J.F., Hanselman, D.H., and Quinn, T.J. 2012. Determining
effective sample size in integrated age-structured assessment models.
ICES Journal of Marine Science: Journal du Conseil \textbf{69}(2):
281--292.
\href{http://doi.org/10.1093/icesjms/fsr189}{doi: 10.1093/icesjms/fsr189}.

\hyperdef{}{ref-ices2010a}{\label{ref-ices2010a}}
ICES. 2010a. Report of the Workshop on Reviews of Recent Advances in
Stock Assessment Models Worldwide: ``Around the World in AD Models''
(WKADSAM), 27 September - 1 October 2010, Nantes, France. ICES CM
2010/SSGSUE:10.

\hyperdef{}{ref-ices2010b}{\label{ref-ices2010b}}
ICES. 2010b. Report of the Working Group on Methods of Fish Stock
Assessment (WGMG), 20--29 October 2009, Nantes, France. ICES CM
2009/RMC:12.

\hyperdef{}{ref-ices2012a}{\label{ref-ices2012a}}
ICES. 2012. Report of the ICES Advisory Committee 2012. ICES Advice
2012, Book 6.

\hyperdef{}{ref-ices2014b}{\label{ref-ices2014b}}
ICES. 2014a. Report of the ICES Advisory Committee 2014. ICES Advice
2014, Book 9.

\hyperdef{}{ref-ices2014a}{\label{ref-ices2014a}}
ICES. 2014b. Report of the ICES Advisory Committee 2014. ICES Advice
2014, Book 3.

\hyperdef{}{ref-kristensen2015a}{\label{ref-kristensen2015a}}
Kristensen, K., Nielsen, A., Berg, C.W., and Bell, H.S.B. 2016. TMB:
Automatic Differentiation and Laplace Approximation. Journal of
Statistical Software \textbf{70}(1): 1--21.
\href{http://doi.org/10.18637/jss.v070.i05}{doi: 10.18637/jss.v070.i05}.

\hyperdef{}{ref-maunder2011a}{\label{ref-maunder2011a}}
Maunder, M.N. 2011. Review and evaluation of likelihood functions for
composition data in stock-assessment models: Estimating the effective
sample size. Fisheries Research \textbf{109}(2--3): 311--319.
\href{http://doi.org/10.1016/j.fishres.2011.02.018}{doi: 10.1016/j.fishres.2011.02.018}.

\hyperdef{}{ref-mcallister1997a}{\label{ref-mcallister1997a}}
McAllister, M.K., and Ianelli, J.N. 1997. Bayesian stock assessment
using catch-age data and the sampling - importance resampling algorithm.
Canadian Journal of Fisheries and Aquatic Sciences \textbf{54}(2):
284--300. \href{http://doi.org/10.1139/f96-285}{doi: 10.1139/f96-285}.

\hyperdef{}{ref-methot2013a}{\label{ref-methot2013a}}
Methot Jr., R.D., and Wetzel, C.R. 2013. Stock synthesis: A biological
and statistical framework for fish stock assessment and fishery
management. Fisheries Research \textbf{142}: 86--99.
\href{http://doi.org/http://dx.doi.org/10.1016/j.fishres.2012.10.012}{doi: http://dx.doi.org/10.1016/j.fishres.2012.10.012}.

\hyperdef{}{ref-myers1995a}{\label{ref-myers1995a}}
Myers, R.A., and Cadigan, N.G. 1995. Statistical analysis of
catch-at-age data with correlated errors. Canadian Journal of Fisheries
and Aquatic Sciences \textbf{52}(6): 1265--1273.
\href{http://doi.org/10.1139/f95-123}{doi: 10.1139/f95-123}.

\hyperdef{}{ref-nielsen2014a}{\label{ref-nielsen2014a}}
Nielsen, A., and Berg, C. 2014. Estimation of time-varying selectivity
in stock assessments using state-space models. Fisheries Research
\textbf{158}: 96--101.
\href{http://doi.org/10.1016/j.fishres.2014.01.014}{doi: 10.1016/j.fishres.2014.01.014}.

\hyperdef{}{ref-pinheiro2000a}{\label{ref-pinheiro2000a}}
Pinheiro, J.C., and Bates, D.M. 2000. Mixed-effects models in S and
S-Plus. Springer, New York, USA.

\hyperdef{}{ref-prentice1974a}{\label{ref-prentice1974a}}
Prentice, R.L. 1974. A Log Gamma Model and Its Maximum Likelihood
Estimation. Biometrika \textbf{61}(3): 539--544.

\hyperdef{}{ref-shepherd1999a}{\label{ref-shepherd1999a}}
Shepherd, J.G. 1999. Extended survivors analysis: An improved method for
the analysis of catch-at-age data and abundance indices. ICES Journal of
Marine Science: Journal du Conseil \textbf{56}(5): 584--591.
\href{http://doi.org/10.1006/jmsc.1999.0498}{doi: 10.1006/jmsc.1999.0498}.

\hyperdef{}{ref-stacey1962a}{\label{ref-stacey1962a}}
Stacy, E.W. 1962. A Generalization of the Gamma Distribution. The Annals
of Mathematical Statistics \textbf{33}(3): 1187--1192. Institute of
Mathematical Statistics.

\hyperdef{}{ref-williams1998a}{\label{ref-williams1998a}}
Williams, E.H., and Quinn, T.J. 1998. A Parametric Bootstrap of
Catch-Age Compositions Using the Dirichlet Distribution. \emph{In}
Fishery Stock Assessment Models. \emph{Edited by} Q. Funk F. Alaska Sea
Grant College Program Report No. AK-SG-98-01, University of Alaska
Fairbanks. pp. 371--384.

\hyperdef{}{ref-williams2015a}{\label{ref-williams2015a}}
Williams, E. H., and Shertzer, K.W. 2015. Technical documentation of the
Beaufort Assessment Model (BAM). NOAA Technical Memorandum, U.S.
Department of Commerce.
\href{http://doi.org/10.7289/V57M05W6}{doi: 10.7289/V57M05W6}.

\section{Appendix A}\label{appendix-a}

\subsection{Process model}\label{process-model-1}

The process model is identical to Nielsen and Berg (2014). For a model
including age groups from 1 to \(A^+\) (where age group \(A^+\) contains
all ages from \(A\) and up), the fishing mortality is modelled by a
multivariate random walk, where the oldest modelled ages may be grouped
together, indicated by using \(A^*\) rather than \(A^+\). Let
\(F_y = (F_{1,y},F_{2,y},\ldots,F_{A^*,y})^T\) be a vector of age
specific fishing mortalities in year \(y\). Then \[
\log F_y = \log F_{y-1} + \epsilon_y,
\] where \(\epsilon_y \sim N(0,\Sigma)\). The covariance matrix,
\(\Sigma\), is parameterized by and AR(1) structure,
\(\Sigma_{i,j} = \rho^{|i-j|}\sigma_i\sigma_j\).

Given the fishing mortalities, the population is modelled by an
exponential decay model, \[
\begin{aligned}
\log N_{1,y} &= \log\left(R\left(w_{1,y-1},\ldots,w_{A^+,y-1},p_{1,y-1},\ldots,p_{A^+,y-1},N_{1,y-1},\ldots,\ldots,N_{A^+,y-1}\right)\right) + \eta_{1,y}, \\
\log N_{a,y} &= \log N_{a-1,y-1} - F_{a-1,y-1} - M_{a-1,y-1} + \eta_{a,y}, 2\leq a < A^+, \\
\log N_{A^+,y} &= \log \left( N_{A^+-1,y-1}e^{-F_{A^+-1,y-1}-M_{A^+-1,y-1}} + N_{A^+,y-1}e^{-F_{A^+,y-1}-M_{A^+,y-1}} \right) + \eta_{A^+,y},
\end{aligned}
\] where all error terms are assumed independent normal distributed. The
natural mortalities (\(M_{a,y}\)), the age specific weight in stock
(\(w_{a,y}\)), and the proportion mature (\(p_{a,y}\)) are all assumed
to be known. The function \(R\) describes the relationship between
recruitment and spawning population. For North Sea Cod, \(R\) was
modelled by a Beverton-Holt curve, \[
R\left(w_{1,y-1},\ldots,w_{A^+,y-1},p_{1,y-1},\ldots,p_{A^+,y-1},N_{1,y-1},\ldots,N_{A^+,y-1}\right) = \frac{a \cdot SSB_{y-1}}{1+b \cdot SSB_{y-1}},\quad a,b>0
\] with \(SSB_{y} = \sum_{a=1}^{A^+} p_{a,y}w_{a,y}N_{a,y}\), whereas
for the other stocks, \(R\) was modelled by a random walk \[
R\left(w_{1,y-1},\ldots,w_{A^+,y-1},p_{1,y-1},\ldots,p_{A^+,y-1},N_{1,y-1},\ldots,N_{A^+,y-1}\right) = N_{1,y-1}.
\]

\subsection{Observational models}\label{observational-models-1}

\subsubsection{Univariate numbers-at-age
models}\label{univariate-numbers-at-age-models}

We consider six univariate observational models for numbers-at-age.
Their densities are listed below for each age each year. The joint
density for the vector of catches-at-age each year is the product of the
age-wise densities. Unless otherwise noted, \(x > 0\) is the observed
catch (or survey index) at a given age, \(\mu > 0\) is the calculated
catch at that age (or survey index) based on the Baranov catch equation
(or proportional to the total abundance), \(\sigma > 0\) is a scale
matrix, and \(\tau\in\mathbb{R}\) is a shape parameter.

\paragraph{Log-normal distribution}\label{log-normal-distribution}

\[
f_1(x;\mu,\sigma) = \frac{1}{\sqrt{2\pi\sigma^2}}\exp\left(\frac{-(\log(x)-\log(\mu))^2}{2\sigma^2} \right)x^{-1}
\] The mean of the log-normal distribution is \(\mu\exp(\sigma^2/2)\),
the median is \(\mu\) and the variance is
\(\left(\exp\left(\sigma^2\right)-1\right)\exp\left(2\log(\mu)+\sigma^2\right)\).

\paragraph{Gamma distribution}\label{gamma-distribution}

\[
f_2(x;\mu,\sigma) = \frac{1}{\Gamma(\sigma)\left(\frac{\mu}{\sigma}\right)^{\sigma}}x^{\sigma-1} \exp\left(-x\sigma/\mu\right)
\]

The gamma distribution has mean \(\mu\) and variance
\(\frac{\mu^2}{\sigma}\).

\paragraph{Generalized gamma
distribution}\label{generalized-gamma-distribution}

\[
f_3(x;\mu,\sigma,\tau) = \left\{\begin{array}{ll} |\tau|(\tau^{-2})^{\tau^{-2}} \exp\left( \tau^{-2}\left(\tau\frac{\log(x)-\log(\mu)}{\sigma}-\exp\left( \tau\frac{\log(x)-\log(\mu)}{\sigma} \right)  \right)  \right)/(\sigma x \Gamma(\tau^{-2})) & \tau \neq 0 \\ (2\pi)^{-1/2}\exp\left(-\frac{\left(\log(x)-\log(\mu)\right)^2}{2\sigma^2}  \right)(\sigma x)^{-1} & \tau = 0 \end{array} \right.
\]

Note that \(f_3(x;\mu,\sigma,0) = f_1(x;\mu,\sigma)\), and
\(f_3(x;\mu,\sigma,\sigma)=f_2(x;\mu,\sigma^{-2})\) for \(\sigma > 0\)
(Cox et al. 2007). The mean of the generalized gamma is \[
\begin{array}{cc}
{\frac {{\mu} \left( {\tau}^{-2} \right) ^{{\tau}^{-2}}}{
\Gamma  \left( {\tau}^{-2} \right) } \left( {\tau}^{2} \right) ^{{
\frac {\sigma\,\tau+1}{{\tau}^{2}}}}\Gamma  \left( {\frac {\sigma\,
\tau+1}{{\tau}^{2}}} \right) } & \tau < 0 \\
\mu\exp(\sigma^2/2) & \tau = 0 \\
{\frac {{\mu}}{\Gamma  \left( {\tau}^{-2} \right) }{\tau}^{2
\,{\frac {\sigma}{\tau}}}\Gamma  \left( {\frac {\sigma\,\tau+1}{{\tau}
^{2}}} \right) } & \tau > 0
\end{array}
\] and the variance is \[
\begin{array}{cc}
-{\frac {{\mu^2}}{ \left( \Gamma  \left( {\tau}^{-2}
 \right)  \right) ^{2}} \left(  \left( {\tau}^{-2} \right) ^{2\,{\tau}
^{-2}} \left( {\tau}^{2} \right) ^{2\,{\frac {\sigma\,\tau+1}{{\tau}^{
2}}}} \left( \Gamma  \left( {\frac {\sigma\,\tau+1}{{\tau}^{2}}}
 \right)  \right) ^{2}- \left( {\tau}^{2} \right) ^{{\frac {2\,\sigma
\,\tau+1}{{\tau}^{2}}}} \left( {\tau}^{-2} \right) ^{{\tau}^{-2}}
\Gamma  \left( {\frac {2\,\sigma\,\tau+1}{{\tau}^{2}}} \right) \Gamma 
 \left( {\tau}^{-2} \right)  \right) }
 & \tau < 0 \\
\left(\exp\left(\sigma^2\right)-1\right)\exp\left(2\log(\mu)+\sigma^2\right) & \tau = 0 \\
{\frac {{\mu^2}}{ \left( \Gamma  \left( {\tau}^{-2}
 \right)  \right) ^{2}}{\tau}^{4\,{\frac {\sigma}{\tau}}} \left( 
\Gamma  \left( {\frac {2\,\sigma\,\tau+1}{{\tau}^{2}}} \right) \Gamma 
 \left( {\tau}^{-2} \right) - \left( \Gamma  \left( {\frac {\sigma\,
\tau+1}{{\tau}^{2}}} \right)  \right) ^{2} \right) }
 & \tau > 0
\end{array}
\]

\paragraph{Normal distribution}\label{normal-distribution}

\[
f_4(x;\mu,\sigma) = \frac{1}{\sqrt{2\pi(\mu\sigma)^2}} \exp\left(\frac{-(x-\mu)^2}{2(\mu\sigma)^2} \right)
\]

For the normal distribution \(x\in\mathbb{R}\) and \(\mu\in\mathbb{R}\).
The mean is \(\mu\) and the variance is \(\mu^2\sigma^2\)

\paragraph{Truncated normal
distribution}\label{truncated-normal-distribution}

\[
f_5(x;\mu,\sigma) = \frac{f_4(x;\mu,\sigma)1_{x\geq 0}(x)}{1-\int_{-\infty}^0 f_4(y;\mu,\sigma) dy}
\]

For the truncated normal distribution \(x \geq 0\) and \(\mu \geq 0\).
The mean of the distribution is \[
\mu + \mu\sigma \frac{f_4(0;\mu,\sigma)}{1-\int_{-\infty}^0 f_4(y;\mu,\sigma) dy}
\] and the variance is \[
\mu^2\sigma^2 - \mu^2\sigma^2 \frac{f_4(0;\mu,\sigma)^2}{\left(1-\int_{-\infty}^0 f_4(y;\mu,\sigma) dy\right)^2}
\]

\paragraph{Student's t-distribution on
log-scale}\label{students-t-distribution-on-log-scale}

\[
f_6(x;\mu,\sigma,\tau) = \frac{\Gamma\left(\frac{\tau+1}{2}\right)}{\sigma x \sqrt{\tau\pi}\Gamma{\left(\frac{\tau}{2}\right)}}\left(1+\frac{((\log(x)-\log(\mu))/\sigma)^2}{\tau}\right)^{-(\tau+1)/2}
\]

For the Student's t-distribution on log-scale, \(\tau > 0\) is the
degrees of freedom. As \(\tau \rightarrow \infty\) the distribution
converges to a log-normal distribution. The distribution does not have
mean and variance.

\subsubsection{Multivariate numbers-at-age
models}\label{multivariate-numbers-at-age-models}

We consider one multivariate observational model for numbers-at-age. In
the density listed below, \(A\) is the number of ages,
\(\mathbf{x} > \mathbf{0}\) is the observed vector of catches (or survey
indices), \(\mathbf{\mu} > \mathbf{0}\) is the calculated catches (or
survey indices) based on the Baranov catch equation (or proportional to
the total abundance), and \(\Sigma\) is an \(A \times A\) symmetric
positive-definite scale matrix.

\paragraph{Multivariate log-normal
distribution}\label{multivariate-log-normal-distribution}

\[
f_7(\mathbf{x};\mathbf{\mu},\mathbf{\Sigma}) = (2\pi)^{-A/2} |\Sigma|^{-1/2}\exp\left(-\frac{1}{2}(\log(\mathbf{x})-\log(\mathbf{\mu}))^T\mathbf{\Sigma}^{-1}(\log(\mathbf{x})-\log(\mathbf{\mu})) \right) (x_1\cdots x_k)^{-1}
\]

For the multivariate log-normal distribution, the scale matrix is the
covariance matrix of the logarithm of the observations. When the scale
matrix is diagonal, the distribution reduces to univariate log-normals.
The marginal means are \(\mu_i\exp\left(\frac{1}{2}\Sigma_{ii}\right)\),
and the variance/covariance is
\(\mu_i\mu_j\exp\left(\frac{1}{2}(\Sigma_{ii}+\Sigma_{jj})\right)(\exp(\Sigma_{ij})-1)\).

\subsubsection{Proportions-at-age
models}\label{proportions-at-age-models}

We consider three multivariate observational model for
proportions-at-age. In the density listed below, \(A\) is the number of
ages, \(\mathbf{x} > \mathbf{0}\) with \(\sum_{i=1}^{A} x_i = 1\) is a
vector of \(A\) observed catch proportions (or survey proportions),
\(\mathbf{\mu} > \mathbf{0}\) with \(\sum_{i=1}^{A} \mu_i = 1\) is a
vector of \(A\) calculated catch proportions (or survey proportions)
based on the Baranov catch equation (or proportional to the total
abundance), and \(\mathbf{\Sigma}\) is an \(A-1 \times A-1\) symmetric
positive-definite scale matrix. A vector with subscript \(-A\) denotes
the vector without the \(A\)th element.

\paragraph{Additive logistic-normal
distribution}\label{additive-logistic-normal-distribution}

\[
f_8(\mathbf{x};\mathbf{\mu},\mathbf{\Sigma}) = (2\pi)^{-(A-1)/2} |\Sigma|^{-1/2}\exp\left(-\frac{1}{2}\left(\operatorname{\alpha}(\mathbf{x})-\operatorname{\alpha}(\mathbf{\mu})\right)^T\mathbf{\Sigma}^{-1}\left(\operatorname{\alpha}(\mathbf{x})-\operatorname{\alpha}(\mathbf{\mu}) \right)\right) (x_1\cdots x_A)^{-1}
\]

Here, \(\operatorname{\alpha}\) is the additive logratio transformation
\(\operatorname{\alpha}(\mathbf{x}) = \log\left(\frac{\mathbf{x}_{-A}}{x_A}\right)\).
The scale matrix is the covariance of the additive logratio transformed
observations. Note that if a numbers-at-age vector \(\mathbf{y}\)
follows a multivariate log-normal, then the proportions
\(\mathbf{y}/\sum_i y_i\) follows an additive logistic-normal
distribution (Aitchison 2003). The mean and variance does not have
simple forms (Aitchison 2003).

\paragraph{Multiplicative logistic-normal
distribution}\label{multiplicative-logistic-normal-distribution}

\[
f_9(\mathbf{x};\mathbf{\mu},\mathbf{\Sigma}) = (2\pi)^{-(A-1)/2} |\Sigma|^{-1/2}\exp\left(-\frac{1}{2}\left(\operatorname{m}(\mathbf{x})-\operatorname{m}(\mathbf{\mu})\right)^T\mathbf{\Sigma}^{-1}\left(\operatorname{m}(\mathbf{x})-\operatorname{m}(\mathbf{\mu}) \right)\right) (x_1\cdots x_A)^{-1}
\]

Here, \(\operatorname{m}\) is the multiplicative logratio transformation

\[
\operatorname{m}(\mathbf{x}) = \left(\log\left(\frac{x_1}{1-x_1} \right), \ldots, \log\left(\frac{x_{A-1}}{1-x_1-\cdots - x_{A-1}} \right) \right).
\]

The scale matrix is the covariance of the multiplicative logratio
transformed observations. The mean and variance does not have simple
forms (Aitchison 2003).

\paragraph{Dirichlet distribution}\label{dirichlet-distribution}

\[
f_{10}(\mathbf{x};\mathbf{\mu}, \sigma) = \frac{\Gamma\left(\sum_{i=1}^A \sigma\mu_i \right)}{\prod_{i=1}^A \Gamma(\sigma\mu_i)} \prod_{i=1}^A x_i ^ {\sigma\mu_i-1}
\]

For the Dirichlet distribution, \(\sigma > 0\) is a shape parameter. The
marginal means of the Dirichlet distribution are \(\mu_i\), the
variances are \(\frac{\mu_i-\mu_i^2}{\sigma+1}\), and the covariances
are \(\frac{-\mu_i\mu_j}{\sigma+1}\). The Dirichlet distribution is
related to the gamma distribution, since if each element of a
numbers-at-age vector follows a gamma distribution where the scale
parameters in the usual parameterization are equal, then the vector of
proportions \(\mathbf{y}/\sum_i y_i\) follow a Dirichlet distribution.
Note that \(f_2\) does not have the same scale parameters for all ages,
as they depend on the mean value.

\subsection*{References}

\hyperdef{}{ref-aitchison2003a}{\label{ref-aitchison2003a}}
Aitchison, J. 2003. The Statistical Analysis of Compositional Data. The
Blackburn Press, Caldwell, N.J.

\hyperdef{}{ref-cox2007a}{\label{ref-cox2007a}}
Cox, C., Chu, H., Schneider, M.F., and Muñoz, A. 2007. Parametric
survival analysis and taxonomy of hazard functions for the generalized
gamma distribution. Statistics in Medicine \textbf{26}(23): 4352--4374. \href{http://doi.org/10.1002/sim.2836}{doi: 10.1002/sim.2836}.

\hyperdef{}{ref-nielsen2014a}{\label{ref-nielsen2014a}}
Nielsen, A., and Berg, C. 2014. Estimation of time-varying selectivity
in stock assessments using state-space models. Fisheries Research
\textbf{158}: 96--101.
\href{http://doi.org/10.1016/j.fishres.2014.01.014}{doi: 10.1016/j.fishres.2014.01.014}.

\section{Appendix B}\label{appendix-b}

\subsection{Transformation of densities for proportions-at-age
models}\label{transformation-of-densities-for-proportions-at-age-models}

To compare the AIC of models for numbers-at-age with models for
proportions-at-age, the data must be on the same scale. We note that we
can transform the numbers-at-age data to proportions-at-age with total
catch in numbers by the function

\[
g\left(\left(x_1,\ldots,x_A \right)^T\right) = \left(\frac{x_1}{\sum_{i=1}^A x_i},\ldots,\frac{x_{A-1}}{\sum_{i=1}^A x_i}, \sum_{i=1}^A x_i \right)^T,
\]

with inverse function

\[
h\left(\left(y_1, \ldots, y_{A-1}, y_{total} \right)^T\right) = \left( y_1\cdot y_{total}, \ldots, y_{A-1}\cdot y_{total}, (1-\sum_{i=1}^{A-1} y_i)\cdot y_{total} \right)^T.
\]

Hence, if \(Y\) is a random vector of proportions-at-age with total
catch in numbers with density \(f\), then the corresponding
numbers-at-age is \(X=h(Y)\). By a change of variable,

\[
\begin{aligned}
P(X \in B) &= P(Y \in g(B)) \\
&= \int_{g(B)} f(y) dy \\
&= \int_A f(g(x)) |\det(Dg)| dy,
\end{aligned}
\]

the density for the numbers-at-age data is \(f(g(y)) |\det(Dg)|\).
Hence, to compare the AIC between the (natively) numbers-at-age and
proportions-at-age models, the log-likelihoods of the proportions-at-age
models must be corrected by the logarithm of the absolute determinant of
the Jacobian of \(g\), \(\log|\det(Dg)|\). The entries of \((Dg)\) are
\((Dg)_{A,j} = 1\) for all \(j\),
\((Dg)_{i,i} = \frac{1}{\sum_{k=1}^A x_k} - \frac{x_i}{\left(\sum_{k=1}^A x_k\right)^2}\)
for all \(i < A\) and
\((Dg)_{i,j} = \frac{-x_i}{\left(\sum_{k=1}^A x_k\right)^2}\) otherwise.
If the total is in weight, where the weight-at-age is assumed to be
known, then \(g\) is adjusted to

\[
g\left(\left(x_1, \ldots, x_A \right)^T\right) = \left(\frac{x_1}{\sum_{i=1}^A x_i}, \ldots, \frac{x_{A-1}}{\sum_{i=1}^A x_i}, \sum_{i=1}^A w_ix_i \right)^T
\]

\end{document}


\maketitle

\renewcommand{\thefigure}{S\arabic{figure}}
\renewcommand{\thetable}{S\arabic{table}}
\renewcommand{\figurename}{Fig.}

\newpage

\section{Estimated coefficients of
variation}\label{estimated-coefficients-of-variation}

\begin{table}[h]
\caption{Estimated marginal coefficients of variation for models $M_1$, $M_2$, $M_4$, $M_7$ for  Blue Whiting. Standard errors of the estimates (in parentheses) are derived by the delta method. $M_5$ and $M_6$ are not included as they do not have constant CV. For $M_3$, $\tau$ is set to 0 in the calculation for $|\hat{\tau}| < 0.1$ (indicated by *) due to numerical instability of the calculations.}
\label{tab:data}
\begin{tabular}{lclllll}
\toprule
Fleet & Age & $M_{1}$ & $M_{2}$ & $M_{3}$ & $M_{4}$ & $M_{7}$ \tabularnewline 
\midrule
Commercial & 1 & 0.40 (0.08) & 0.37 (0.07) & 0.35 (0.07) & 0.32 (0.06) & 0.50 (0.11) \tabularnewline 
Commercial & 2 & 0.25 (0.06) & 0.24 (0.06) & 0.23 (0.02)* & 0.21 (0.05) & 0.38 (0.09) \tabularnewline 
Commercial & 3 & 0.07 (0.11) & 0.06 (0.13) & 0.01 (0.15)* & 0.00 (0.01) & 0.15 (0.07) \tabularnewline 
Commercial & 4 & 0.18 (0.05) & 0.18 (0.05) & 0.20 (0.01)* & 0.17 (0.05) & 0.01 (0.09) \tabularnewline 
Commercial & 5 & 0.14 (0.06) & 0.14 (0.06) & 0.11 (0.04)* & 0.12 (0.06) & 0.23 (0.07) \tabularnewline 
Commercial & 6 & 0.12 (0.05) & 0.12 (0.05) & 0.13 (0.02)* & 0.12 (0.05) & 0.25 (0.05) \tabularnewline 
Commercial & 7 & 0.06 (0.10) & 0.06 (0.11) & 0.07 (0.12)* & 0.00 (0.09) & 0.23 (0.05) \tabularnewline 
Commercial & 8 & 0.05 (0.16) & 0.05 (0.16) & 0.04 (0.71)* & 0.00 (0.03) & 0.28 (0.06) \tabularnewline 
Commercial & 9 & 0.31 (0.08) & 0.29 (0.07) & 0.31 (0.02)* & 0.26 (0.06) & 0.42 (0.08) \tabularnewline 
Commercial & 10 & 0.38 (0.08) & 0.37 (0.07) &  -  ( - ) & 0.32 (0.06) & 0.48 (0.08) \tabularnewline 
Survey 1 & 3 & 0.40 (0.11) & 0.35 (0.09) & 0.29 (0.80) & 0.31 (0.08) & 0.44 (0.12) \tabularnewline 
Survey 1 & 4 & 0.23 (0.07) & 0.22 (0.07) & 0.21 (0.58) & 0.21 (0.06) & 0.22 (0.07) \tabularnewline 
Survey 1 & 5 & 0.31 (0.09) & 0.29 (0.08) & 0.22 (8.23) & 0.24 (0.07) & 0.29 (0.08) \tabularnewline 
Survey 1 & 6 & 0.17 (0.06) & 0.16 (0.06) & 0.18 (0.12) & 0.17 (0.06) & 0.20 (0.07) \tabularnewline 
Survey 1 & 7 & 0.25 (0.07) & 0.25 (0.07) & 0.26 (0.02)* & 0.26 (0.08) & 0.24 (0.07) \tabularnewline 
Survey 1 & 8 & 0.39 (0.10) & 0.37 (0.09) & 0.39 (0.02)* & 0.37 (0.10) & 0.38 (0.10) \tabularnewline 
\bottomrule
\end{tabular}

\end{table}

\clearpage

\begin{table}[h]
\caption{Estimated marginal coefficients of variation for models $M_1$, $M_2$, $M_4$, $M_7$ for  North-East Arctic Haddock. Standard errors of the estimates (in parentheses) are derived by the delta method. $M_5$ and $M_6$ are not included as they do not have constant CV. For $M_3$, $\tau$ is set to 0 in the calculation for $|\hat{\tau}| < 0.1$ (indicated by *) due to numerical instability of the calculations.}
\label{tab:data2}
\begin{tabular}{lclllll}
\toprule
Fleet & Age & $M_{1}$ & $M_{2}$ & $M_{3}$ & $M_{4}$ & $M_{7}$ \tabularnewline 
\midrule
Commercial & 3 & 0.60 (0.08) & 0.51 (0.03) & 0.49 (0.02) & 0.43 (0.06) & 0.68 (0.09) \tabularnewline 
Commercial & 4 & 0.30 (0.05) & 0.28 ( - ) & 0.29 (0.01)* & 0.25 (0.04) & 0.35 (0.05) \tabularnewline 
Commercial & 5 & 0.25 (0.04) & 0.24 ( - ) & 0.23 (0.04) & 0.22 (0.03) & 0.28 (0.04) \tabularnewline 
Commercial & 6 & 0.20 (0.04) & 0.20 (0.03) & 0.20 (0.01)* & 0.18 (0.04) & 0.24 (0.04) \tabularnewline 
Commercial & 7 & 0.24 (0.04) & 0.23 (0.04) & 0.23 (0.01)* & 0.21 (0.03) & 0.27 (0.04) \tabularnewline 
Commercial & 8 & 0.00 (0.01) & 0.00 ( - ) & 0.00 (0.00)* & 0.00 (0.03) & 0.01 (0.07) \tabularnewline 
Commercial & 9 & 0.73 (0.09) & 0.50 (0.05) & 0.40 (0.03) & 0.35 (0.05) & 0.73 (0.09) \tabularnewline 
Commercial & 10 & 0.39 (0.07) & 0.35 (0.05) & 0.32 (0.04) & 0.29 (0.05) & 0.38 (0.07) \tabularnewline 
Commercial & 11 & 0.36 (0.06) & 0.31 (0.05) &  -  ( - ) & 0.25 (0.04) & 0.35 (0.06) \tabularnewline 
Survey 1 & 3 & 0.35 (0.07) & 0.35 (0.06) & 0.32 (0.04) & 0.29 (0.06) & 0.48 (0.10) \tabularnewline 
Survey 1 & 4 & 0.42 (0.08) & 0.38 (0.07) &  -  ( - ) & 0.47 (0.11) & 0.42 (0.07) \tabularnewline 
Survey 1 & 5 & 0.52 (0.10) & 0.47 (0.07) & 0.45 (0.03) & 0.41 (0.07) & 0.53 (0.10) \tabularnewline 
Survey 1 & 6 & 0.35 (0.06) & 0.32 (0.06) & 0.32 (0.05) & 0.32 (0.06) & 0.35 (0.06) \tabularnewline 
Survey 1 & 7 & 0.58 (0.11) & 0.46 (0.07) & 0.41 (0.05) & 0.41 (0.08) & 0.52 (0.09) \tabularnewline 
Survey 2 & 3 & 0.24 (0.06) & 0.24 (0.05) & 0.21 (0.59) & 0.23 (0.06) & 0.26 (0.06) \tabularnewline 
Survey 2 & 4 & 0.28 (0.06) & 0.27 (0.06) &  -  ( - ) & 0.28 (0.07) & 0.29 (0.06) \tabularnewline 
Survey 2 & 5 & 0.32 (0.08) & 0.29 (0.06) & 0.29 (0.02)* & 0.28 (0.08) & 0.33 (0.07) \tabularnewline 
Survey 2 & 6 & 0.37 (0.09) & 0.36 (0.08) & 0.37 (0.05) & 0.34 (0.08) & 0.35 (0.08) \tabularnewline 
Survey 2 & 7 & 0.52 (0.13) & 0.49 (0.10) & 0.44 (1.65) & 0.43 (0.09) & 0.57 (0.13) \tabularnewline 
Survey 3 & 3 & 0.31 (0.06) & 0.29 (0.05) & 0.27 (0.05) & 0.30 (0.07) & 0.28 (0.06) \tabularnewline 
Survey 3 & 4 & 0.34 (0.07) & 0.34 (0.06) & 0.32 (0.04) & 0.32 (0.06) & 0.28 (0.05) \tabularnewline 
Survey 3 & 5 & 0.42 (0.09) & 0.41 (0.08) &  -  ( - ) & 0.48 (0.12) & 0.36 (0.07) \tabularnewline 
Survey 3 & 6 & 0.40 (0.09) & 0.39 (0.07) & 0.41 (0.02)* & 0.40 (0.08) & 0.35 (0.07) \tabularnewline 
Survey 3 & 7 & 0.36 (0.09) & 0.36 (0.08) & 0.28 (1.16) & 0.32 (0.07) & 0.51 (0.14) \tabularnewline 
Survey 3 & 8 & 0.71 (0.15) & 0.59 (0.10) & 0.63 (0.06) & 0.56 (0.13) & 0.77 (0.18) \tabularnewline 
Survey 4 & 3 & 0.41 (0.12) & 0.39 (0.10) & 0.42 (0.04)* & 0.38 (0.12) & 0.34 (0.08) \tabularnewline 
Survey 4 & 4 & 0.29 (0.08) & 0.29 (0.07) & 0.28 (0.02)* & 0.27 (0.08) & 0.24 (0.06) \tabularnewline 
Survey 4 & 5 & 0.12 (0.06) & 0.12 (0.05) &  -  ( - ) & 0.13 (0.07) & 0.12 (0.04) \tabularnewline 
Survey 4 & 6 & 0.09 (0.07) & 0.09 (0.07) & 0.08 (0.06)* & 0.10 (0.06) & 0.17 (0.09) \tabularnewline 
Survey 4 & 7 & 0.32 (0.09) & 0.31 (0.09) & 0.33 (0.03)* & 0.32 (0.10) & 0.39 (0.13) \tabularnewline 
Survey 4 & 8 & 0.40 (0.12) & 0.39 (0.10) & 0.37 (0.13) & 0.40 (0.12) & 0.56 (0.19) \tabularnewline 
\bottomrule
\end{tabular}

\end{table}

\clearpage

\begin{table}[h]
\caption{Estimated marginal coefficients of variation for models $M_1$, $M_2$, $M_4$, $M_7$ for  North Sea Cod. Standard errors of the estimates (in parentheses) are derived by the delta method. $M_5$ and $M_6$ are not included as they do not have constant CV. For $M_3$, $\tau$ is set to 0 in the calculation for $|\hat{\tau}| < 0.1$ (indicated by *) due to numerical instability of the calculations.}
\label{tab:data3}
\begin{tabular}{lclllll}
\toprule
Fleet & Age & $M_{1}$ & $M_{2}$ & $M_{3}$ & $M_{4}$ & $M_{7}$ \tabularnewline 
\midrule
Commercial & 1 & 0.81 (0.12) & 0.65 (0.07) & 0.80 (0.01)* & 0.58 (0.09) & 0.82 (0.12) \tabularnewline 
Commercial & 2 & 0.22 (0.04) & 0.21 (0.03) & 0.20 (0.05) & 0.19 (0.03) & 0.20 (0.04) \tabularnewline 
Commercial & 3 & 0.08 (0.03) & 0.08 (0.03) & 0.09 (0.08) & 0.07 (0.03) & 0.10 (0.03) \tabularnewline 
Commercial & 4 & 0.08 (0.02) & 0.08 (0.02) & 0.07 (0.01)* & 0.07 (0.03) & 0.09 (0.02) \tabularnewline 
Commercial & 5 & 0.07 (0.03) & 0.07 (0.03) &  -  ( - ) & 0.05 (0.04) & 0.08 (0.03) \tabularnewline 
Commercial & 6 & 0.09 (0.03) & 0.09 (0.03) & 0.08 (0.01)* & 0.09 (0.04) & 0.10 (0.03) \tabularnewline 
Commercial & 7 & 0.15 (0.03) & 0.15 (0.03) & 0.15 (0.00)* & 0.13 (0.03) & 0.15 (0.03) \tabularnewline 
Survey 1 & 1 & 0.71 (0.12) & 0.60 (0.08) & 0.70 (0.02)* & 0.61 (0.11) & 0.71 (0.12) \tabularnewline 
Survey 1 & 2 & 0.29 (0.05) & 0.28 (0.04) & 0.29 (0.01)* & 0.26 (0.04) & 0.28 (0.04) \tabularnewline 
Survey 1 & 3 & 0.23 (0.03) & 0.23 (0.03) & 0.23 (0.00)* & 0.22 (0.03) & 0.23 (0.03) \tabularnewline 
Survey 1 & 4 & 0.30 (0.04) & 0.29 (0.04) &  -  ( - ) & 0.30 (0.05) & 0.29 (0.04) \tabularnewline 
Survey 1 & 5 & 0.33 (0.05) & 0.32 (0.04) &  -  ( - ) & 0.35 (0.05) & 0.31 (0.04) \tabularnewline 
\bottomrule
\end{tabular}

\end{table}

\clearpage

\begin{table}[h]
\caption{Estimated marginal coefficients of variation for models $M_1$, $M_2$, $M_4$, $M_7$ for  Northern Shelf Haddock. Standard errors of the estimates (in parentheses) are derived by the delta method. $M_5$ and $M_6$ are not included as they do not have constant CV. For $M_3$, $\tau$ is set to 0 in the calculation for $|\hat{\tau}| < 0.1$ (indicated by *) due to numerical instability of the calculations.}
\label{tab:data4}
\begin{tabular}{lclllll}
\toprule
Fleet & Age & $M_{1}$ & $M_{2}$ & $M_{3}$ & $M_{4}$ & $M_{7}$ \tabularnewline 
\midrule
Commercial & 0 & 1.71 (0.35) & 1.01 (0.10) & 1.73 ( 0.00)* & 0.84 (0.14) & 1.86 (0.39) \tabularnewline 
Commercial & 1 & 0.71 (0.12) & 0.56 (0.07) & 0.76 ( 0.01)* & 0.38 (0.06) & 0.79 (0.13) \tabularnewline 
Commercial & 2 & 0.34 (0.05) & 0.31 (0.05) & 0.34 ( 0.01) & 0.26 (0.05) & 0.32 (0.05) \tabularnewline 
Commercial & 3 & 0.25 (0.05) & 0.24 (0.05) & 0.24 ( 0.03) & 0.20 (0.04) & 0.27 (0.05) \tabularnewline 
Commercial & 4 & 0.18 (0.05) & 0.20 (0.04) & 0.20 ( 0.00) & 0.21 (0.04) & 0.21 (0.04) \tabularnewline 
Commercial & 5 & 0.00 (0.00) & 0.00 ( - ) & 0.00 ( 0.00)* & 0.00 (0.27) & 0.07 (0.09) \tabularnewline 
Commercial & 6 & 0.00 ( - ) & 0.00 (0.09) & 0.08 ( 0.09)* & 0.00 (0.00) & 0.13 (0.07) \tabularnewline 
Commercial & 7 & 0.27 (0.06) & 0.23 (0.06) &  -  ( - ) & 0.17 (0.07) & 0.30 (0.05) \tabularnewline 
Commercial & 8 & 0.19 (0.08) & 0.19 (0.08) & 0.19 ( 0.02)* & 0.00 (0.01) & 0.22 (0.06) \tabularnewline 
Survey 1 & 0 & 0.53 (0.14) & 0.46 (0.11) &  -  ( - ) & 0.36 (0.10) & 0.53 (0.13) \tabularnewline 
Survey 1 & 1 & 0.24 (0.06) & 0.23 (0.06) &  -  ( - ) & 0.23 (0.06) & 0.25 (0.06) \tabularnewline 
Survey 1 & 2 & 0.16 (0.04) & 0.16 (0.04) & 0.14 ( 1.10) & 0.17 (0.05) & 0.24 (0.05) \tabularnewline 
Survey 1 & 3 & 0.36 (0.09) & 0.32 (0.07) & 0.33 ( 0.08) & 0.27 (0.07) & 0.36 (0.07) \tabularnewline 
Survey 1 & 4 & 0.23 (0.06) & 0.20 (0.06) &  -  ( - ) & 0.15 (0.06) & 0.28 (0.07) \tabularnewline 
Survey 1 & 5 & 0.40 (0.08) & 0.38 (0.07) &  -  ( - ) & 0.39 (0.09) & 0.48 (0.12) \tabularnewline 
Survey 1 & 6 & 0.94 (0.24) & 0.74 (0.13) & 0.91 ( 0.04)* & 0.74 (0.17) & 1.15 (0.34) \tabularnewline 
Survey 2 & 0 & 0.78 (0.17) & 0.63 (0.10) & 0.51 ( - ) & 0.55 (0.12) & 0.67 (0.13) \tabularnewline 
Survey 2 & 1 & 0.15 (0.05) & 0.14 (0.06) &  -  ( - ) & 0.13 (0.06) & 0.14 (0.04) \tabularnewline 
Survey 2 & 2 & 0.34 (0.06) & 0.33 (0.06) & 0.30 ( 0.00)* & 0.32 (0.07) & 0.31 (0.06) \tabularnewline 
Survey 2 & 3 & 0.36 (0.07) & 0.34 (0.06) & 0.35 ( 0.00)* & 0.31 (0.06) & 0.38 (0.07) \tabularnewline 
Survey 2 & 4 & 0.40 (0.08) & 0.36 (0.07) &  -  ( - ) & 0.30 (0.07) & 0.42 (0.08) \tabularnewline 
Survey 2 & 5 & 0.51 (0.10) & 0.49 (0.08) &  -  ( - ) & 0.47 (0.10) & 0.56 (0.10) \tabularnewline 
Survey 2 & 6 & 1.22 (0.31) & 0.75 (0.12) & 1.25 ( 0.05)* & 0.58 (0.12) & 1.53 (0.43) \tabularnewline 
Survey 3 & 0 & 0.45 (0.10) & 0.42 (0.09) &  -  ( - ) & 0.40 (0.11) & 0.56 (0.13) \tabularnewline 
Survey 3 & 1 & 0.19 (0.05) & 0.19 (0.05) & 0.19 ( 5.99) & 0.17 (0.05) & 0.29 (0.07) \tabularnewline 
Survey 3 & 2 & 0.00 (0.00) & 0.00 ( - ) &  -  ( - ) & 0.08 (0.06) & 0.14 (0.05) \tabularnewline 
Survey 3 & 3 & 0.27 (0.06) & 0.25 (0.06) & 0.25 ( 0.03) & 0.23 (0.06) & 0.29 (0.06) \tabularnewline 
Survey 3 & 4 & 0.30 (0.07) & 0.28 (0.06) & 0.24 ( - ) & 0.25 (0.06) & 0.33 (0.07) \tabularnewline 
Survey 3 & 5 & 0.33 (0.07) & 0.31 (0.06) & 0.33 ( 0.01)* & 0.31 (0.07) & 0.36 (0.07) \tabularnewline 
Survey 3 & 6 & 0.44 (0.09) & 0.42 (0.08) & 0.44 ( 0.02)* & 0.39 (0.08) & 0.43 (0.08) \tabularnewline 
Survey 4 & 0 & 1.01 (0.28) & 0.64 (0.12) & 0.42 ( - ) & 0.44 (0.10) & 0.99 (0.27) \tabularnewline 
Survey 4 & 1 & 0.27 (0.06) & 0.27 (0.06) & 0.24 ( 0.00)* & 0.28 (0.06) & 0.29 (0.06) \tabularnewline 
Survey 4 & 2 & 0.21 (0.06) & 0.21 (0.06) & 0.20 ( - ) & 0.20 (0.06) & 0.26 (0.06) \tabularnewline 
Survey 4 & 3 & 0.19 (0.06) & 0.19 (0.06) &  -  ( - ) & 0.22 (0.06) & 0.21 (0.06) \tabularnewline 
Survey 4 & 4 & 0.37 (0.09) & 0.32 (0.07) & 0.36 (-0.09)* & 0.25 (0.06) & 0.37 (0.09) \tabularnewline 
Survey 4 & 5 & 0.45 (0.10) & 0.42 (0.08) &  -  ( - ) & 0.49 (0.13) & 0.46 (0.10) \tabularnewline 
Survey 4 & 6 & 0.50 (0.12) & 0.48 (0.10) & 0.53 ( 0.03)* & 0.44 (0.10) & 0.53 (0.13) \tabularnewline 
Survey 5 & 0 & 0.28 (0.07) & 0.28 (0.06) & 0.26 ( - ) & 0.32 (0.06) & 0.28 (0.06) \tabularnewline 
Survey 5 & 1 & 0.34 (0.05) & 0.33 (0.05) &  -  ( - ) & 0.33 (0.06) & 0.30 (0.04) \tabularnewline 
Survey 5 & 2 & 0.34 (0.05) & 0.32 (0.05) &  -  ( - ) & 0.30 (0.05) & 0.34 (0.05) \tabularnewline 
Survey 5 & 3 & 0.31 (0.05) & 0.29 (0.05) &  -  ( - ) & 0.28 (0.05) & 0.36 (0.06) \tabularnewline 
Survey 5 & 4 & 0.36 (0.06) & 0.33 (0.05) & 0.34 ( 0.00)* & 0.31 (0.05) & 0.47 (0.08) \tabularnewline 
\bottomrule
\end{tabular}

\end{table}

\clearpage

\begin{table}[h]
\caption{Estimated coefficient of variation for totals with models $M_8-M_{13}$. Standard errors of the estimates (in parentheses) are derived by the delta method from the estimated logarithm of the scale parameter.}
\label{tab:data5}
\begin{tabular}{lcccccc}
\toprule
Fleet &  $M_{8}$ & $M_{9}$ & $M_{10}$ & $M_{11}$ & $M_{12}$ & $M_{13}$ \tabularnewline 
\midrule
\multicolumn{7}{l}{\textbf{Blue Whiting}} \tabularnewline
Com.  &  0.15 (0.11) & 0.16 (0.09) & 0.16 (0.09) & 0.14 (0.12) & 0.15 (0.10) & 0.16 (0.10) \tabularnewline 
Surv. 1  &  0.13 (0.04) & 0.13 (0.04) & 0.13 (0.04) & 0.13 (0.04) & 0.12 (0.04) & 0.13 (0.04) \tabularnewline 
\multicolumn{7}{l}{\textbf{North-East Arctic Haddock}} \tabularnewline
Com.  &  0.11 (0.05) & 0.11 (0.05) & 0.05 (0.07) & 0.10 (0.05) & 0.09 (0.05) & 0.06 (0.06) \tabularnewline 
Surv. 1  &  0.37 (0.06) & 0.34 (0.06) & 0.39 (0.07) & 0.37 (0.06) & 0.35 (0.06) & 0.39 (0.07) \tabularnewline 
Surv. 2  &  0.22 (0.05) & 0.21 (0.05) & 0.25 (0.06) & 0.23 (0.05) & 0.22 (0.05) & 0.25 (0.06) \tabularnewline 
Surv. 3  &  0.22 (0.05) & 0.26 (0.06) & 0.25 (0.06) & 0.22 (0.05) & 0.26 (0.06) & 0.27 (0.06) \tabularnewline 
Surv. 4  &  0.27 (0.07) & 0.28 (0.08) & 0.25 (0.08) & 0.22 (0.06) & 0.23 (0.07) & 0.21 (0.07) \tabularnewline 
\multicolumn{7}{l}{\textbf{North Sea Cod}} \tabularnewline
Com.  &  0.32 (0.05) & 0.32 (0.05) & 0.03 (0.05) & 0.15 (0.03) & 0.15 (0.02) & 0.04 (0.04) \tabularnewline 
Surv. 1  &  0.27 (0.05) & 0.27 (0.05) & 0.34 (0.05) & 0.20 (0.04) & 0.20 (0.04) & 0.21 (0.03) \tabularnewline 
\multicolumn{7}{l}{\textbf{Northen Shelf Haddock}} \tabularnewline
Com.  &  0.26 (0.05) & 0.28 (0.05) & 0.13 (0.06) & 0.18 (0.06) & 0.17 (0.05) & 0.15 (0.06) \tabularnewline 
Surv. 1  &  0.34 (0.08) & 0.34 (0.08) & 0.29 (0.07) & 0.23 (0.06) & 0.25 (0.06) & 0.25 (0.06) \tabularnewline 
Surv. 2  &  0.22 (0.05) & 0.21 (0.05) & 0.24 (0.05) & 0.19 (0.04) & 0.19 (0.04) & 0.19 (0.04) \tabularnewline 
Surv. 3  &  0.24 (0.06) & 0.24 (0.06) & 0.26 (0.06) & 0.19 (0.05) & 0.20 (0.05) & 0.17 (0.05) \tabularnewline 
Surv. 4  &  0.15 (0.05) & 0.17 (0.05) & 0.10 (0.04) & 0.11 (0.03) & 0.12 (0.04) & 0.07 (0.03) \tabularnewline 
Surv. 5  &  0.29 (0.05) & 0.30 (0.05) & 0.29 (0.05) & 0.27 (0.04) & 0.27 (0.04) & 0.25 (0.04) \tabularnewline 
\bottomrule
\end{tabular}

\end{table}

\clearpage

\newpage

\section{Estimated fishing mortality and spawning stock
biomass}\label{estimated-fishing-mortality-and-spawning-stock-biomass}

\subsection{Blue Whiting}\label{blue-whiting}

\subsubsection{log-Normal}\label{log-normal}

\begin{figure}[htbp]
\centering
\includegraphics{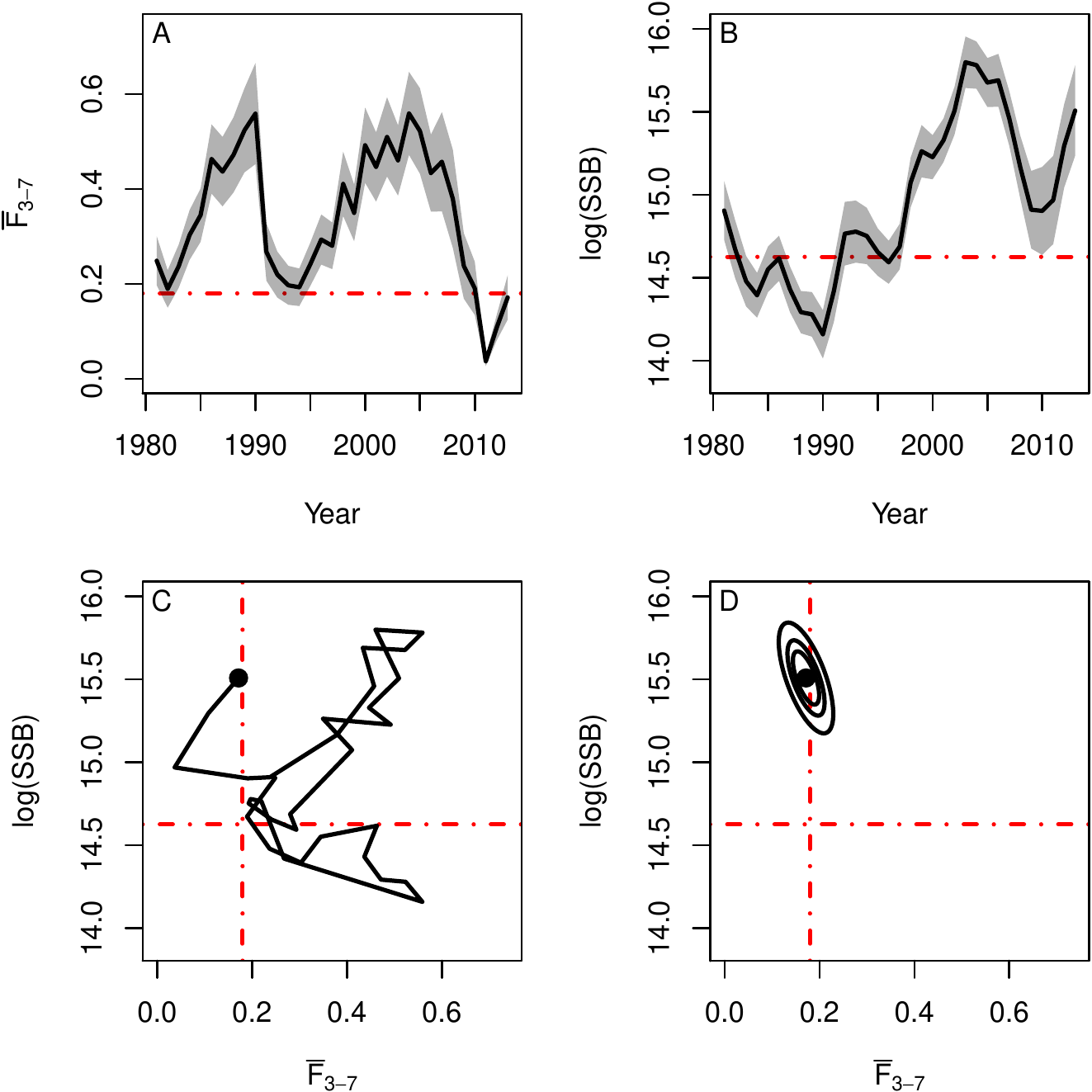}
\caption{Estimated average fishing mortality (A) and log spawning stock
biomass (B) with the log-Normal model for Blue Whiting including 95 \%
pointwise confidence intervals (grey area); their estimated trajectory
(C); and confidence ellipses in the final year (D) at 50 \%, 75 \% and
95 \% levels. The red lines indicate the management plan reference
points while the black point is the estimated value in the final year.}
\end{figure}

\clearpage

\subsubsection{Gamma}\label{gamma}

\begin{figure}[htbp]
\centering
\includegraphics{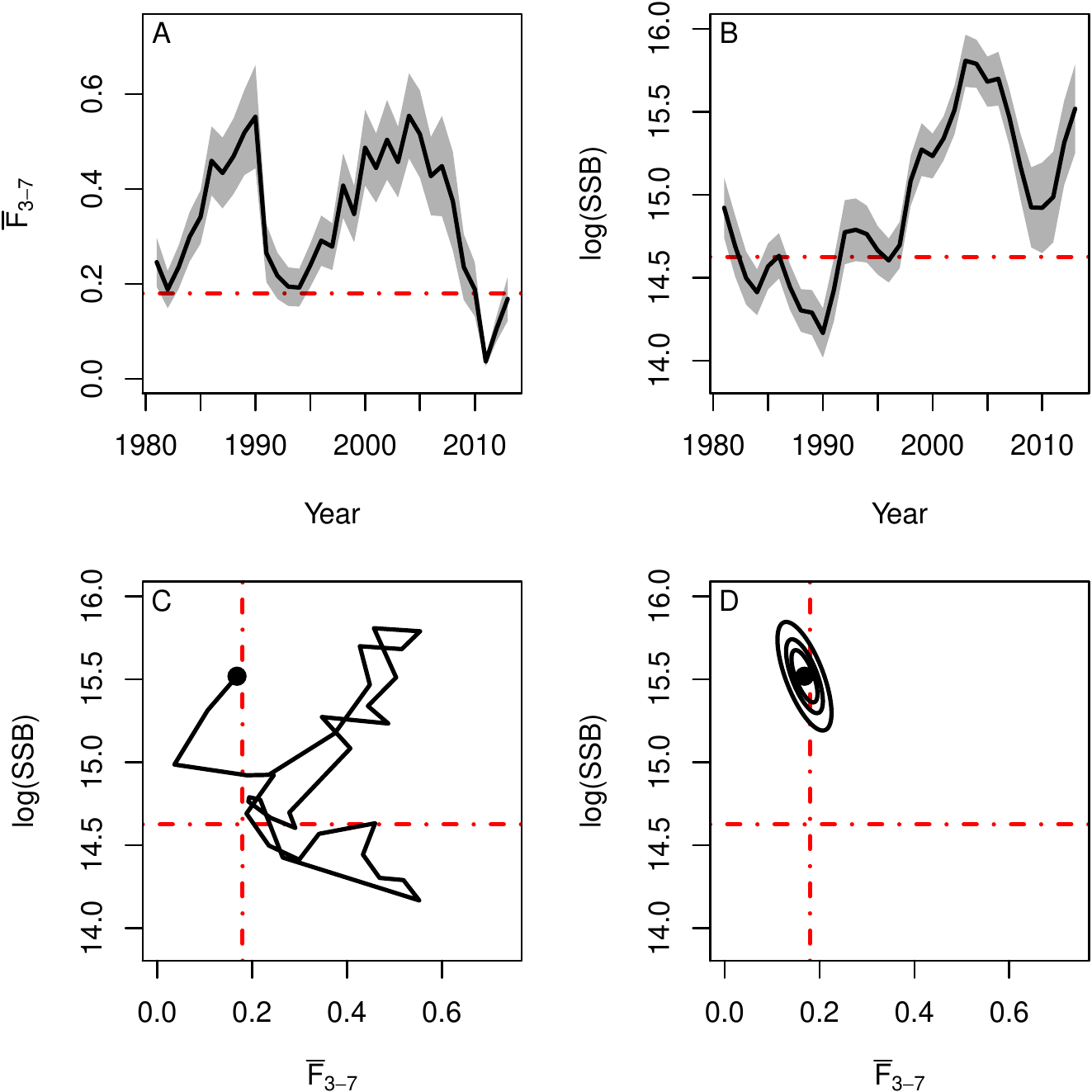}
\caption{Estimated average fishing mortality (A) and log spawning stock
biomass (B) with the Gamma model for Blue Whiting including 95 \%
pointwise confidence intervals (grey area); their estimated trajectory
(C); and confidence ellipses in the final year (D) at 50 \%, 75 \% and
95 \% levels. The red lines indicate the management plan reference
points while the black point is the estimated value in the final year.}
\end{figure}

\clearpage

\subsubsection{Generalized Gamma}\label{generalized-gamma}

\begin{figure}[htbp]
\centering
\includegraphics{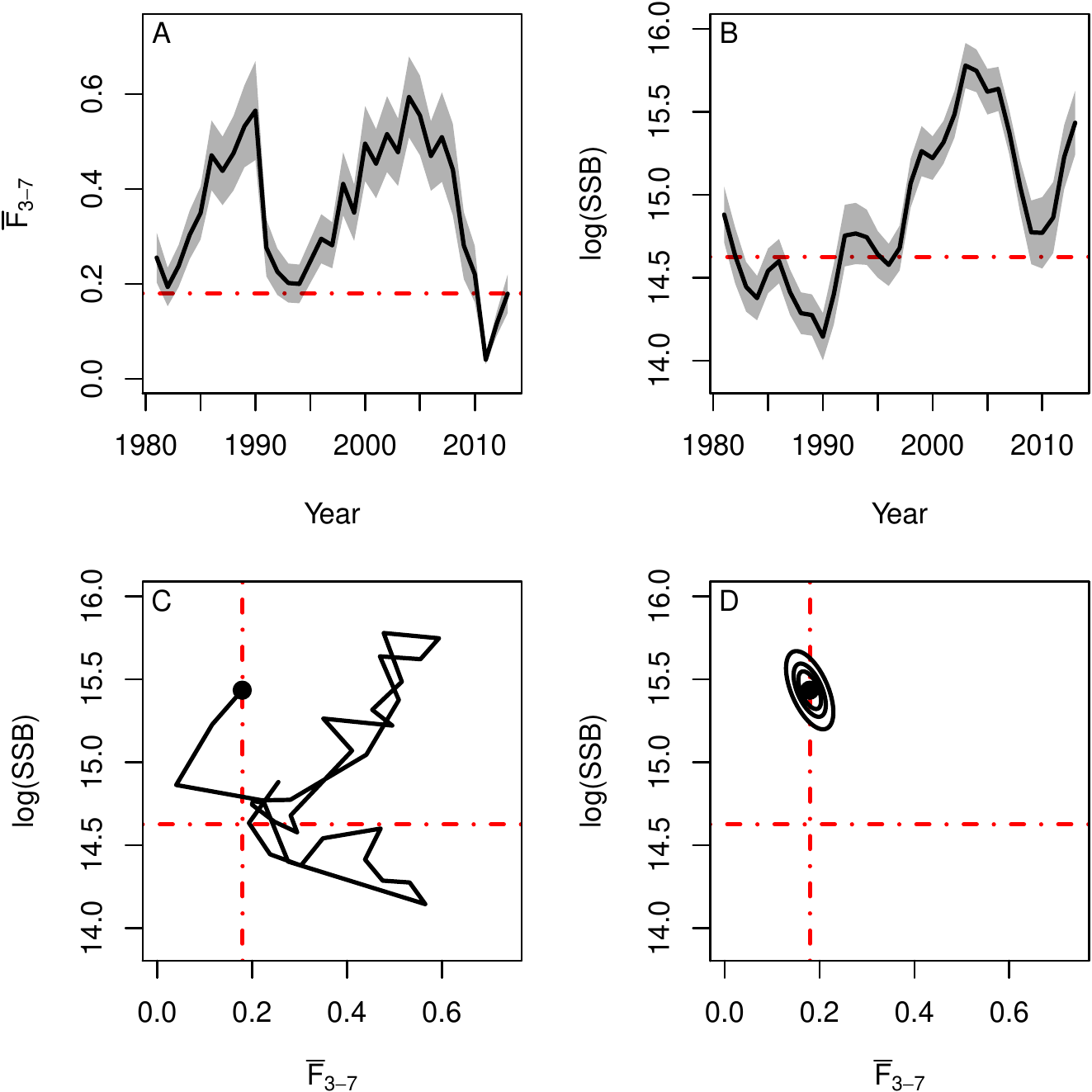}
\caption{Estimated average fishing mortality (A) and log spawning stock
biomass (B) with the Generalized Gamma model for Blue Whiting including
95 \% pointwise confidence intervals (grey area); their estimated
trajectory (C); and confidence ellipses in the final year (D) at 50 \%,
75 \% and 95 \% levels. The red lines indicate the management plan
reference points while the black point is the estimated value in the
final year.}
\end{figure}

\clearpage

\subsubsection{Normal}\label{normal}

\begin{figure}[htbp]
\centering
\includegraphics{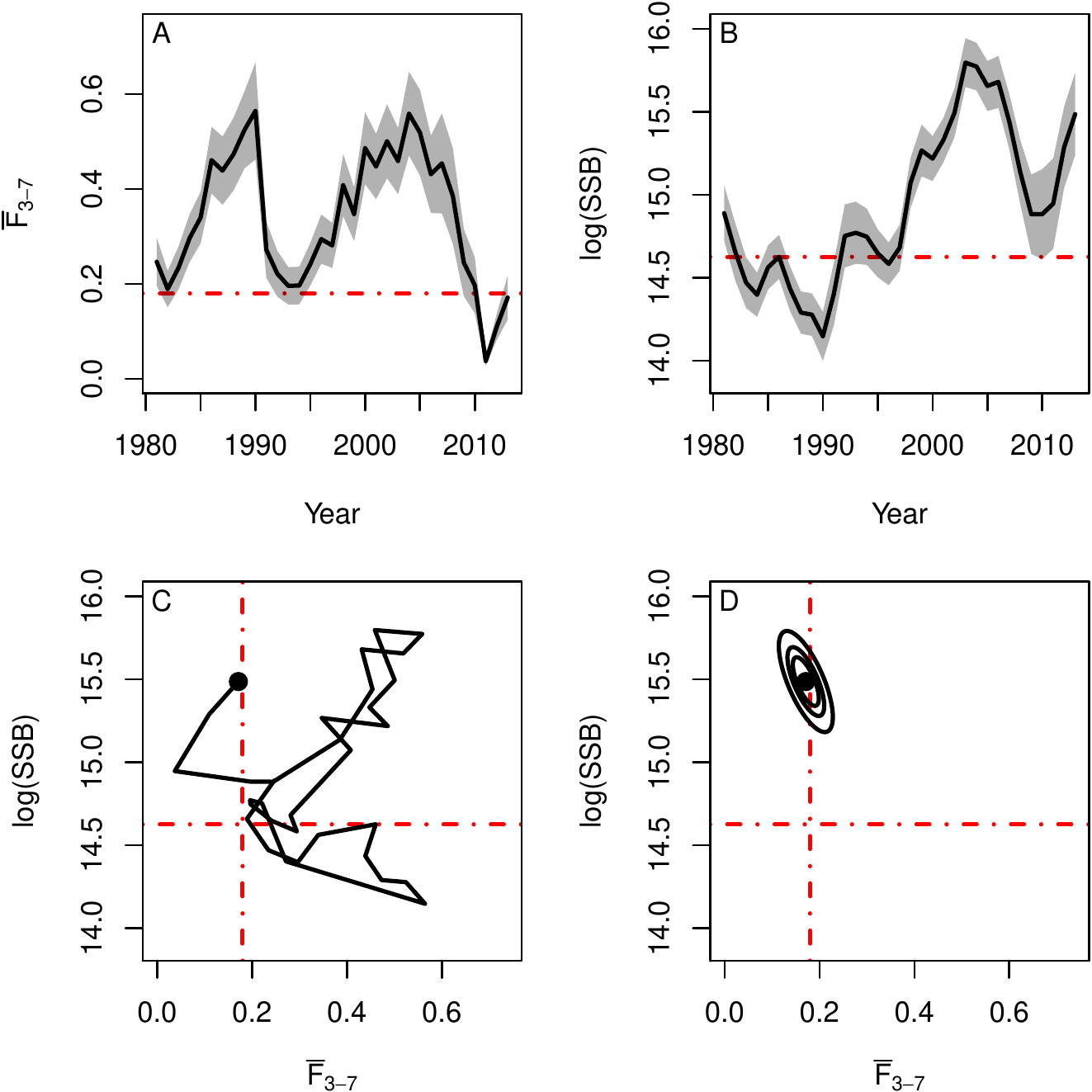}
\caption{Estimated average fishing mortality (A) and log spawning stock
biomass (B) with the Normal model for Blue Whiting including 95 \%
pointwise confidence intervals (grey area); their estimated trajectory
(C); and confidence ellipses in the final year (D) at 50 \%, 75 \% and
95 \% levels. The red lines indicate the management plan reference
points while the black point is the estimated value in the final year.}
\end{figure}

\clearpage

\subsubsection{Left Truncated Normal}\label{left-truncated-normal}

\begin{figure}[htbp]
\centering
\includegraphics{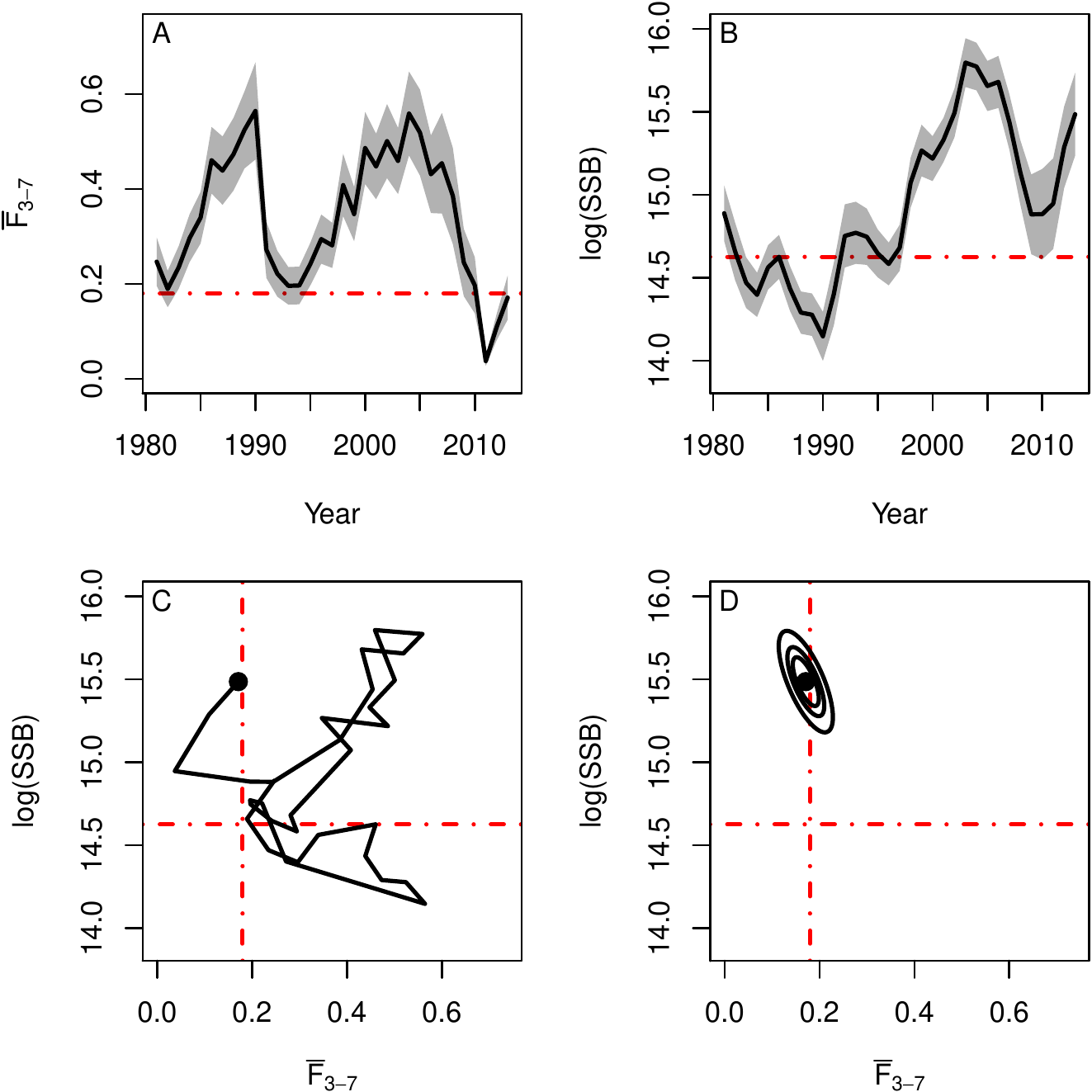}
\caption{Estimated average fishing mortality (A) and log spawning stock
biomass (B) with the Left Truncated Normal model for Blue Whiting
including 95 \% pointwise confidence intervals (grey area); their
estimated trajectory (C); and confidence ellipses in the final year (D)
at 50 \%, 75 \% and 95 \% levels. The red lines indicate the management
plan reference points while the black point is the estimated value in
the final year.}
\end{figure}

\clearpage

\subsubsection{log-Students t}\label{log-students-t}

\begin{figure}[htbp]
\centering
\includegraphics{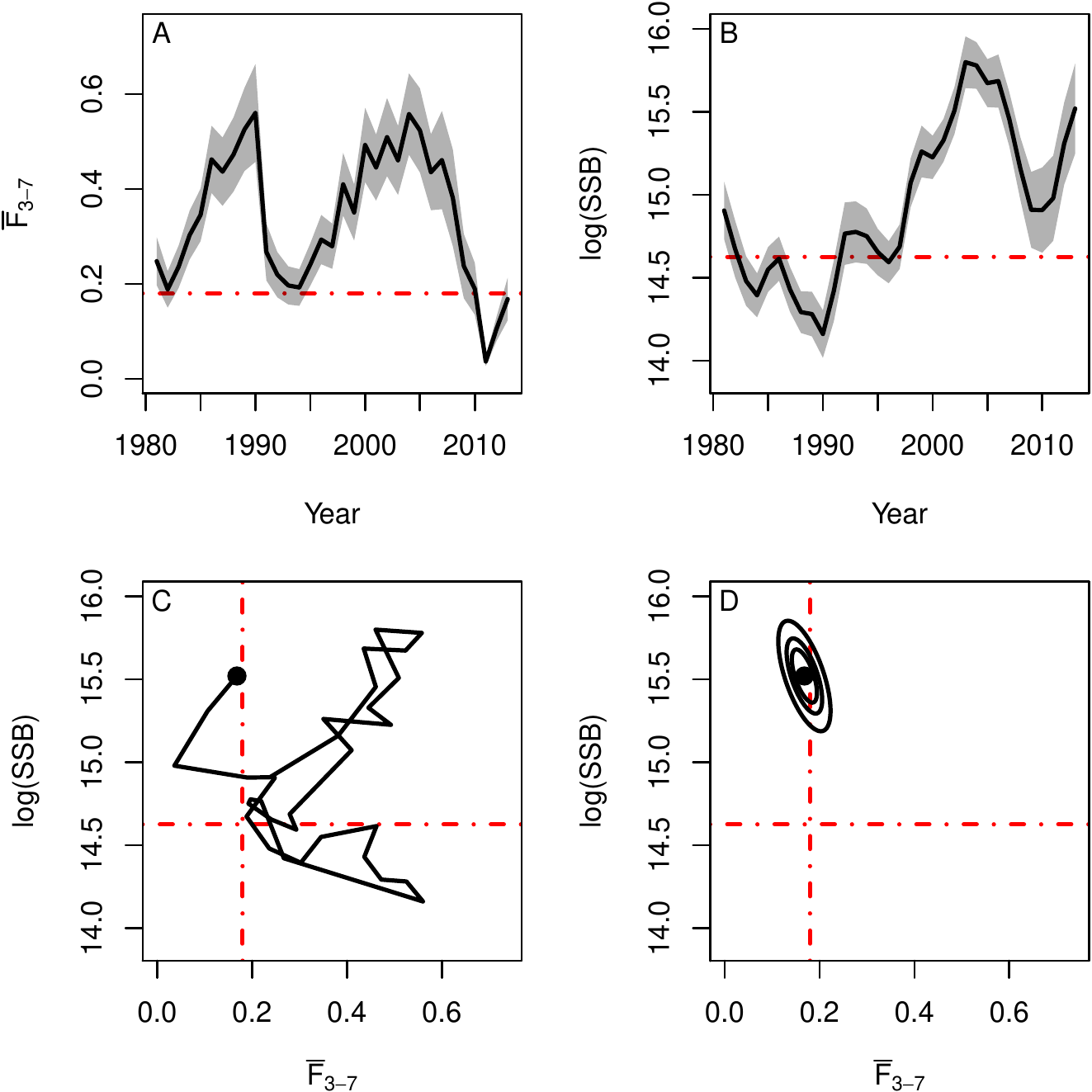}
\caption{Estimated average fishing mortality (A) and log spawning stock
biomass (B) with the log-Students t model for Blue Whiting including 95
\% pointwise confidence intervals (grey area); their estimated
trajectory (C); and confidence ellipses in the final year (D) at 50 \%,
75 \% and 95 \% levels. The red lines indicate the management plan
reference points while the black point is the estimated value in the
final year.}
\end{figure}

\clearpage

\subsubsection{Multivariate log-Normal}\label{multivariate-log-normal}

\begin{figure}[htbp]
\centering
\includegraphics{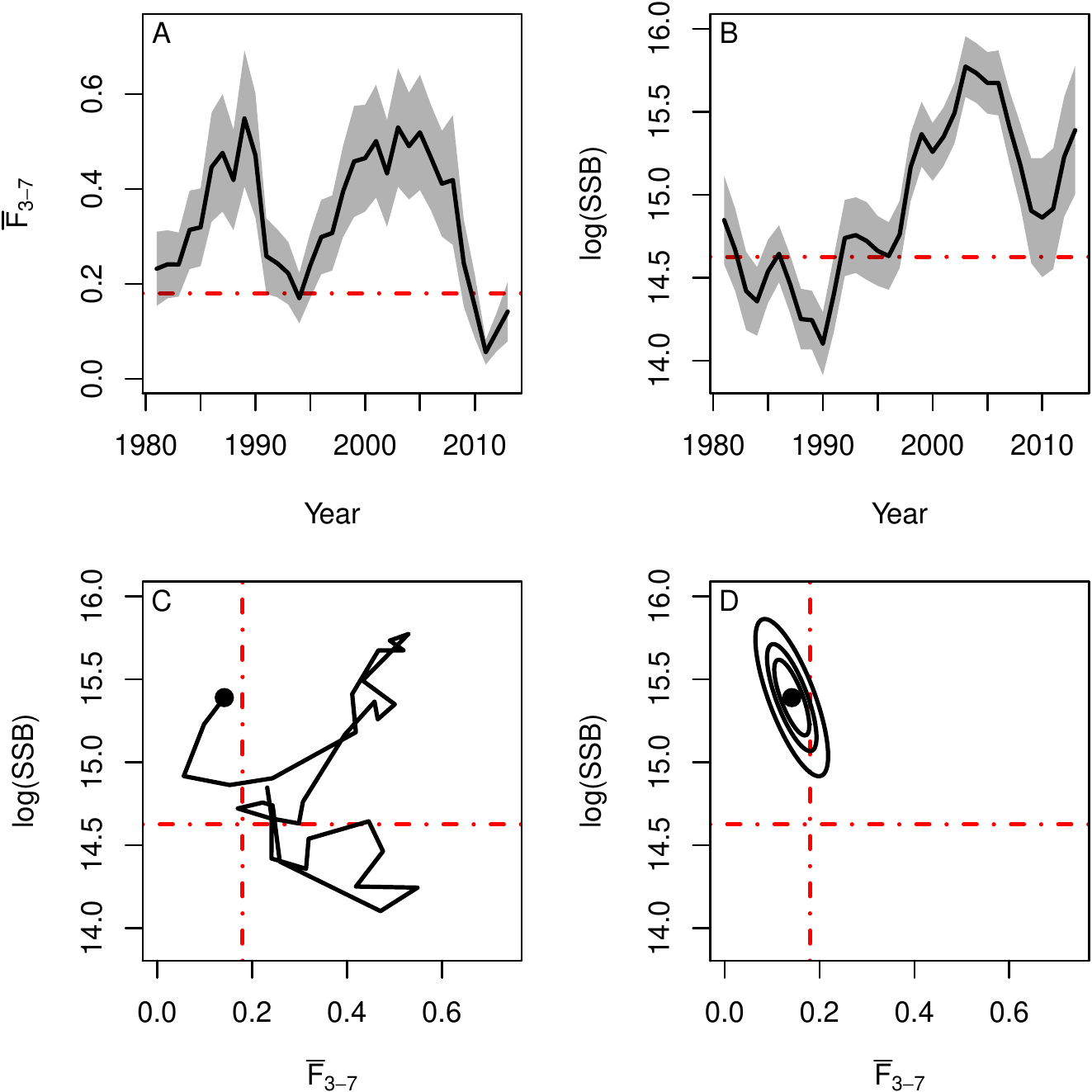}
\caption{Estimated average fishing mortality (A) and log spawning stock
biomass (B) with the Multivariate log-Normal model for Blue Whiting
including 95 \% pointwise confidence intervals (grey area); their
estimated trajectory (C); and confidence ellipses in the final year (D)
at 50 \%, 75 \% and 95 \% levels. The red lines indicate the management
plan reference points while the black point is the estimated value in
the final year.}
\end{figure}

\clearpage

\subsubsection{Additive Logistic Normal with log-Normal
Numbers}\label{additive-logistic-normal-with-log-normal-numbers}

\begin{figure}[htbp]
\centering
\includegraphics{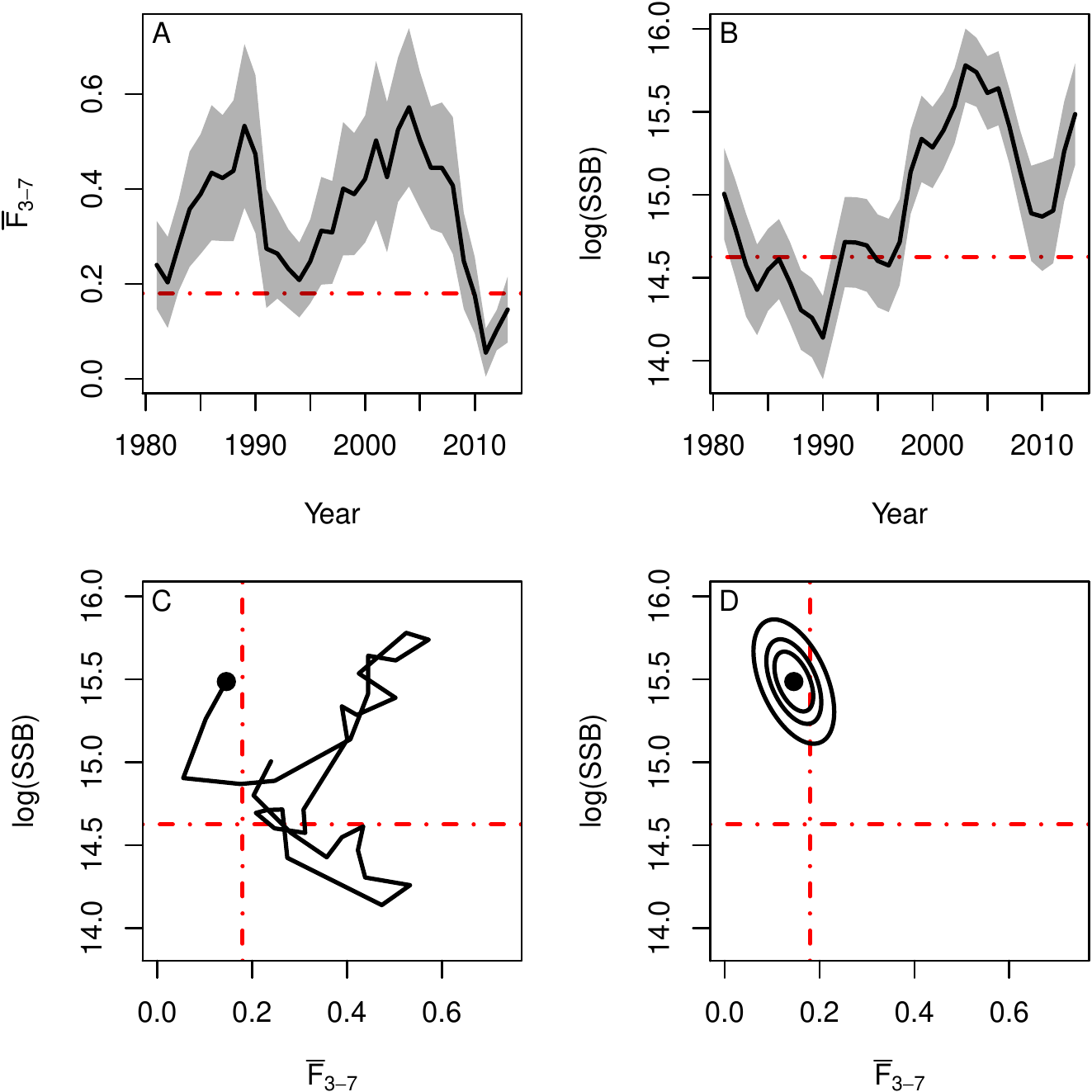}
\caption{Estimated average fishing mortality (A) and log spawning stock
biomass (B) with the Additive Logistic Normal with log-Normal Numbers
model for Blue Whiting including 95 \% pointwise confidence intervals
(grey area); their estimated trajectory (C); and confidence ellipses in
the final year (D) at 50 \%, 75 \% and 95 \% levels. The red lines
indicate the management plan reference points while the black point is
the estimated value in the final year.}
\end{figure}

\clearpage

\subsubsection{Multiplicative Logistic Normal with log-Normal
Numbers}\label{multiplicative-logistic-normal-with-log-normal-numbers}

\begin{figure}[htbp]
\centering
\includegraphics{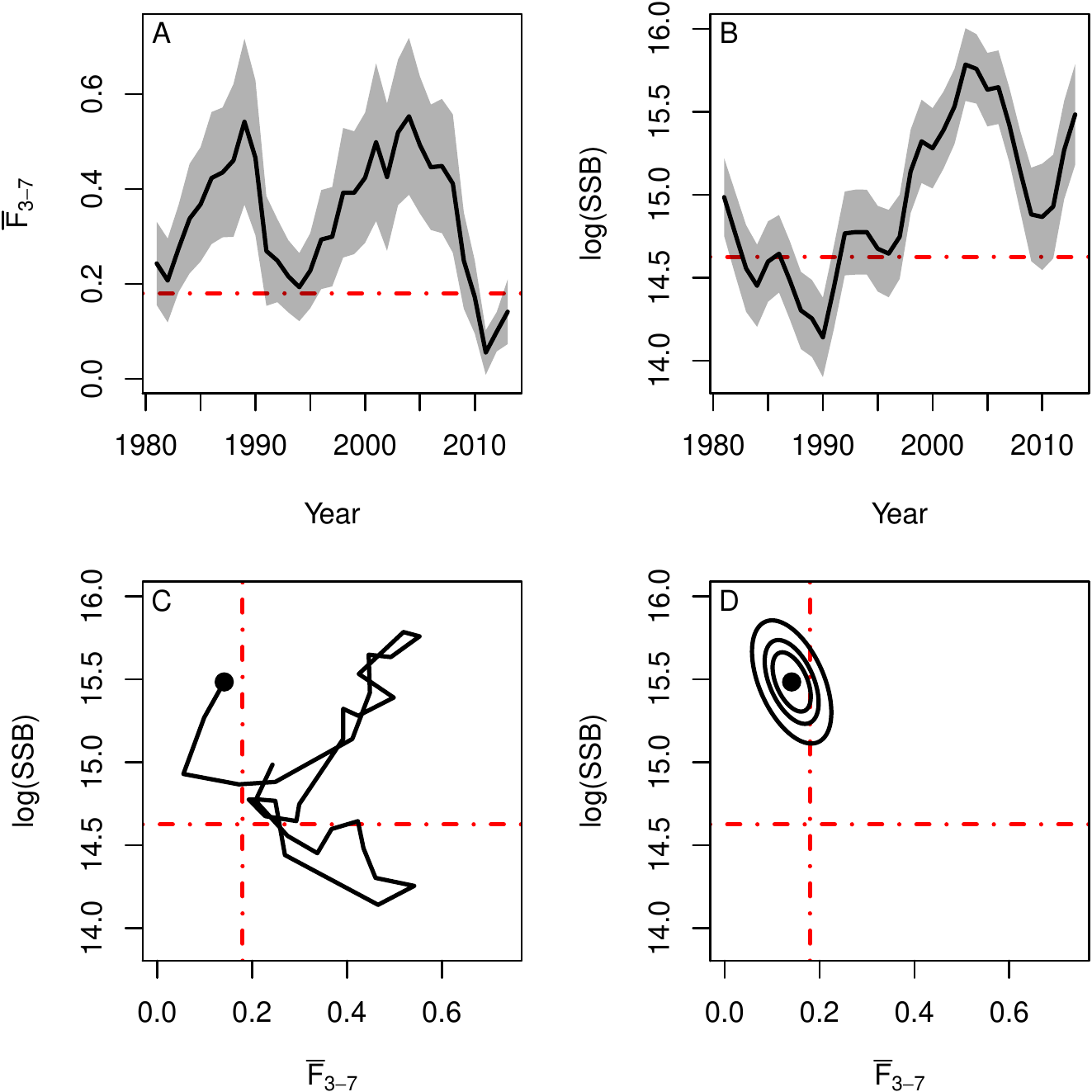}
\caption{Estimated average fishing mortality (A) and log spawning stock
biomass (B) with the Multiplicative Logistic Normal with log-Normal
Numbers model for Blue Whiting including 95 \% pointwise confidence
intervals (grey area); their estimated trajectory (C); and confidence
ellipses in the final year (D) at 50 \%, 75 \% and 95 \% levels. The red
lines indicate the management plan reference points while the black
point is the estimated value in the final year.}
\end{figure}

\clearpage

\subsubsection{Dirichlet with log-Normal
Numbers}\label{dirichlet-with-log-normal-numbers}

\begin{figure}[htbp]
\centering
\includegraphics{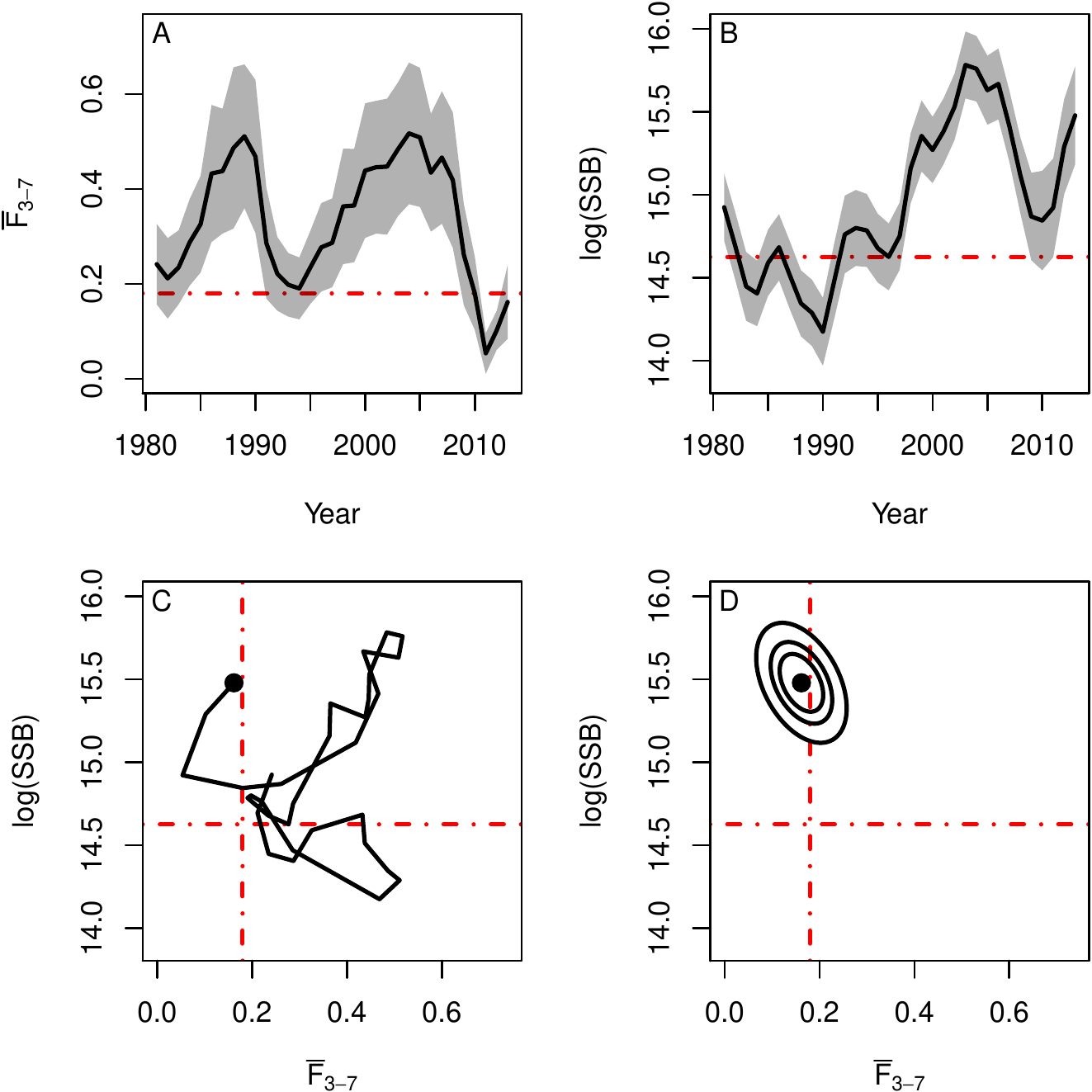}
\caption{Estimated average fishing mortality (A) and log spawning stock
biomass (B) with the Dirichlet with log-Normal Numbers model for Blue
Whiting including 95 \% pointwise confidence intervals (grey area);
their estimated trajectory (C); and confidence ellipses in the final
year (D) at 50 \%, 75 \% and 95 \% levels. The red lines indicate the
management plan reference points while the black point is the estimated
value in the final year.}
\end{figure}

\clearpage

\subsubsection{Additive Logisitc Normal with log-Normal
Weight}\label{additive-logisitc-normal-with-log-normal-weight}

\begin{figure}[htbp]
\centering
\includegraphics{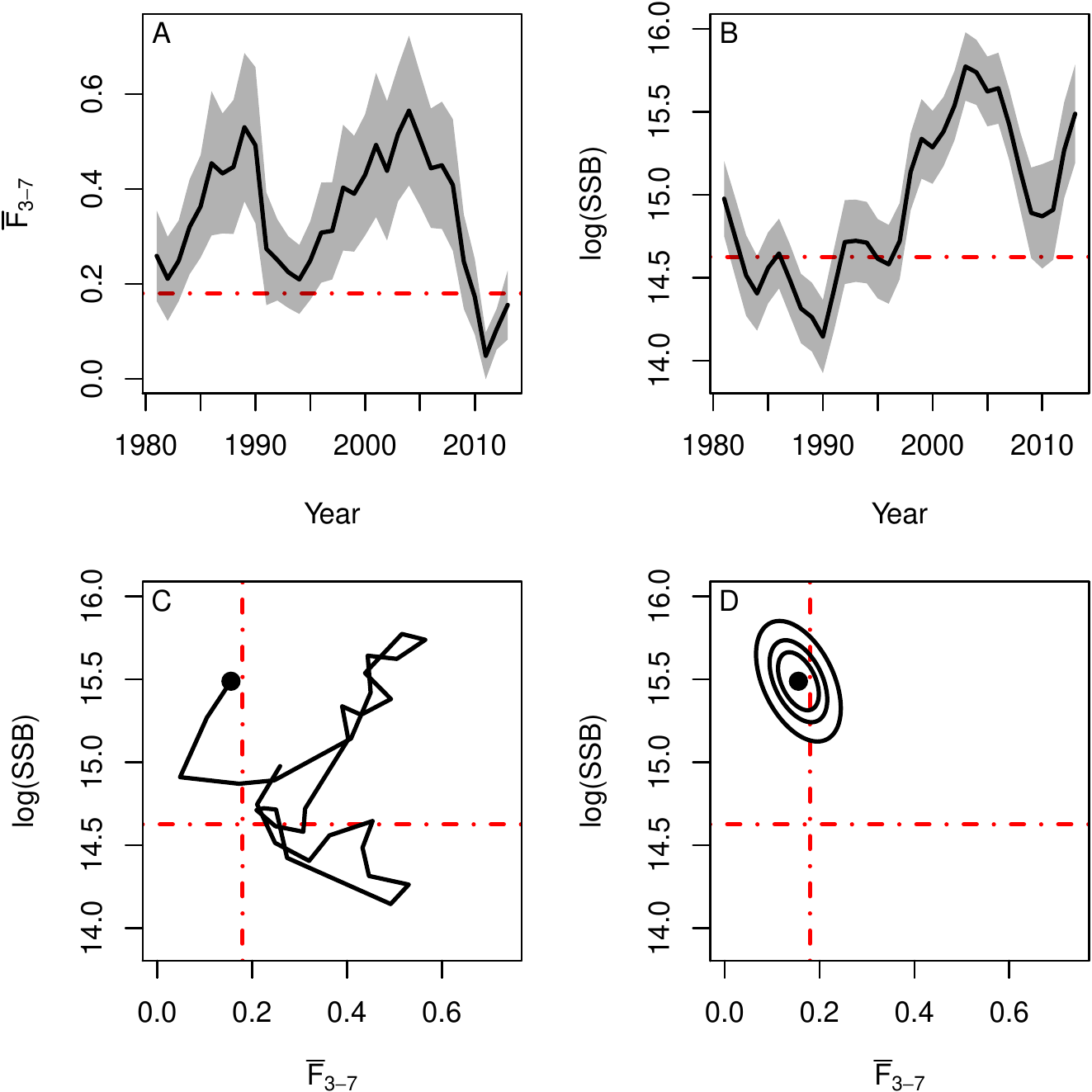}
\caption{Estimated average fishing mortality (A) and log spawning stock
biomass (B) with the Additive Logisitc Normal with log-Normal Weight
model for Blue Whiting including 95 \% pointwise confidence intervals
(grey area); their estimated trajectory (C); and confidence ellipses in
the final year (D) at 50 \%, 75 \% and 95 \% levels. The red lines
indicate the management plan reference points while the black point is
the estimated value in the final year.}
\end{figure}

\clearpage

\subsubsection{Multiplicative Logistic Normal with log-Normal
Weight}\label{multiplicative-logistic-normal-with-log-normal-weight}

\begin{figure}[htbp]
\centering
\includegraphics{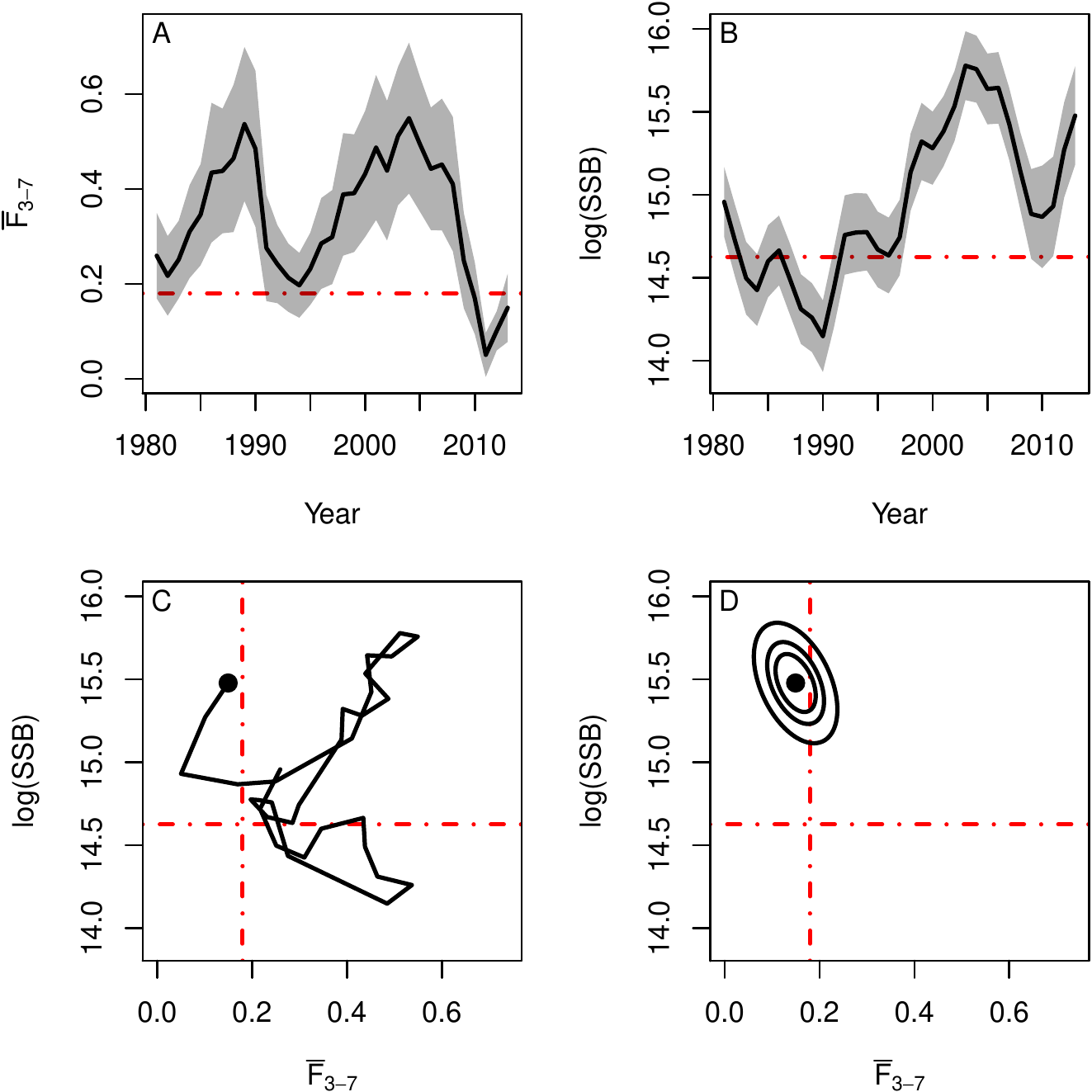}
\caption{Estimated average fishing mortality (A) and log spawning stock
biomass (B) with the Multiplicative Logistic Normal with log-Normal
Weight model for Blue Whiting including 95 \% pointwise confidence
intervals (grey area); their estimated trajectory (C); and confidence
ellipses in the final year (D) at 50 \%, 75 \% and 95 \% levels. The red
lines indicate the management plan reference points while the black
point is the estimated value in the final year.}
\end{figure}

\clearpage

\subsubsection{Dirichlet with log-Normal
Weight}\label{dirichlet-with-log-normal-weight}

\begin{figure}[htbp]
\centering
\includegraphics{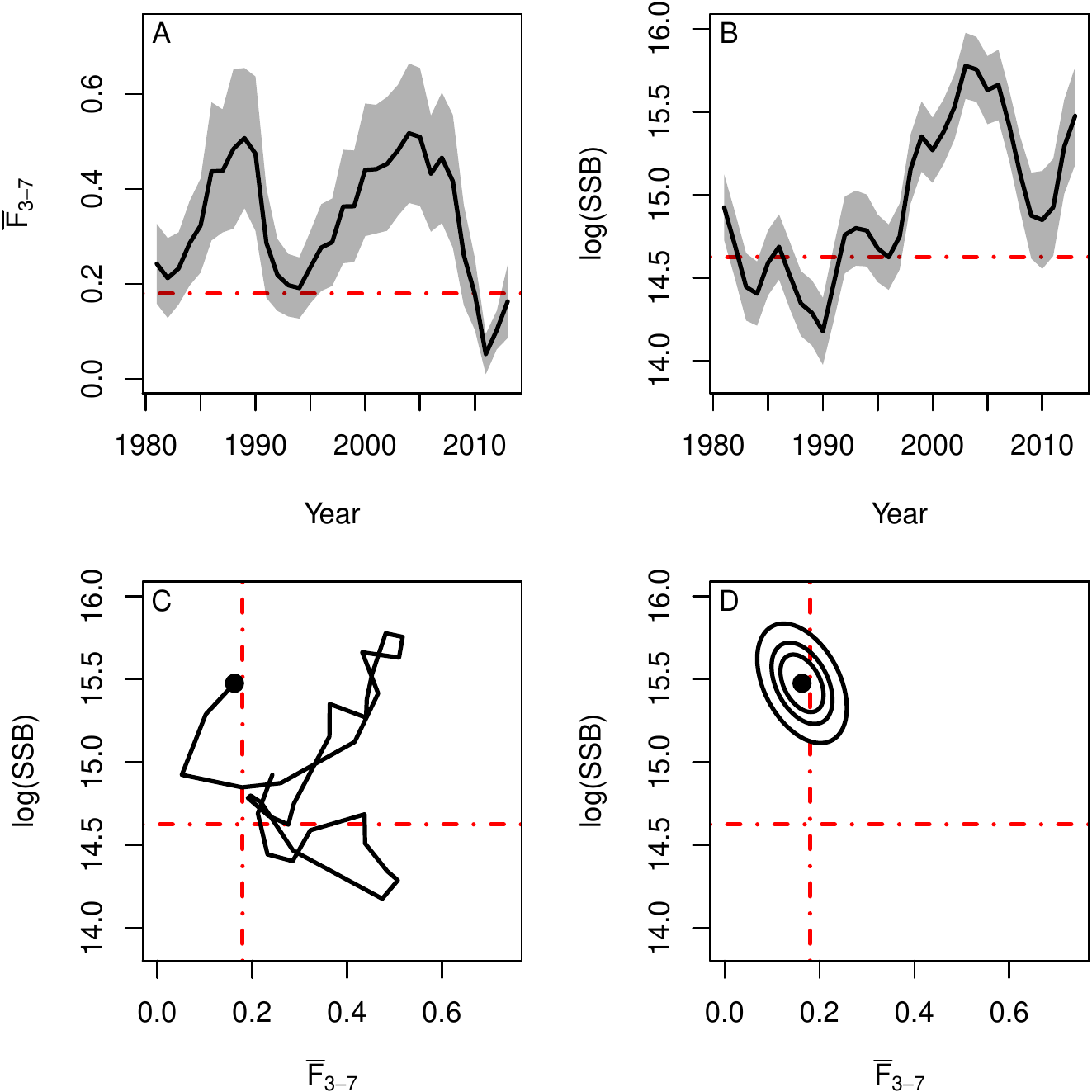}
\caption{Estimated average fishing mortality (A) and log spawning stock
biomass (B) with the Dirichlet with log-Normal Weight model for Blue
Whiting including 95 \% pointwise confidence intervals (grey area);
their estimated trajectory (C); and confidence ellipses in the final
year (D) at 50 \%, 75 \% and 95 \% levels. The red lines indicate the
management plan reference points while the black point is the estimated
value in the final year.}
\end{figure}

\clearpage

\subsection{North-East Arctic Haddock}\label{north-east-arctic-haddock}

\subsubsection{log-Normal}\label{log-normal-1}

\begin{figure}[htbp]
\centering
\includegraphics{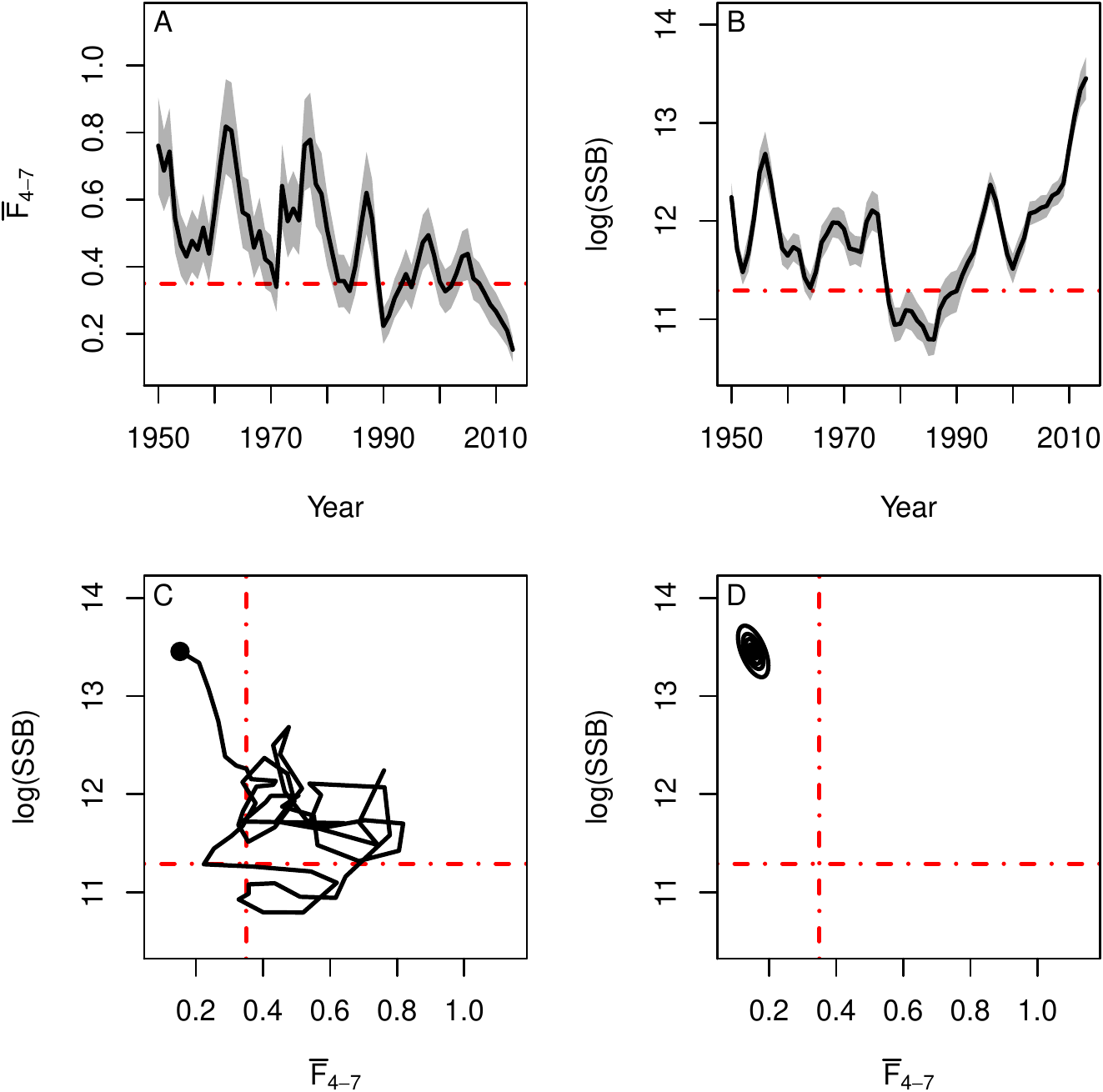}
\caption{Estimated average fishing mortality (A) and log spawning stock
biomass (B) with the log-Normal model for North-East Arctic Haddock
including 95 \% pointwise confidence intervals (grey area); their
estimated trajectory (C); and confidence ellipses in the final year (D)
at 50 \%, 75 \% and 95 \% levels. The red lines indicate the management
plan reference points while the black point is the estimated value in
the final year.}
\end{figure}

\clearpage

\subsubsection{Gamma}\label{gamma-1}

\begin{figure}[htbp]
\centering
\includegraphics{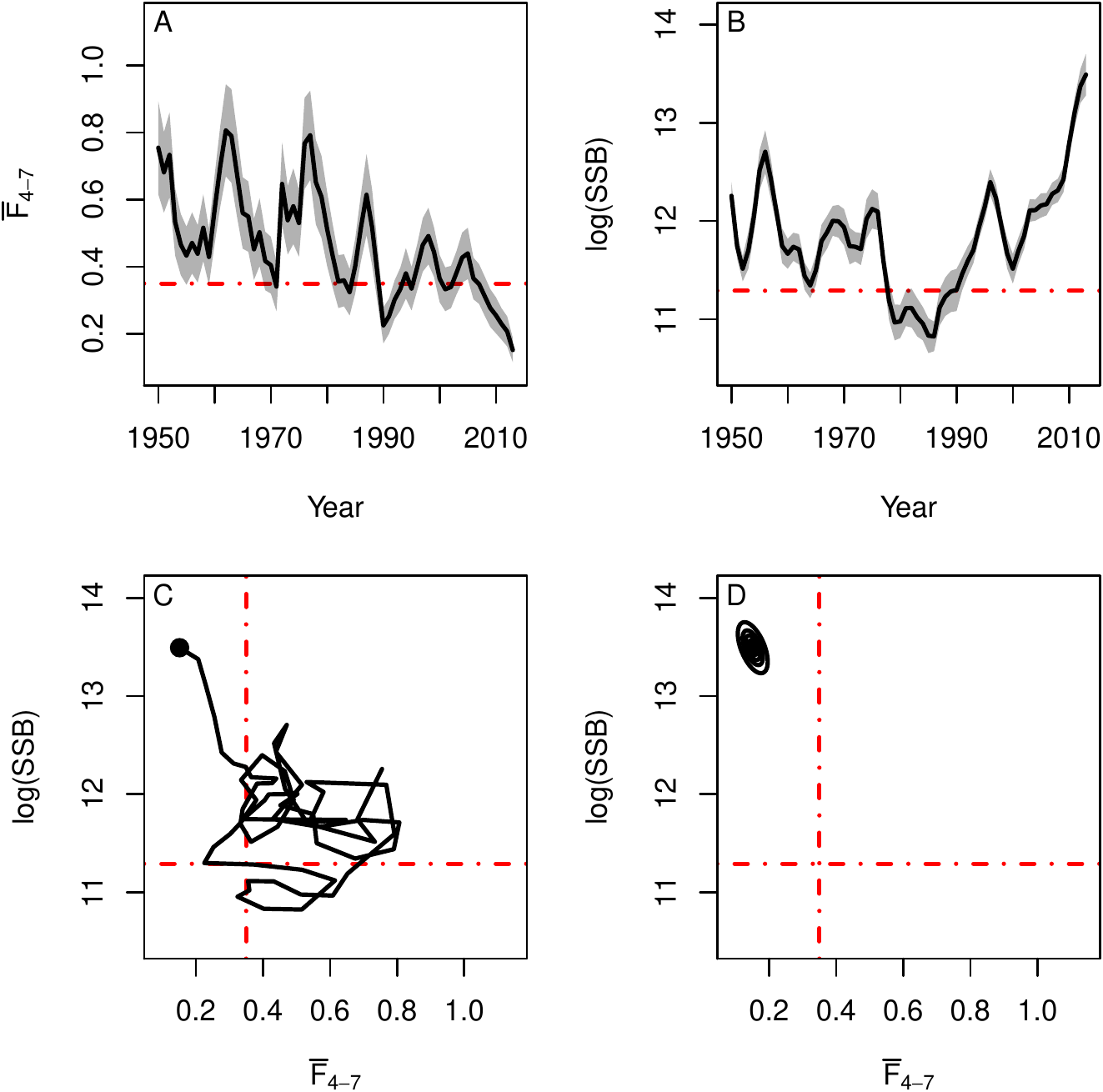}
\caption{Estimated average fishing mortality (A) and log spawning stock
biomass (B) with the Gamma model for North-East Arctic Haddock including
95 \% pointwise confidence intervals (grey area); their estimated
trajectory (C); and confidence ellipses in the final year (D) at 50 \%,
75 \% and 95 \% levels. The red lines indicate the management plan
reference points while the black point is the estimated value in the
final year.}
\end{figure}

\clearpage

\subsubsection{Generalized Gamma}\label{generalized-gamma-1}

\begin{figure}[htbp]
\centering
\includegraphics{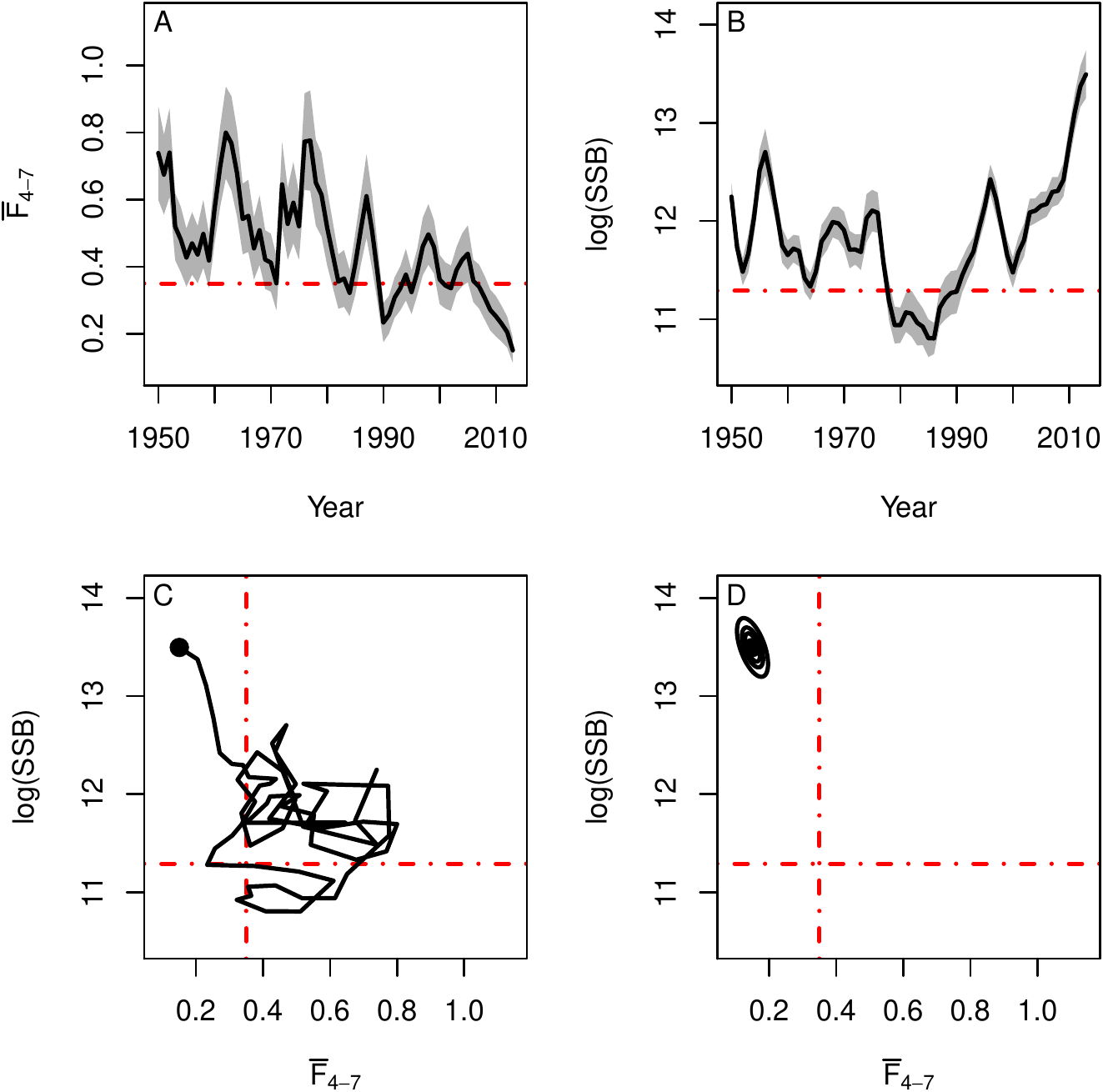}
\caption{Estimated average fishing mortality (A) and log spawning stock
biomass (B) with the Generalized Gamma model for North-East Arctic
Haddock including 95 \% pointwise confidence intervals (grey area);
their estimated trajectory (C); and confidence ellipses in the final
year (D) at 50 \%, 75 \% and 95 \% levels. The red lines indicate the
management plan reference points while the black point is the estimated
value in the final year.}
\end{figure}

\clearpage

\subsubsection{Normal}\label{normal-1}

\begin{figure}[htbp]
\centering
\includegraphics{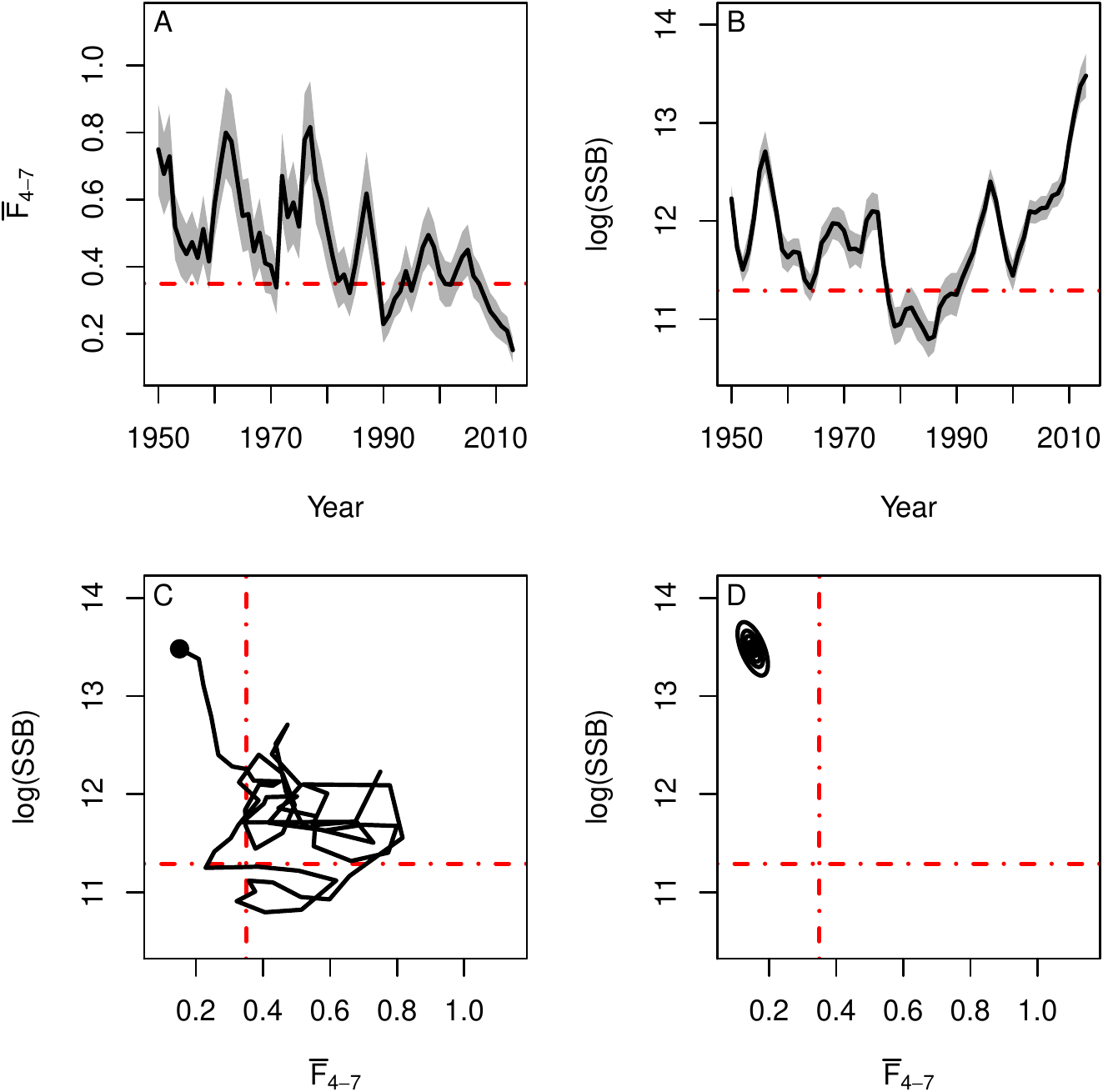}
\caption{Estimated average fishing mortality (A) and log spawning stock
biomass (B) with the Normal model for North-East Arctic Haddock
including 95 \% pointwise confidence intervals (grey area); their
estimated trajectory (C); and confidence ellipses in the final year (D)
at 50 \%, 75 \% and 95 \% levels. The red lines indicate the management
plan reference points while the black point is the estimated value in
the final year.}
\end{figure}

\clearpage

\subsubsection{Left Truncated Normal}\label{left-truncated-normal-1}

\begin{figure}[htbp]
\centering
\includegraphics{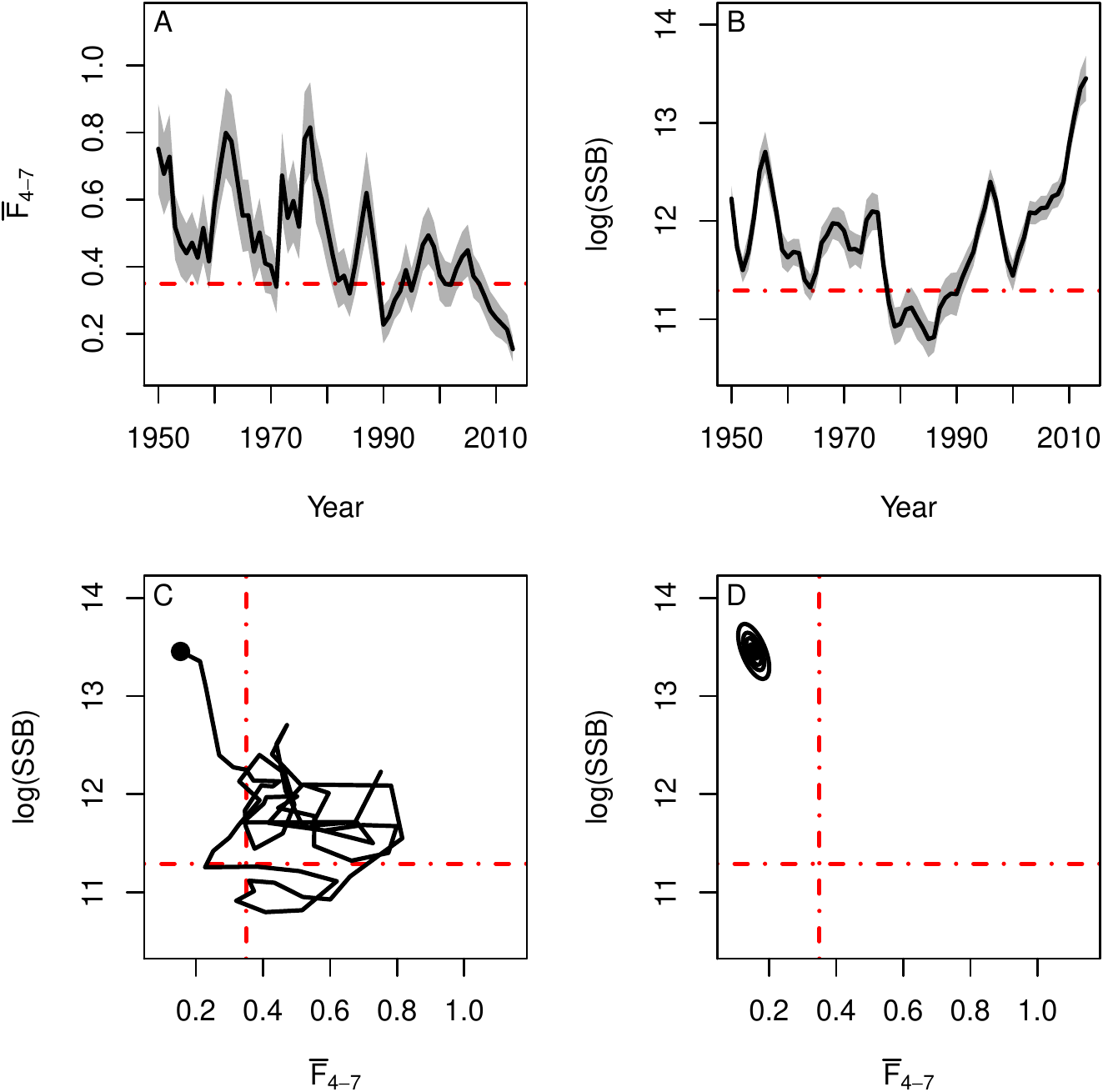}
\caption{Estimated average fishing mortality (A) and log spawning stock
biomass (B) with the Left Truncated Normal model for North-East Arctic
Haddock including 95 \% pointwise confidence intervals (grey area);
their estimated trajectory (C); and confidence ellipses in the final
year (D) at 50 \%, 75 \% and 95 \% levels. The red lines indicate the
management plan reference points while the black point is the estimated
value in the final year.}
\end{figure}

\clearpage

\subsubsection{log-Students t}\label{log-students-t-1}

\begin{figure}[htbp]
\centering
\includegraphics{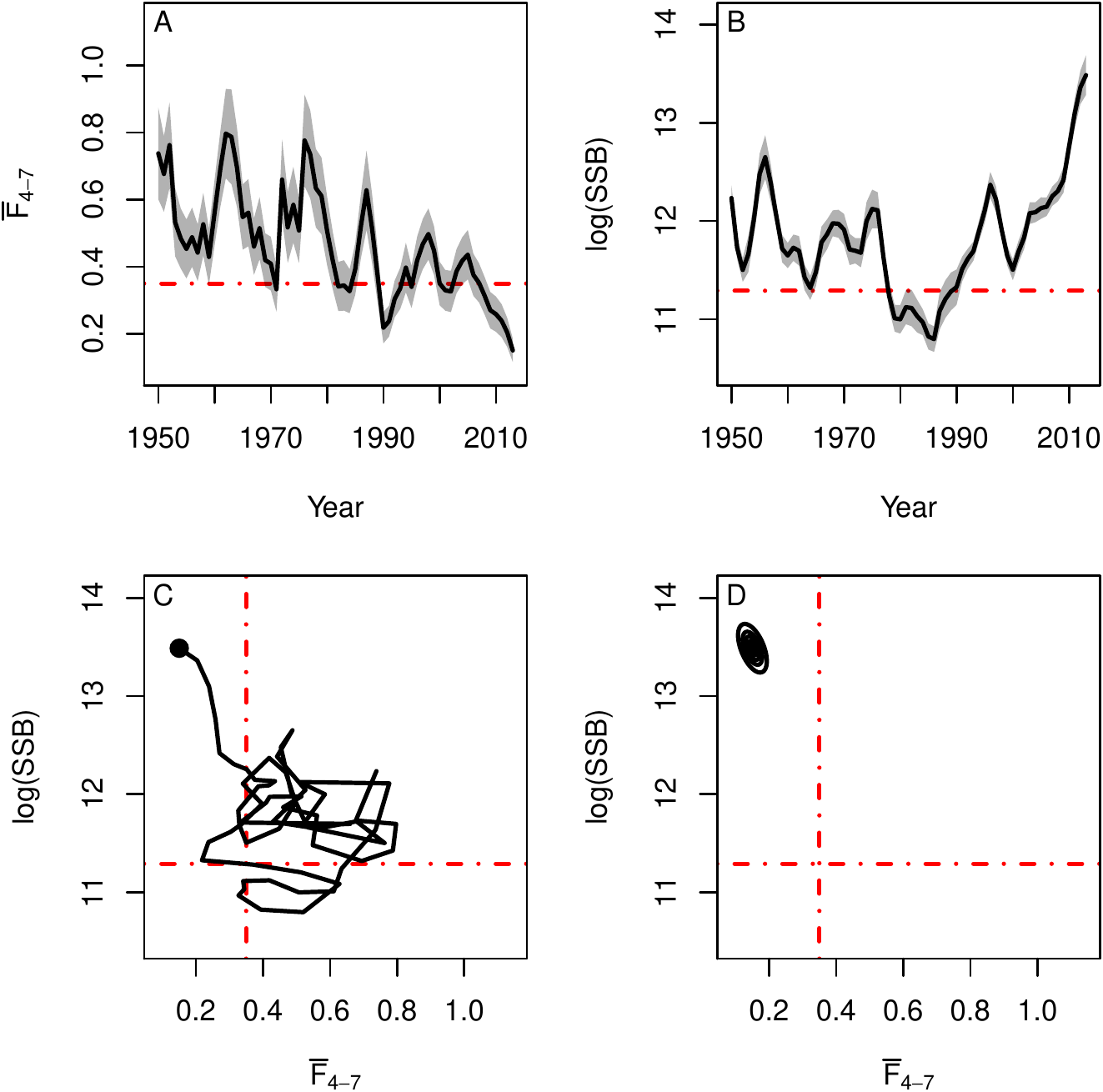}
\caption{Estimated average fishing mortality (A) and log spawning stock
biomass (B) with the log-Students t model for North-East Arctic Haddock
including 95 \% pointwise confidence intervals (grey area); their
estimated trajectory (C); and confidence ellipses in the final year (D)
at 50 \%, 75 \% and 95 \% levels. The red lines indicate the management
plan reference points while the black point is the estimated value in
the final year.}
\end{figure}

\clearpage

\subsubsection{Multivariate log-Normal}\label{multivariate-log-normal-1}

\begin{figure}[htbp]
\centering
\includegraphics{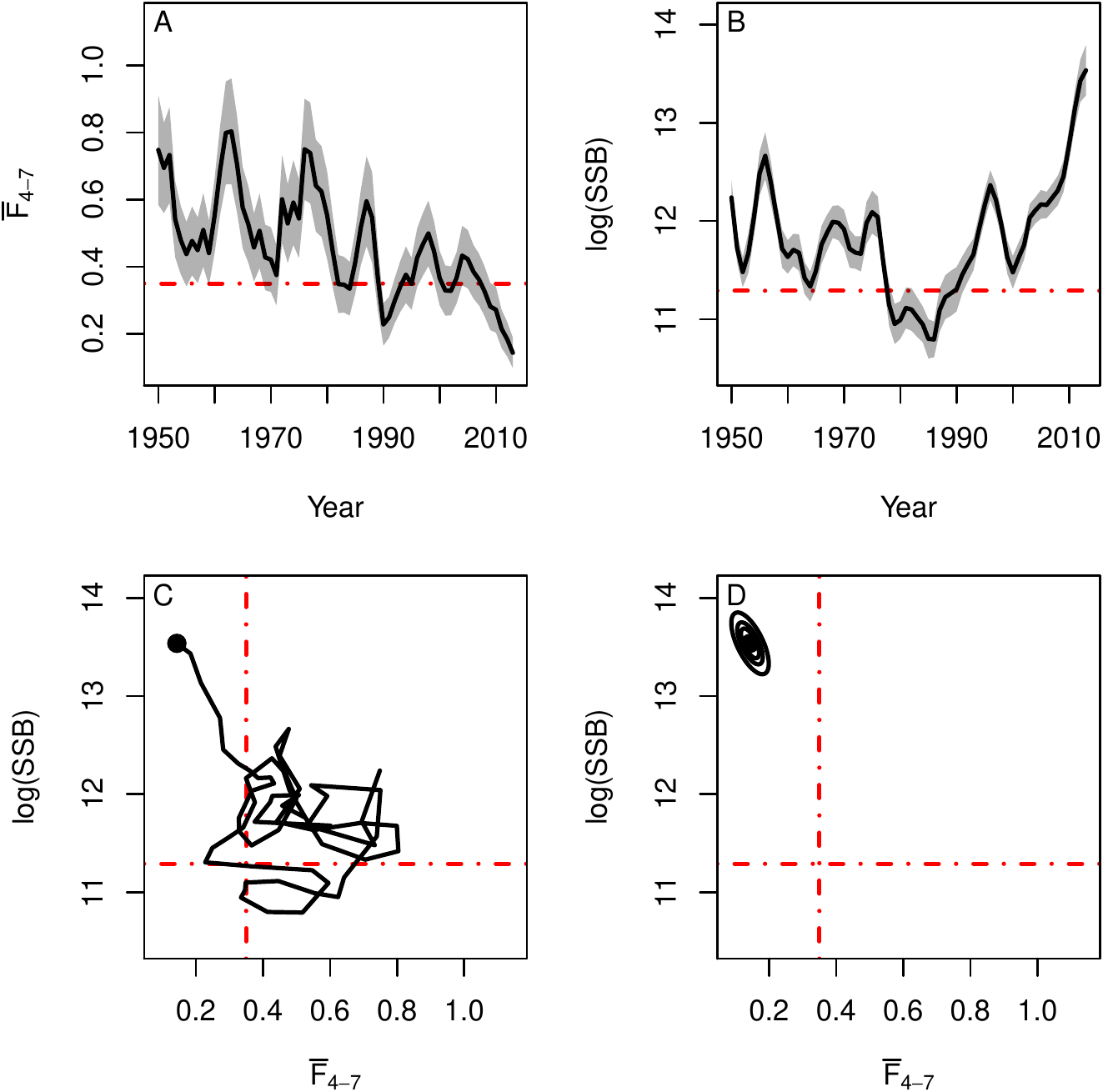}
\caption{Estimated average fishing mortality (A) and log spawning stock
biomass (B) with the Multivariate log-Normal model for North-East Arctic
Haddock including 95 \% pointwise confidence intervals (grey area);
their estimated trajectory (C); and confidence ellipses in the final
year (D) at 50 \%, 75 \% and 95 \% levels. The red lines indicate the
management plan reference points while the black point is the estimated
value in the final year.}
\end{figure}

\clearpage

\subsubsection{Additive Logistic Normal with log-Normal
Numbers}\label{additive-logistic-normal-with-log-normal-numbers-1}

\begin{figure}[htbp]
\centering
\includegraphics{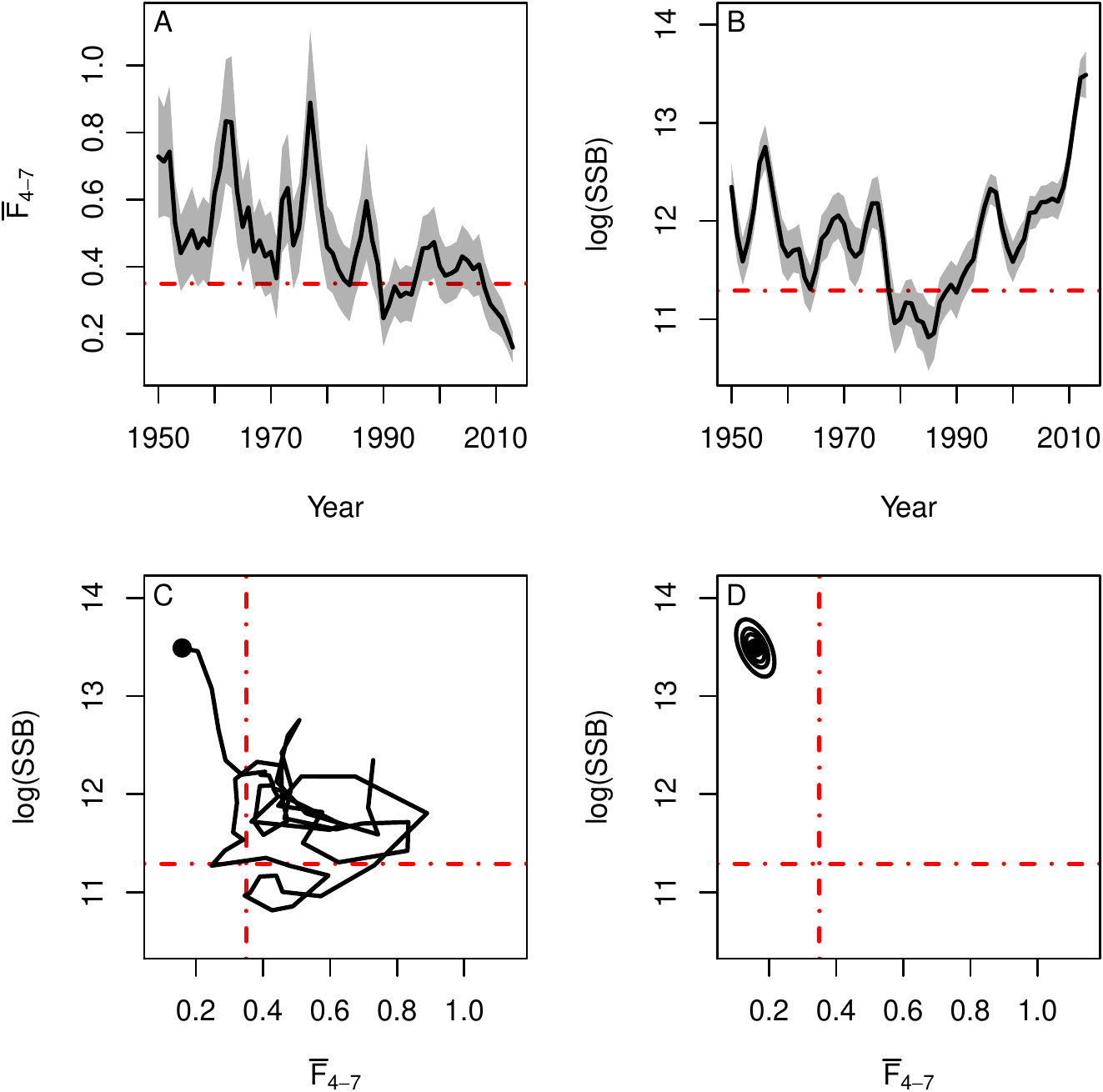}
\caption{Estimated average fishing mortality (A) and log spawning stock
biomass (B) with the Additive Logistic Normal with log-Normal Numbers
model for North-East Arctic Haddock including 95 \% pointwise confidence
intervals (grey area); their estimated trajectory (C); and confidence
ellipses in the final year (D) at 50 \%, 75 \% and 95 \% levels. The red
lines indicate the management plan reference points while the black
point is the estimated value in the final year.}
\end{figure}

\clearpage

\subsubsection{Multiplicative Logistic Normal with log-Normal
Numbers}\label{multiplicative-logistic-normal-with-log-normal-numbers-1}

\begin{figure}[htbp]
\centering
\includegraphics{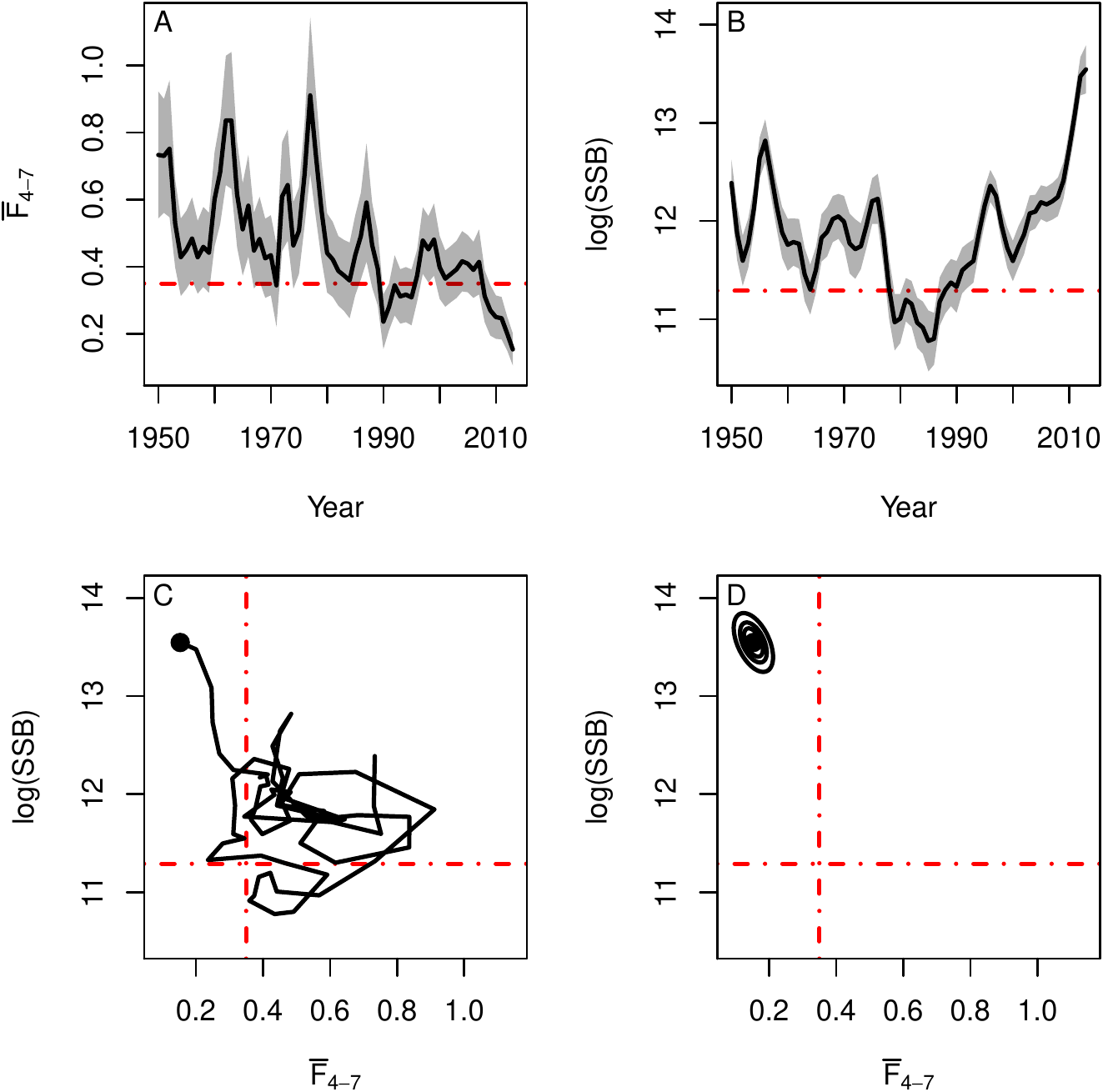}
\caption{Estimated average fishing mortality (A) and log spawning stock
biomass (B) with the Multiplicative Logistic Normal with log-Normal
Numbers model for North-East Arctic Haddock including 95 \% pointwise
confidence intervals (grey area); their estimated trajectory (C); and
confidence ellipses in the final year (D) at 50 \%, 75 \% and 95 \%
levels. The red lines indicate the management plan reference points
while the black point is the estimated value in the final year.}
\end{figure}

\clearpage

\subsubsection{Dirichlet with log-Normal
Numbers}\label{dirichlet-with-log-normal-numbers-1}

\begin{figure}[htbp]
\centering
\includegraphics{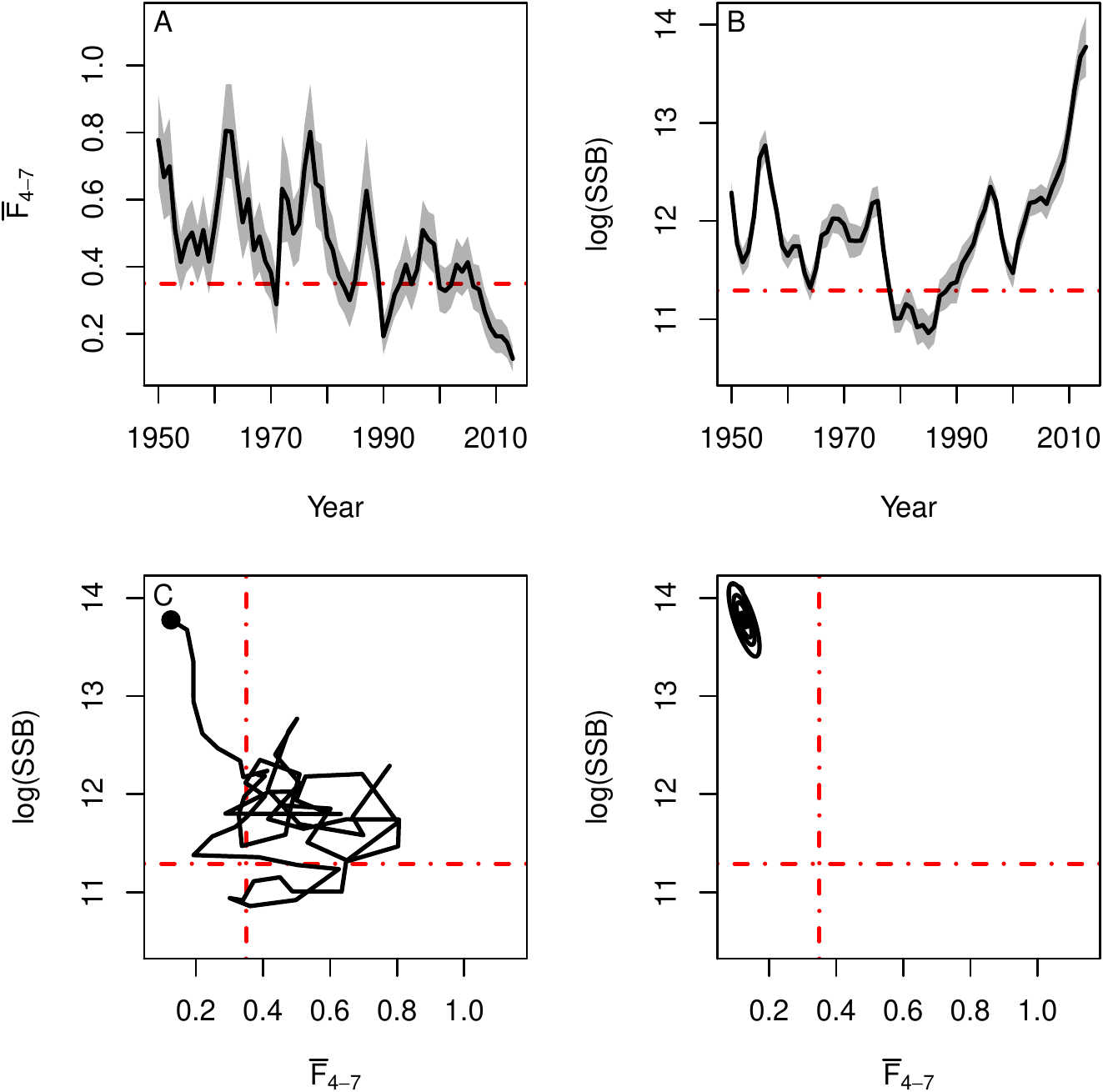}
\caption{Estimated average fishing mortality (A) and log spawning stock
biomass (B) with the Dirichlet with log-Normal Numbers model for
North-East Arctic Haddock including 95 \% pointwise confidence intervals
(grey area); their estimated trajectory (C); and confidence ellipses in
the final year (D) at 50 \%, 75 \% and 95 \% levels. The red lines
indicate the management plan reference points while the black point is
the estimated value in the final year.}
\end{figure}

\clearpage

\subsubsection{Additive Logisitc Normal with log-Normal
Weight}\label{additive-logisitc-normal-with-log-normal-weight-1}

\begin{figure}[htbp]
\centering
\includegraphics{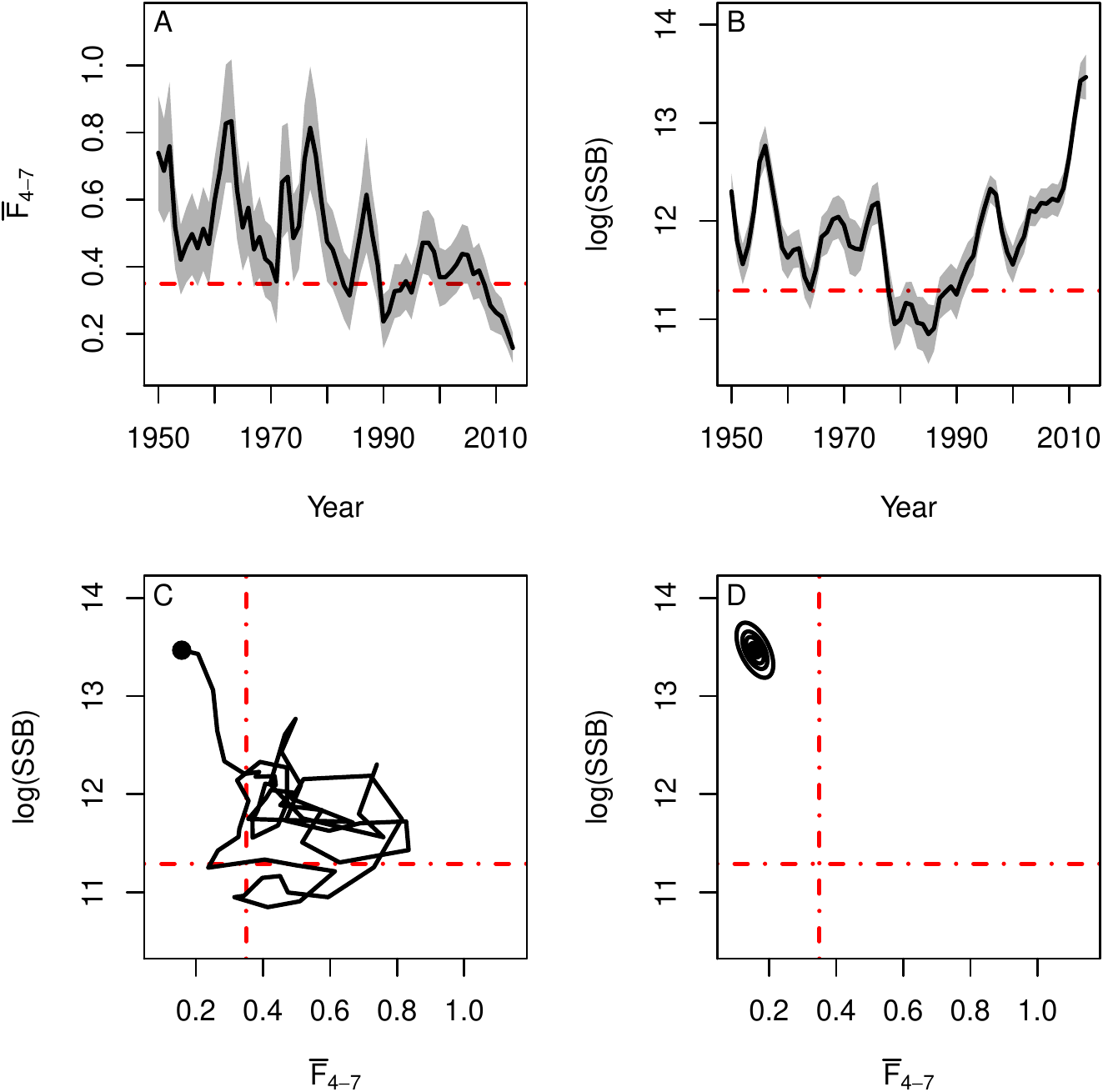}
\caption{Estimated average fishing mortality (A) and log spawning stock
biomass (B) with the Additive Logisitc Normal with log-Normal Weight
model for North-East Arctic Haddock including 95 \% pointwise confidence
intervals (grey area); their estimated trajectory (C); and confidence
ellipses in the final year (D) at 50 \%, 75 \% and 95 \% levels. The red
lines indicate the management plan reference points while the black
point is the estimated value in the final year.}
\end{figure}

\clearpage

\subsubsection{Multiplicative Logistic Normal with log-Normal
Weight}\label{multiplicative-logistic-normal-with-log-normal-weight-1}

\begin{figure}[htbp]
\centering
\includegraphics{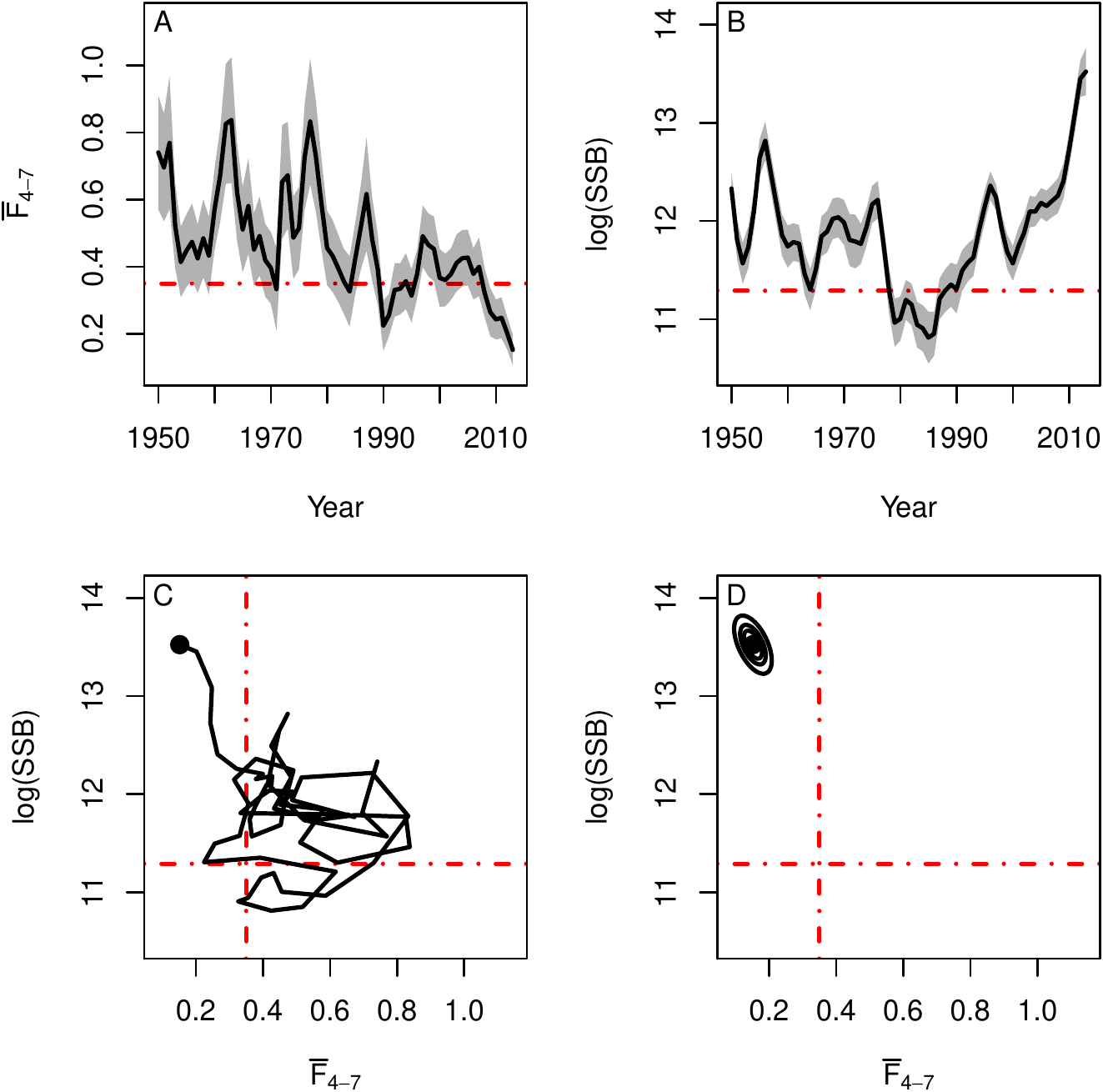}
\caption{Estimated average fishing mortality (A) and log spawning stock
biomass (B) with the Multiplicative Logistic Normal with log-Normal
Weight model for North-East Arctic Haddock including 95 \% pointwise
confidence intervals (grey area); their estimated trajectory (C); and
confidence ellipses in the final year (D) at 50 \%, 75 \% and 95 \%
levels. The red lines indicate the management plan reference points
while the black point is the estimated value in the final year.}
\end{figure}

\clearpage

\subsubsection{Dirichlet with log-Normal
Weight}\label{dirichlet-with-log-normal-weight-1}

\begin{figure}[htbp]
\centering
\includegraphics{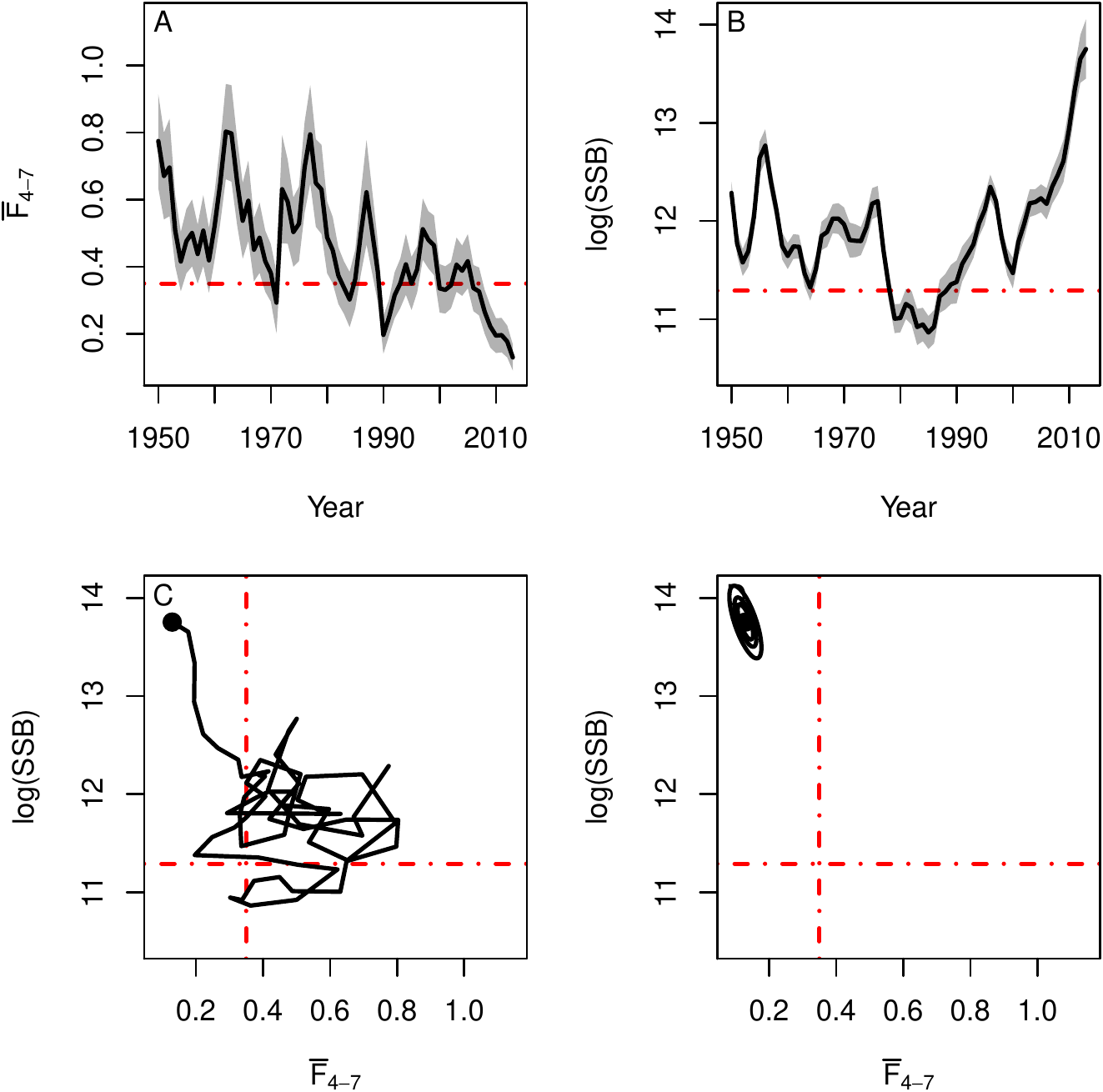}
\caption{Estimated average fishing mortality (A) and log spawning stock
biomass (B) with the Dirichlet with log-Normal Weight model for
North-East Arctic Haddock including 95 \% pointwise confidence intervals
(grey area); their estimated trajectory (C); and confidence ellipses in
the final year (D) at 50 \%, 75 \% and 95 \% levels. The red lines
indicate the management plan reference points while the black point is
the estimated value in the final year.}
\end{figure}

\clearpage

\subsection{North Sea Cod}\label{north-sea-cod}

\subsubsection{log-Normal}\label{log-normal-2}

\begin{figure}[htbp]
\centering
\includegraphics{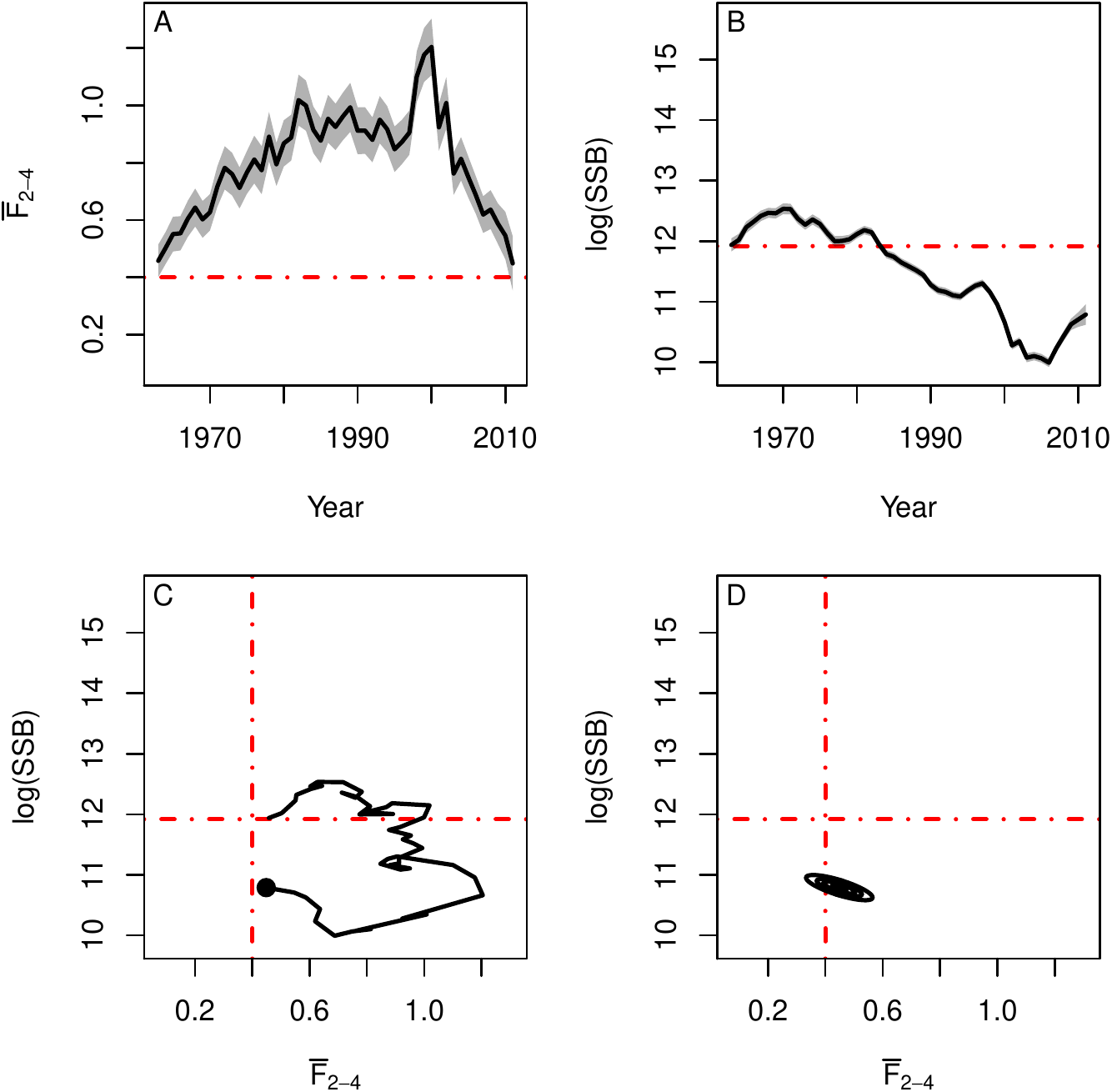}
\caption{Estimated average fishing mortality (A) and log spawning stock
biomass (B) with the log-Normal model for North Sea Cod including 95 \%
pointwise confidence intervals (grey area); their estimated trajectory
(C); and confidence ellipses in the final year (D) at 50 \%, 75 \% and
95 \% levels. The red lines indicate the management plan reference
points while the black point is the estimated value in the final year.}
\end{figure}

\clearpage

\subsubsection{Gamma}\label{gamma-2}

\begin{figure}[htbp]
\centering
\includegraphics{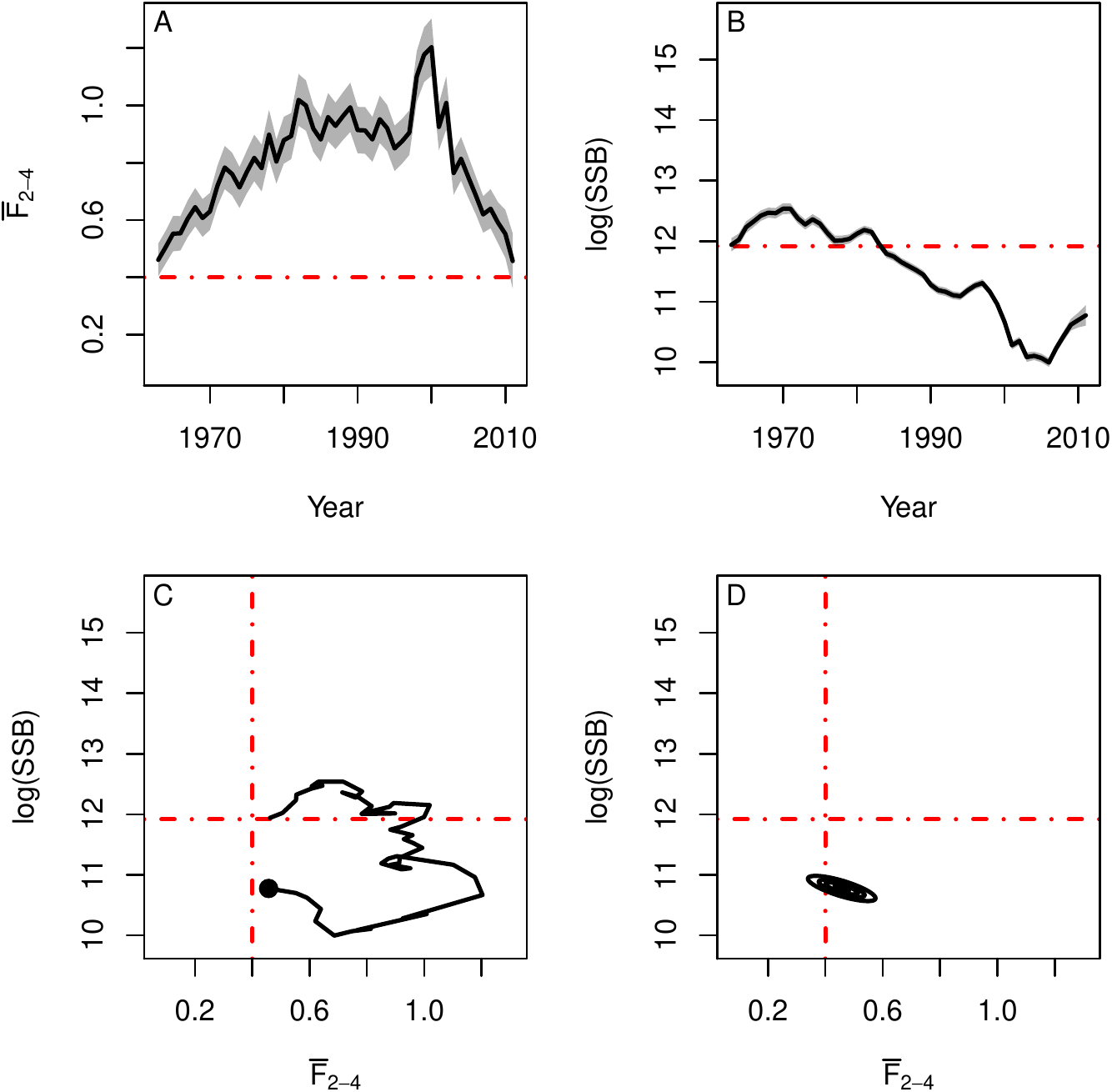}
\caption{Estimated average fishing mortality (A) and log spawning stock
biomass (B) with the Gamma model for North Sea Cod including 95 \%
pointwise confidence intervals (grey area); their estimated trajectory
(C); and confidence ellipses in the final year (D) at 50 \%, 75 \% and
95 \% levels. The red lines indicate the management plan reference
points while the black point is the estimated value in the final year.}
\end{figure}

\clearpage

\subsubsection{Generalized Gamma}\label{generalized-gamma-2}

\begin{figure}[htbp]
\centering
\includegraphics{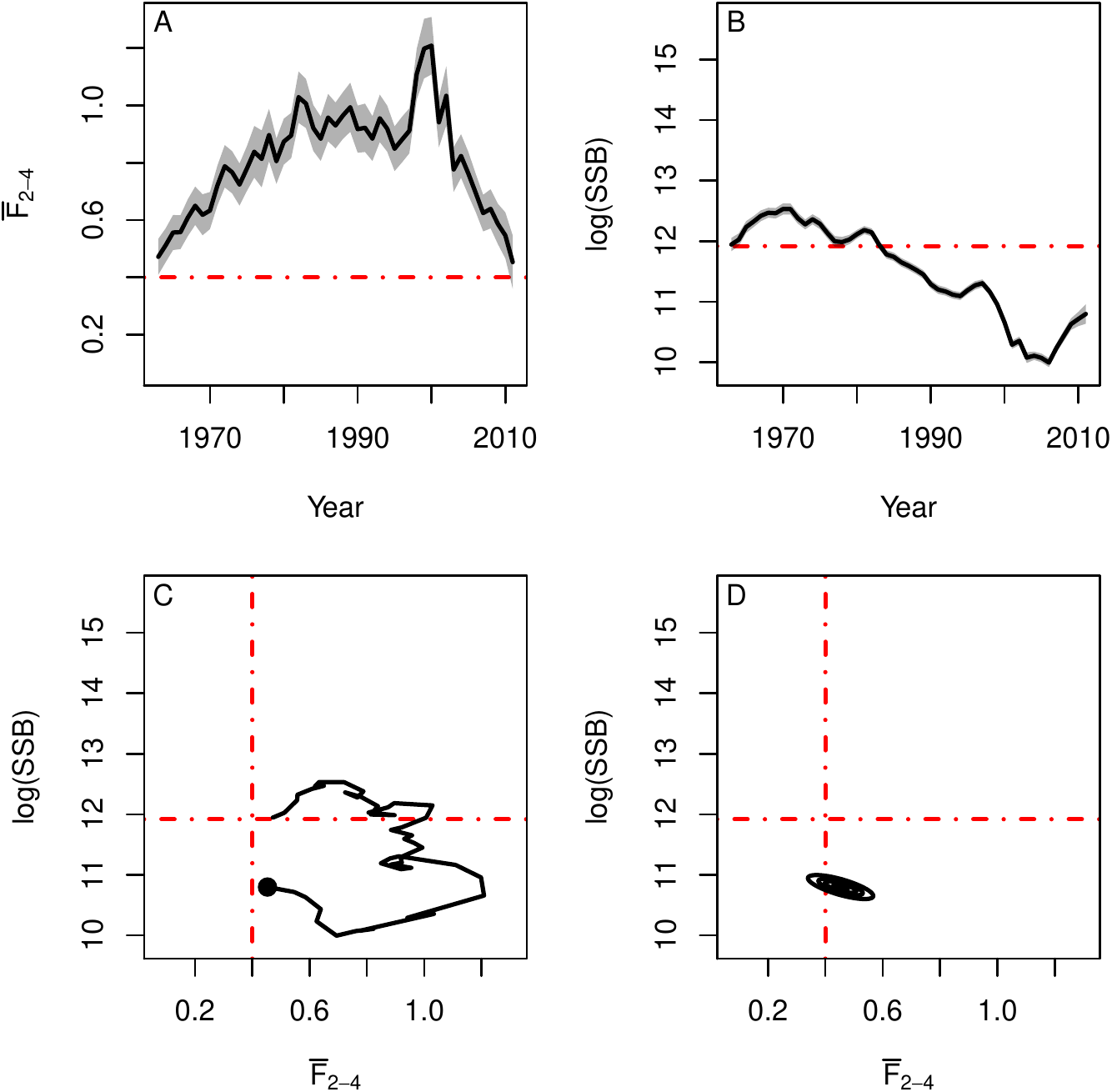}
\caption{Estimated average fishing mortality (A) and log spawning stock
biomass (B) with the Generalized Gamma model for North Sea Cod including
95 \% pointwise confidence intervals (grey area); their estimated
trajectory (C); and confidence ellipses in the final year (D) at 50 \%,
75 \% and 95 \% levels. The red lines indicate the management plan
reference points while the black point is the estimated value in the
final year.}
\end{figure}

\clearpage

\subsubsection{Normal}\label{normal-2}

\begin{figure}[htbp]
\centering
\includegraphics{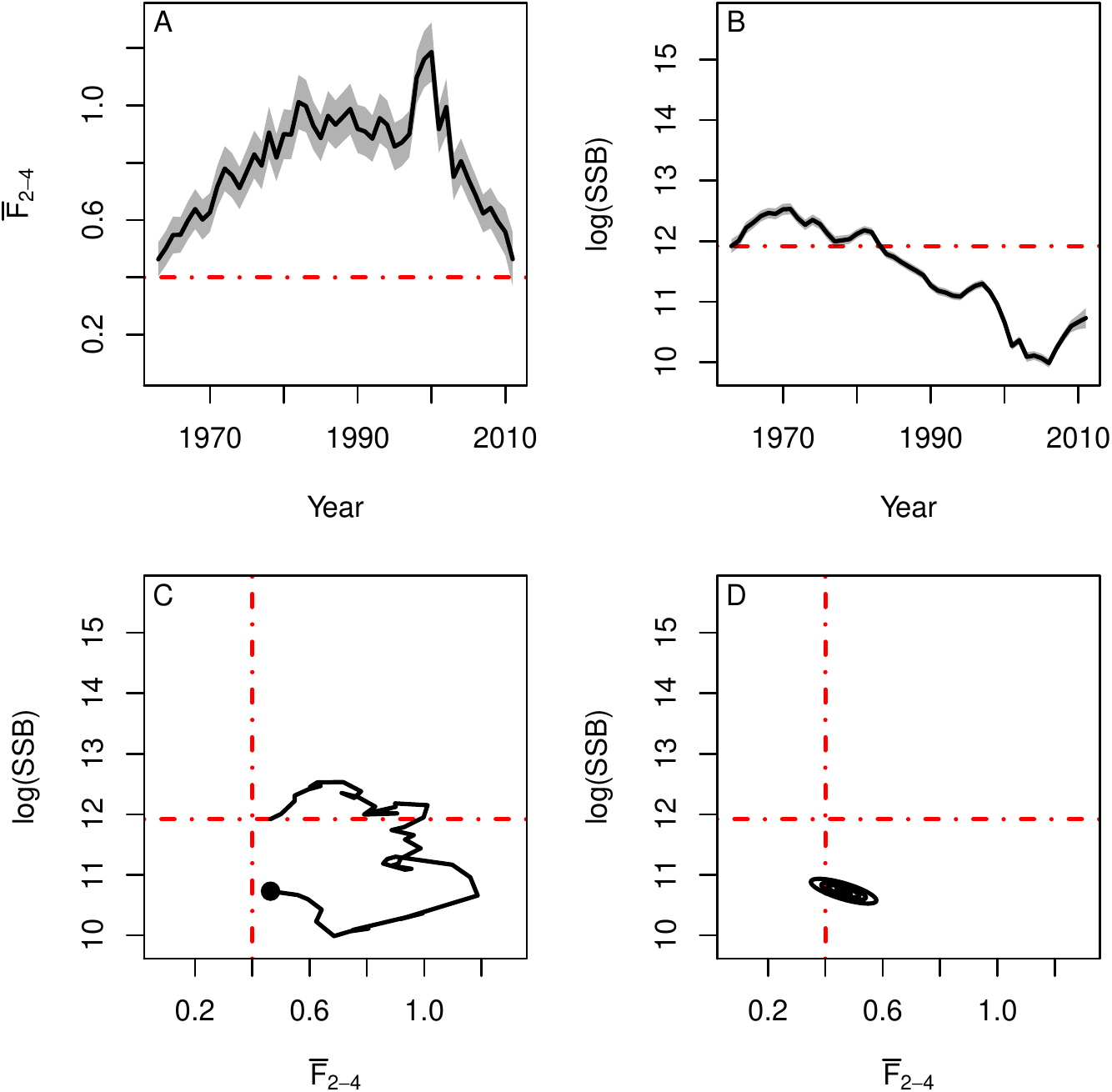}
\caption{Estimated average fishing mortality (A) and log spawning stock
biomass (B) with the Normal model for North Sea Cod including 95 \%
pointwise confidence intervals (grey area); their estimated trajectory
(C); and confidence ellipses in the final year (D) at 50 \%, 75 \% and
95 \% levels. The red lines indicate the management plan reference
points while the black point is the estimated value in the final year.}
\end{figure}

\clearpage

\subsubsection{Left Truncated Normal}\label{left-truncated-normal-2}

\begin{figure}[htbp]
\centering
\includegraphics{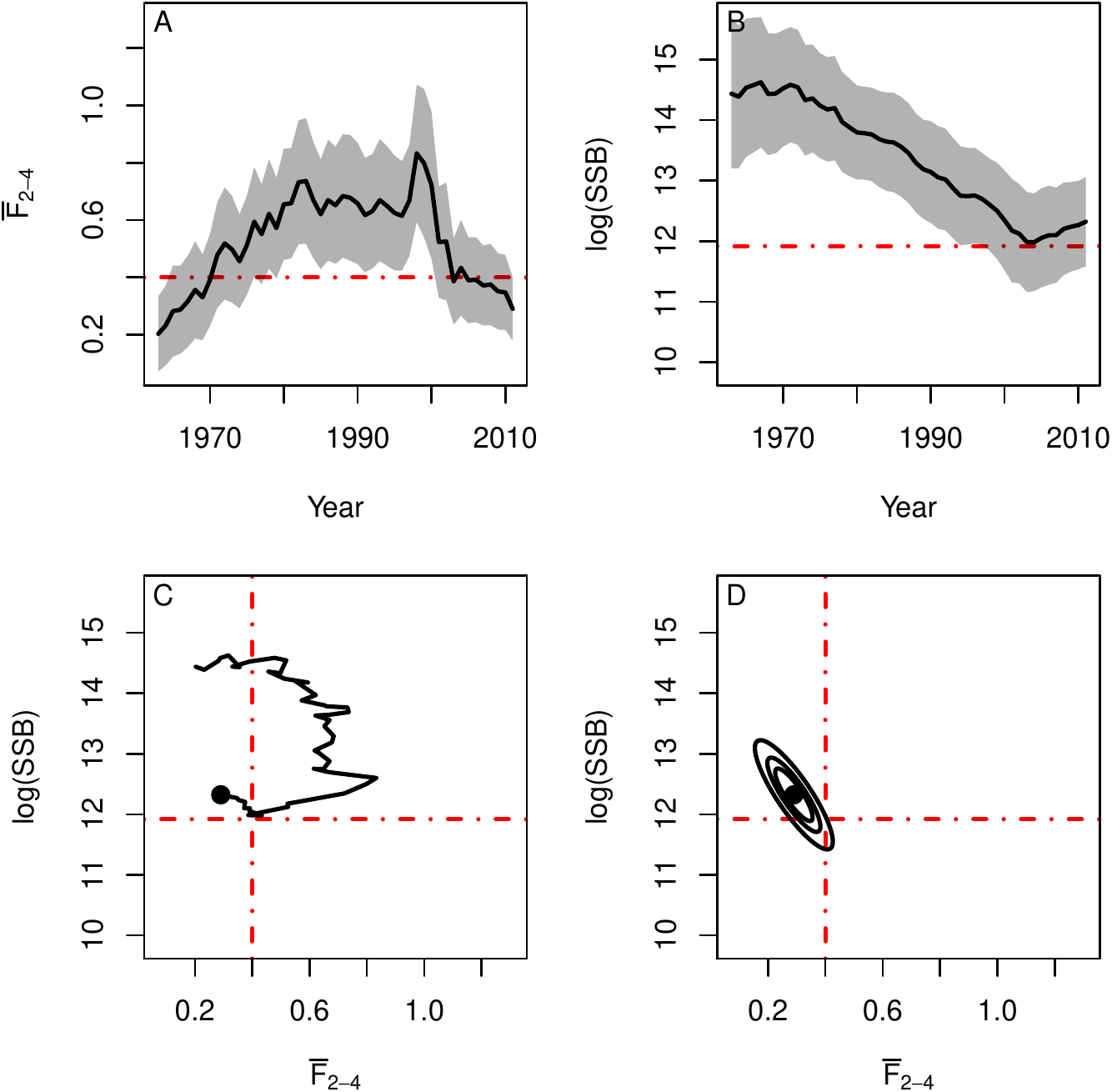}
\caption{Estimated average fishing mortality (A) and log spawning stock
biomass (B) with the Left Truncated Normal model for North Sea Cod
including 95 \% pointwise confidence intervals (grey area); their
estimated trajectory (C); and confidence ellipses in the final year (D)
at 50 \%, 75 \% and 95 \% levels. The red lines indicate the management
plan reference points while the black point is the estimated value in
the final year.}
\end{figure}

\clearpage

\subsubsection{log-Students t}\label{log-students-t-2}

\begin{figure}[htbp]
\centering
\includegraphics{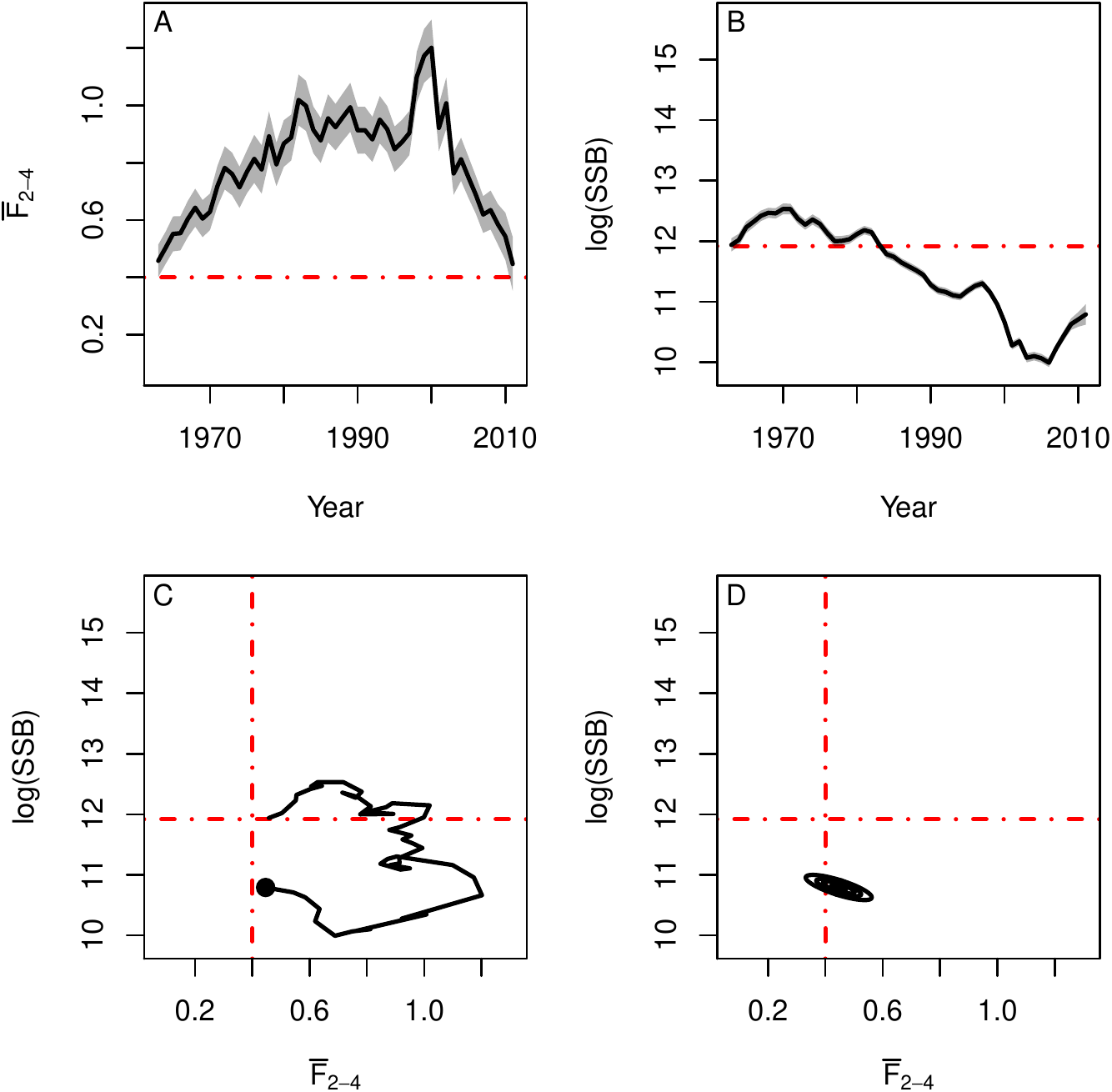}
\caption{Estimated average fishing mortality (A) and log spawning stock
biomass (B) with the log-Students t model for North Sea Cod including 95
\% pointwise confidence intervals (grey area); their estimated
trajectory (C); and confidence ellipses in the final year (D) at 50 \%,
75 \% and 95 \% levels. The red lines indicate the management plan
reference points while the black point is the estimated value in the
final year.}
\end{figure}

\clearpage

\subsubsection{Multivariate log-Normal}\label{multivariate-log-normal-2}

\begin{figure}[htbp]
\centering
\includegraphics{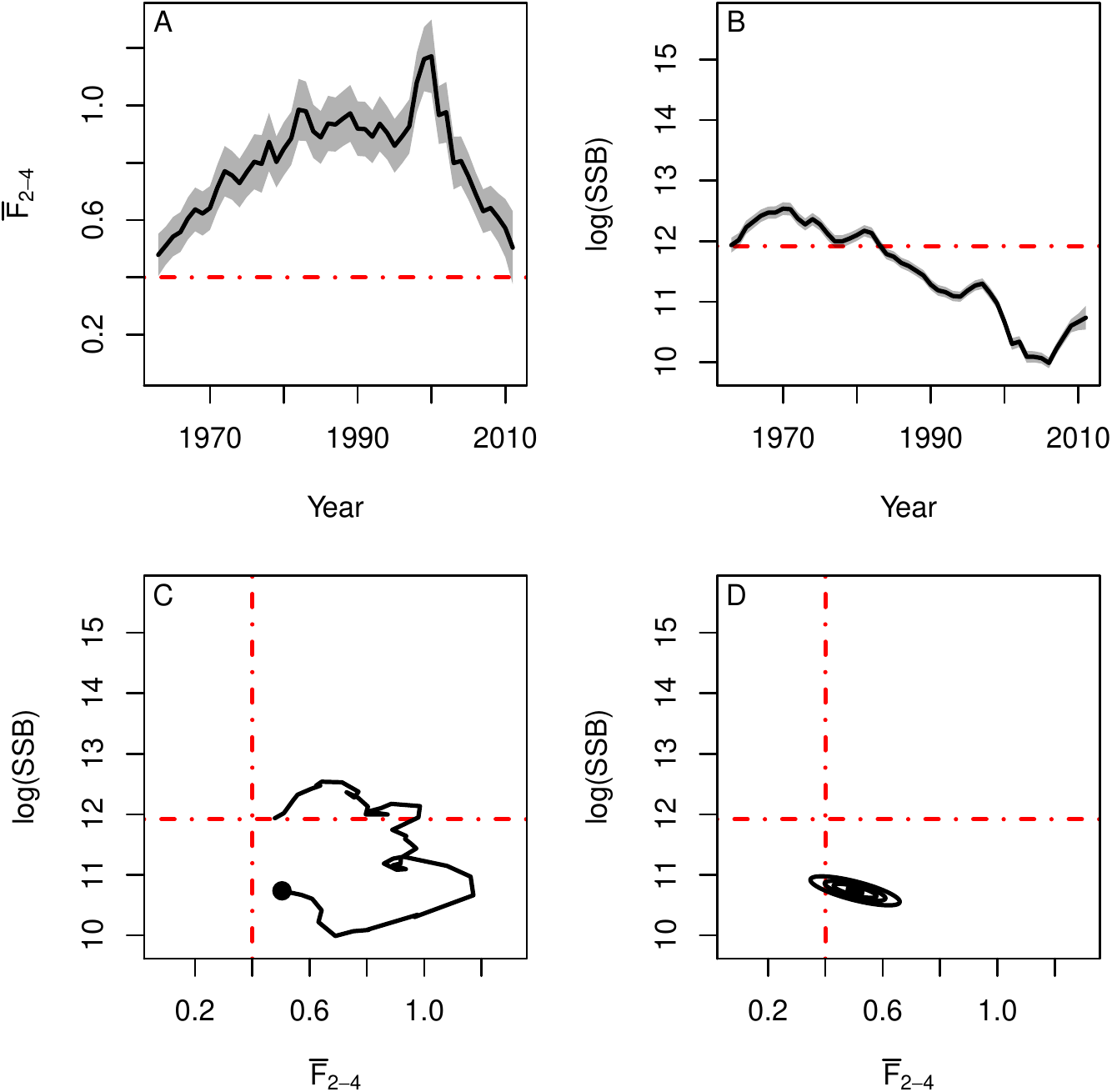}
\caption{Estimated average fishing mortality (A) and log spawning stock
biomass (B) with the Multivariate log-Normal model for North Sea Cod
including 95 \% pointwise confidence intervals (grey area); their
estimated trajectory (C); and confidence ellipses in the final year (D)
at 50 \%, 75 \% and 95 \% levels. The red lines indicate the management
plan reference points while the black point is the estimated value in
the final year.}
\end{figure}

\clearpage

\subsubsection{Additive Logistic Normal with log-Normal
Numbers}\label{additive-logistic-normal-with-log-normal-numbers-2}

\begin{figure}[htbp]
\centering
\includegraphics{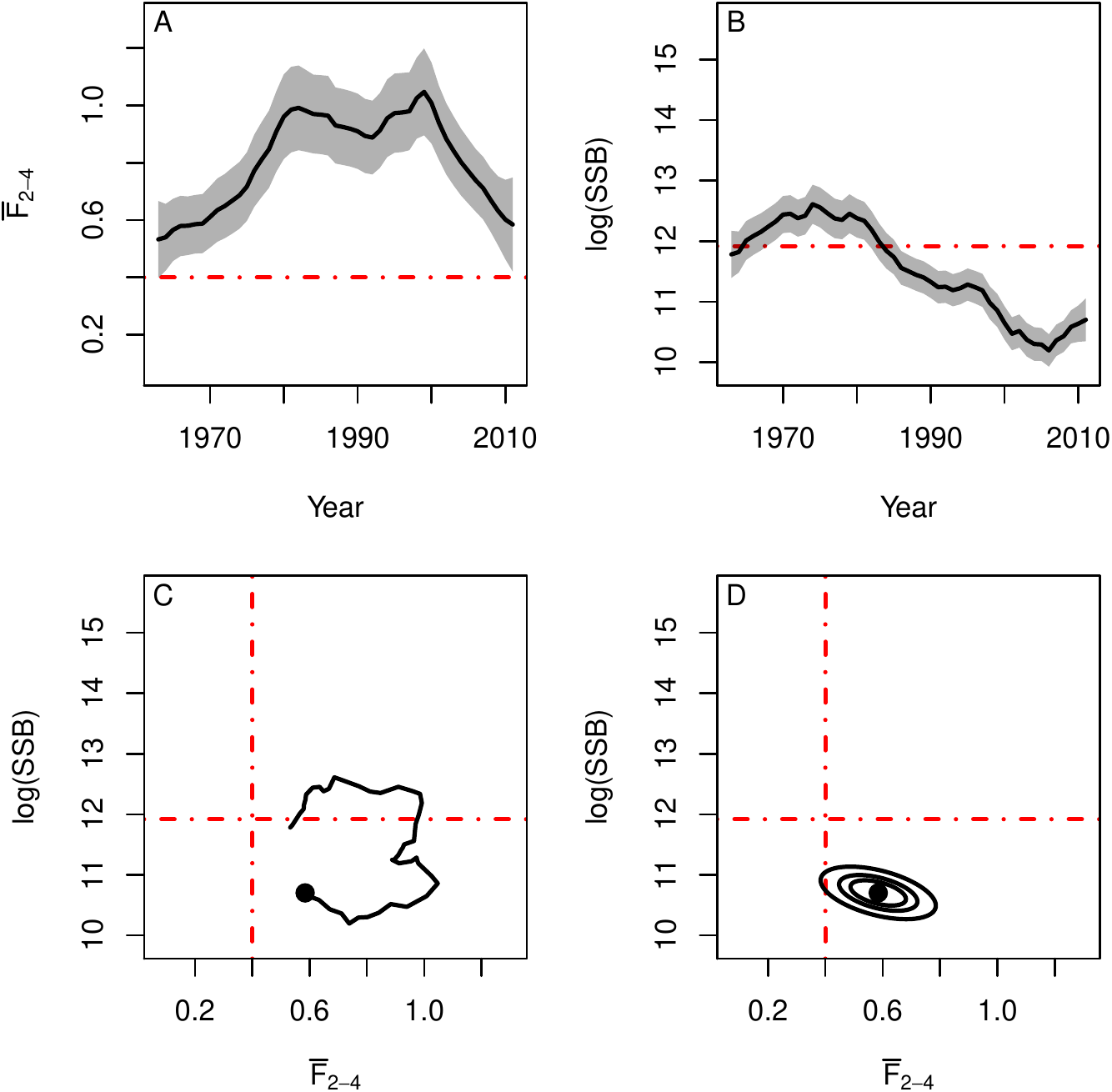}
\caption{Estimated average fishing mortality (A) and log spawning stock
biomass (B) with the Additive Logistic Normal with log-Normal Numbers
model for North Sea Cod including 95 \% pointwise confidence intervals
(grey area); their estimated trajectory (C); and confidence ellipses in
the final year (D) at 50 \%, 75 \% and 95 \% levels. The red lines
indicate the management plan reference points while the black point is
the estimated value in the final year.}
\end{figure}

\clearpage

\subsubsection{Multiplicative Logistic Normal with log-Normal
Numbers}\label{multiplicative-logistic-normal-with-log-normal-numbers-2}

\begin{figure}[htbp]
\centering
\includegraphics{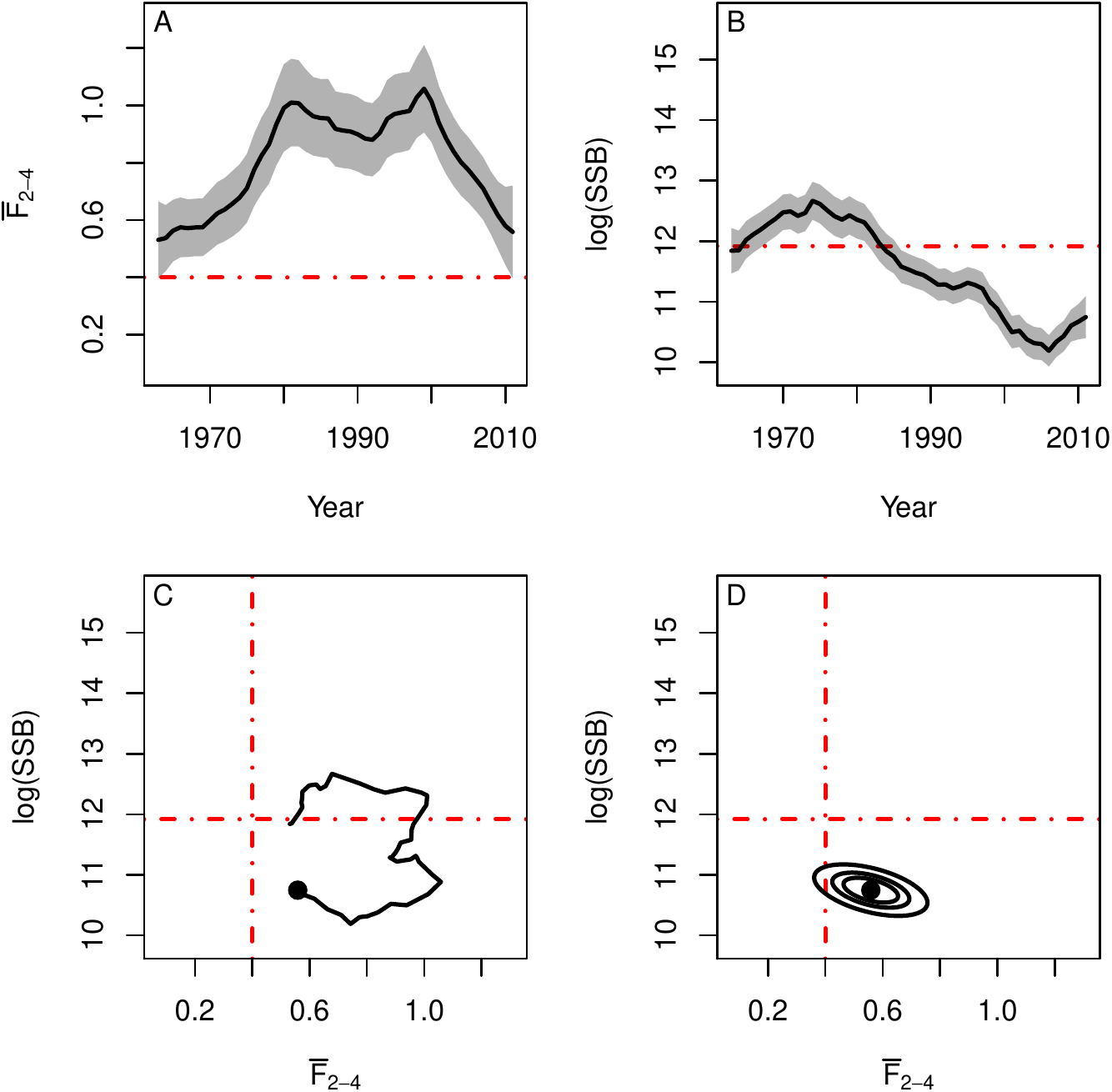}
\caption{Estimated average fishing mortality (A) and log spawning stock
biomass (B) with the Multiplicative Logistic Normal with log-Normal
Numbers model for North Sea Cod including 95 \% pointwise confidence
intervals (grey area); their estimated trajectory (C); and confidence
ellipses in the final year (D) at 50 \%, 75 \% and 95 \% levels. The red
lines indicate the management plan reference points while the black
point is the estimated value in the final year.}
\end{figure}

\clearpage

\subsubsection{Dirichlet with log-Normal
Numbers}\label{dirichlet-with-log-normal-numbers-2}

\begin{figure}[htbp]
\centering
\includegraphics{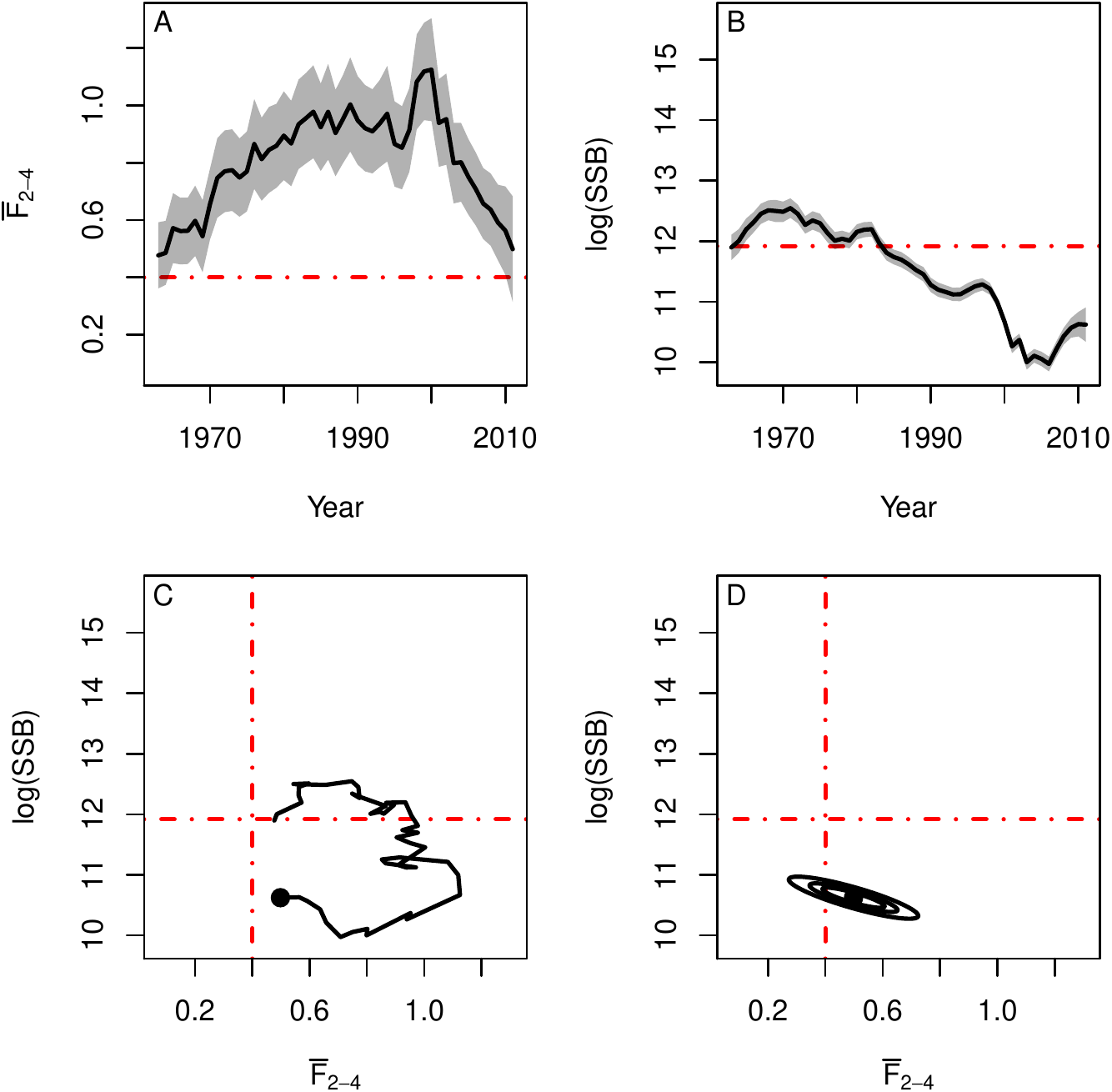}
\caption{Estimated average fishing mortality (A) and log spawning stock
biomass (B) with the Dirichlet with log-Normal Numbers model for North
Sea Cod including 95 \% pointwise confidence intervals (grey area);
their estimated trajectory (C); and confidence ellipses in the final
year (D) at 50 \%, 75 \% and 95 \% levels. The red lines indicate the
management plan reference points while the black point is the estimated
value in the final year.}
\end{figure}

\clearpage

\subsubsection{Additive Logisitc Normal with log-Normal
Weight}\label{additive-logisitc-normal-with-log-normal-weight-2}

\begin{figure}[htbp]
\centering
\includegraphics{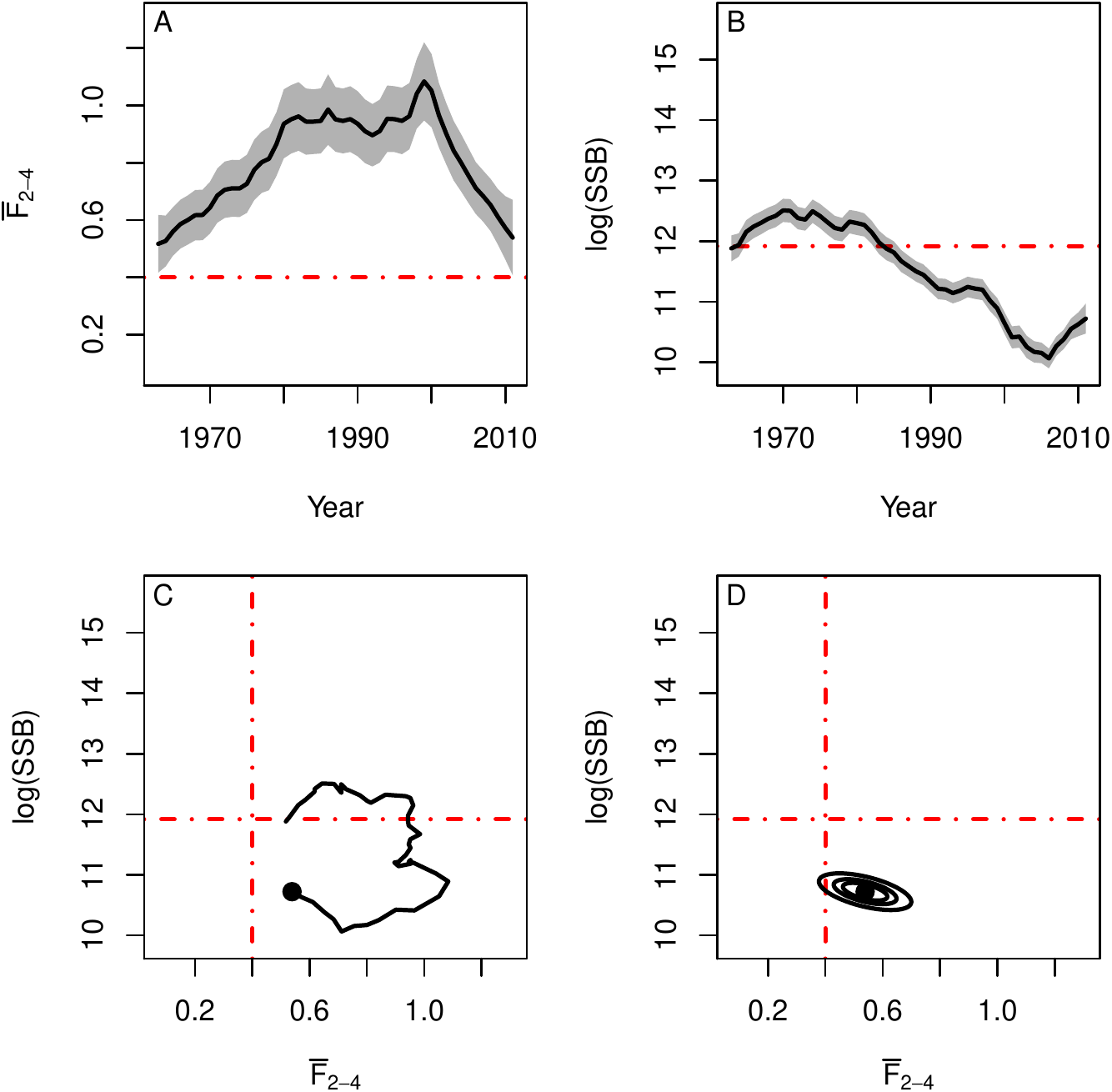}
\caption{Estimated average fishing mortality (A) and log spawning stock
biomass (B) with the Additive Logisitc Normal with log-Normal Weight
model for North Sea Cod including 95 \% pointwise confidence intervals
(grey area); their estimated trajectory (C); and confidence ellipses in
the final year (D) at 50 \%, 75 \% and 95 \% levels. The red lines
indicate the management plan reference points while the black point is
the estimated value in the final year.}
\end{figure}

\clearpage

\subsubsection{Multiplicative Logistic Normal with log-Normal
Weight}\label{multiplicative-logistic-normal-with-log-normal-weight-2}

\begin{figure}[htbp]
\centering
\includegraphics{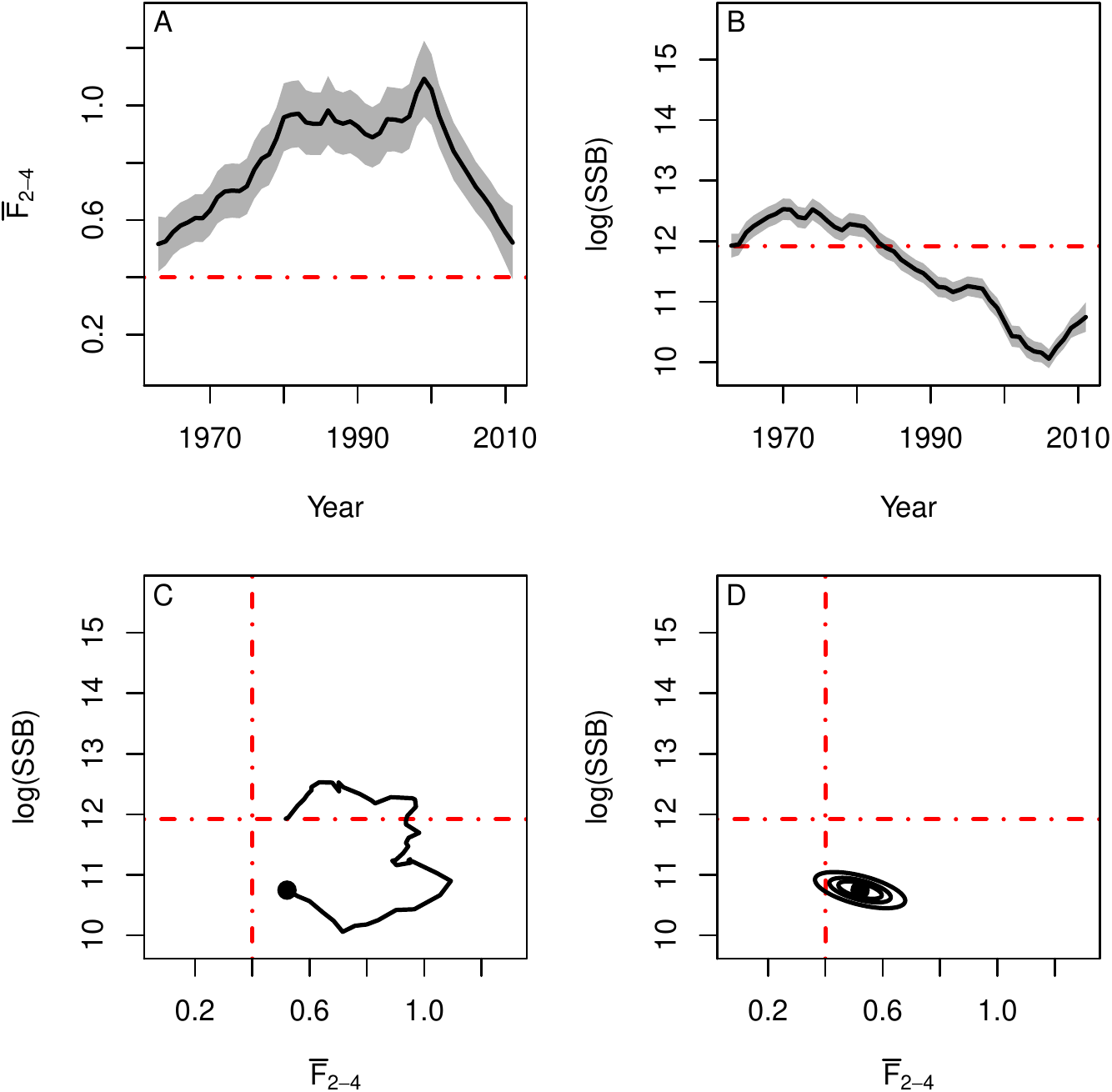}
\caption{Estimated average fishing mortality (A) and log spawning stock
biomass (B) with the Multiplicative Logistic Normal with log-Normal
Weight model for North Sea Cod including 95 \% pointwise confidence
intervals (grey area); their estimated trajectory (C); and confidence
ellipses in the final year (D) at 50 \%, 75 \% and 95 \% levels. The red
lines indicate the management plan reference points while the black
point is the estimated value in the final year.}
\end{figure}

\clearpage

\subsubsection{Dirichlet with log-Normal
Weight}\label{dirichlet-with-log-normal-weight-2}

\begin{figure}[htbp]
\centering
\includegraphics{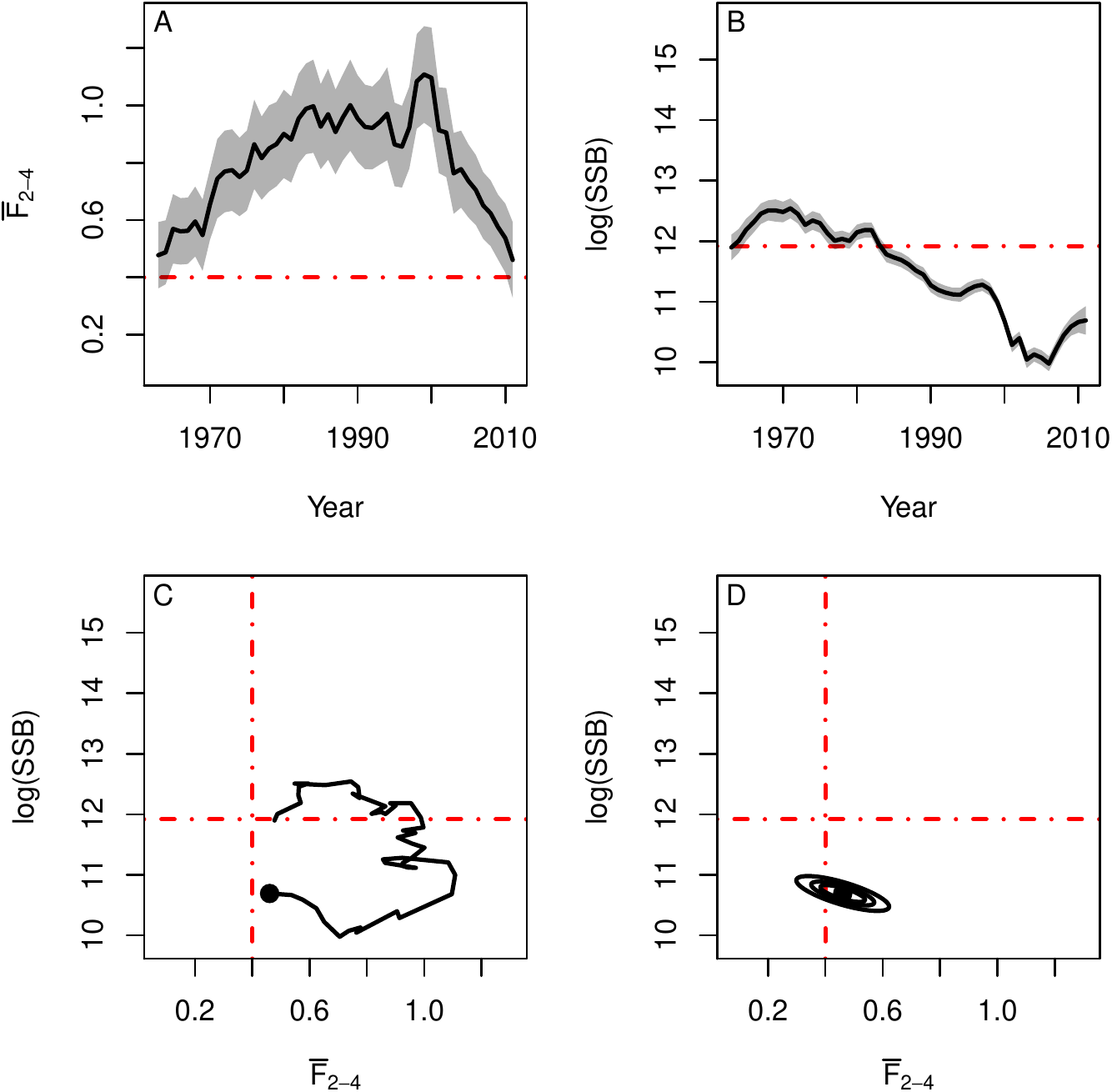}
\caption{Estimated average fishing mortality (A) and log spawning stock
biomass (B) with the Dirichlet with log-Normal Weight model for North
Sea Cod including 95 \% pointwise confidence intervals (grey area);
their estimated trajectory (C); and confidence ellipses in the final
year (D) at 50 \%, 75 \% and 95 \% levels. The red lines indicate the
management plan reference points while the black point is the estimated
value in the final year.}
\end{figure}

\clearpage

\subsection{Northern Shelf Haddock}\label{northern-shelf-haddock}

\subsubsection{log-Normal}\label{log-normal-3}

\begin{figure}[htbp]
\centering
\includegraphics{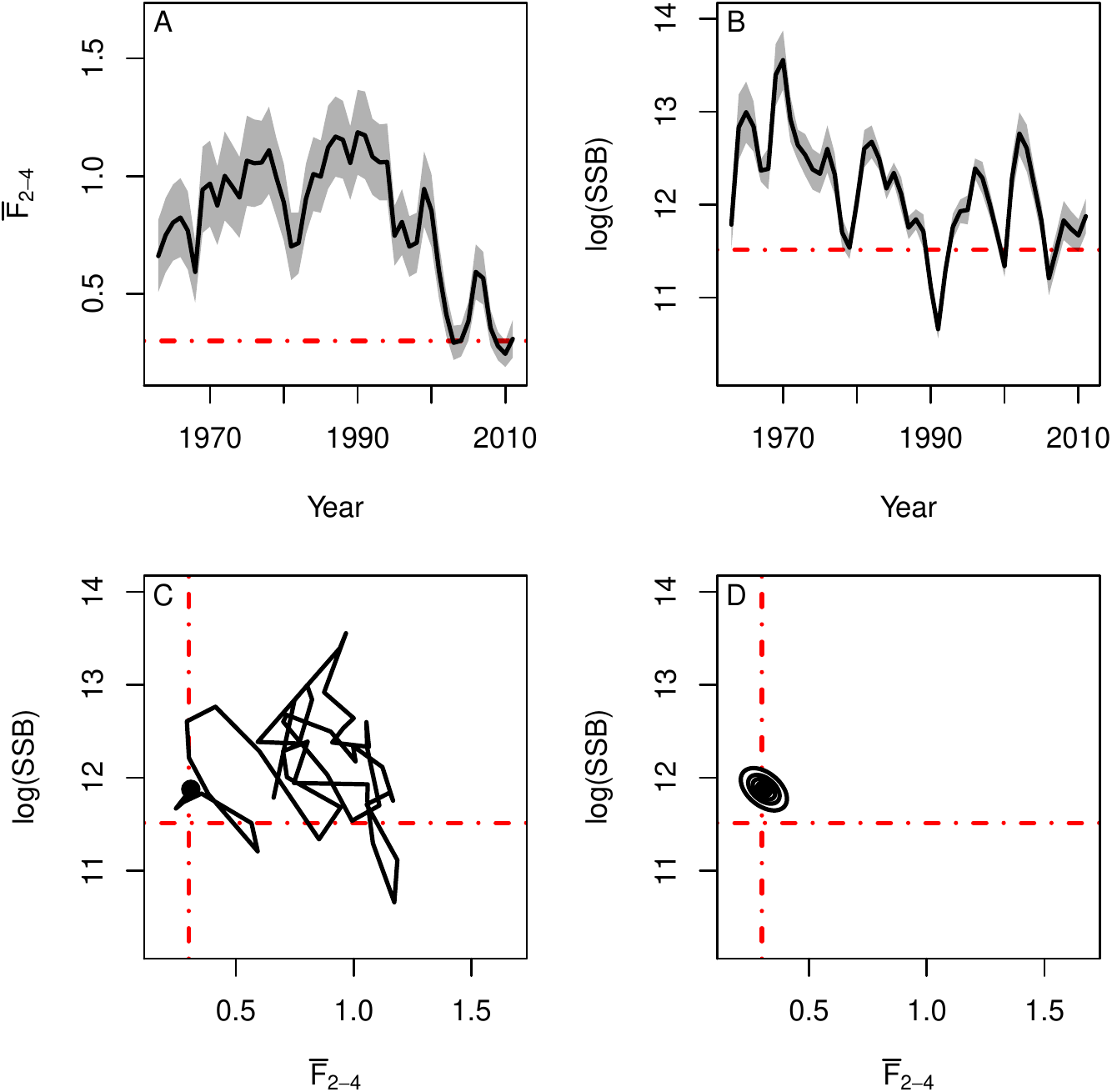}
\caption{Estimated average fishing mortality (A) and log spawning stock
biomass (B) with the log-Normal model for Northern Shelf Haddock
including 95 \% pointwise confidence intervals (grey area); their
estimated trajectory (C); and confidence ellipses in the final year (D)
at 50 \%, 75 \% and 95 \% levels. The red lines indicate the management
plan reference points while the black point is the estimated value in
the final year.}
\end{figure}

\clearpage

\subsubsection{Gamma}\label{gamma-3}

\begin{figure}[htbp]
\centering
\includegraphics{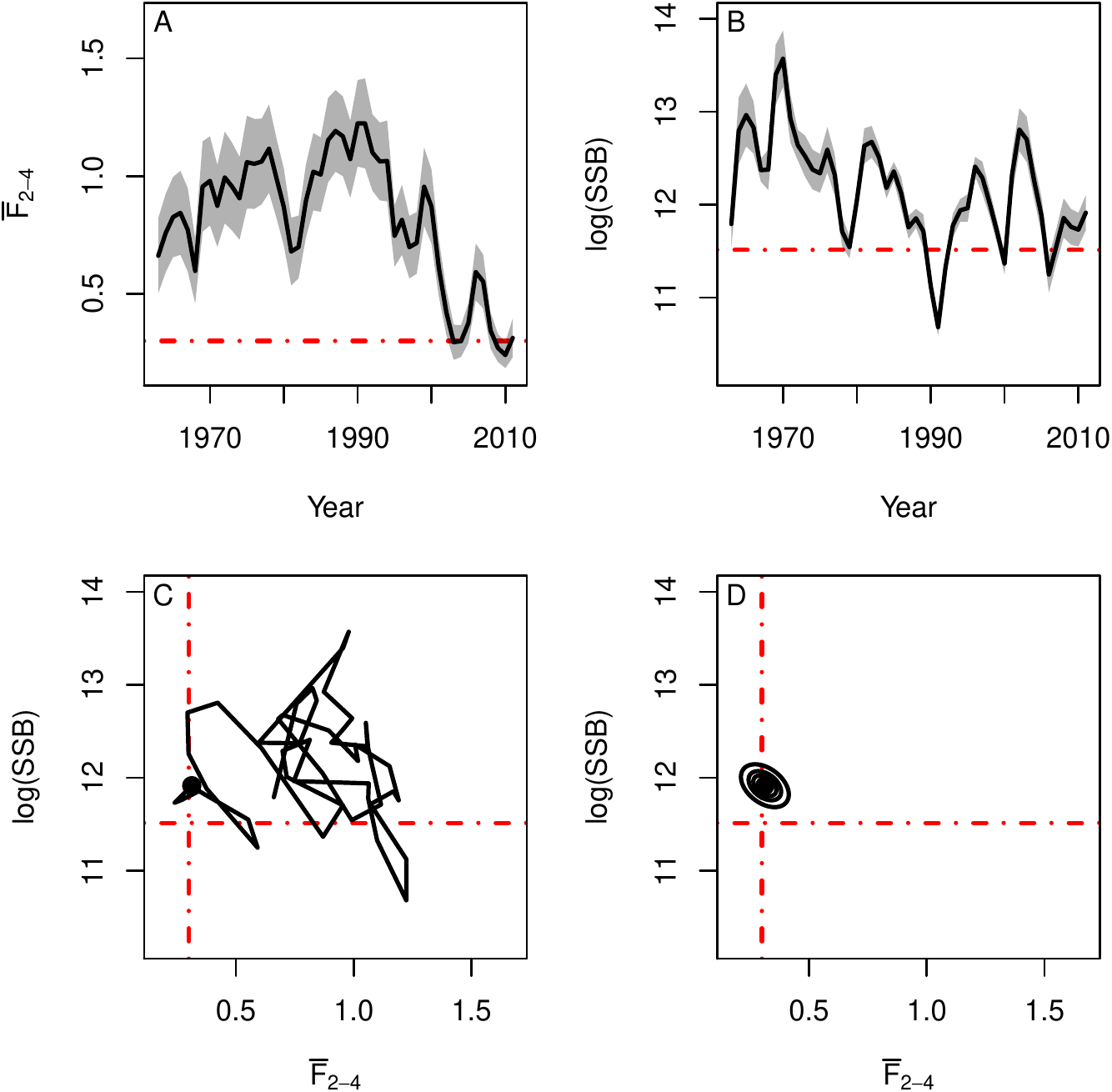}
\caption{Estimated average fishing mortality (A) and log spawning stock
biomass (B) with the Gamma model for Northern Shelf Haddock including 95
\% pointwise confidence intervals (grey area); their estimated
trajectory (C); and confidence ellipses in the final year (D) at 50 \%,
75 \% and 95 \% levels. The red lines indicate the management plan
reference points while the black point is the estimated value in the
final year.}
\end{figure}

\clearpage

\subsubsection{Generalized Gamma}\label{generalized-gamma-3}

\begin{figure}[htbp]
\centering
\includegraphics{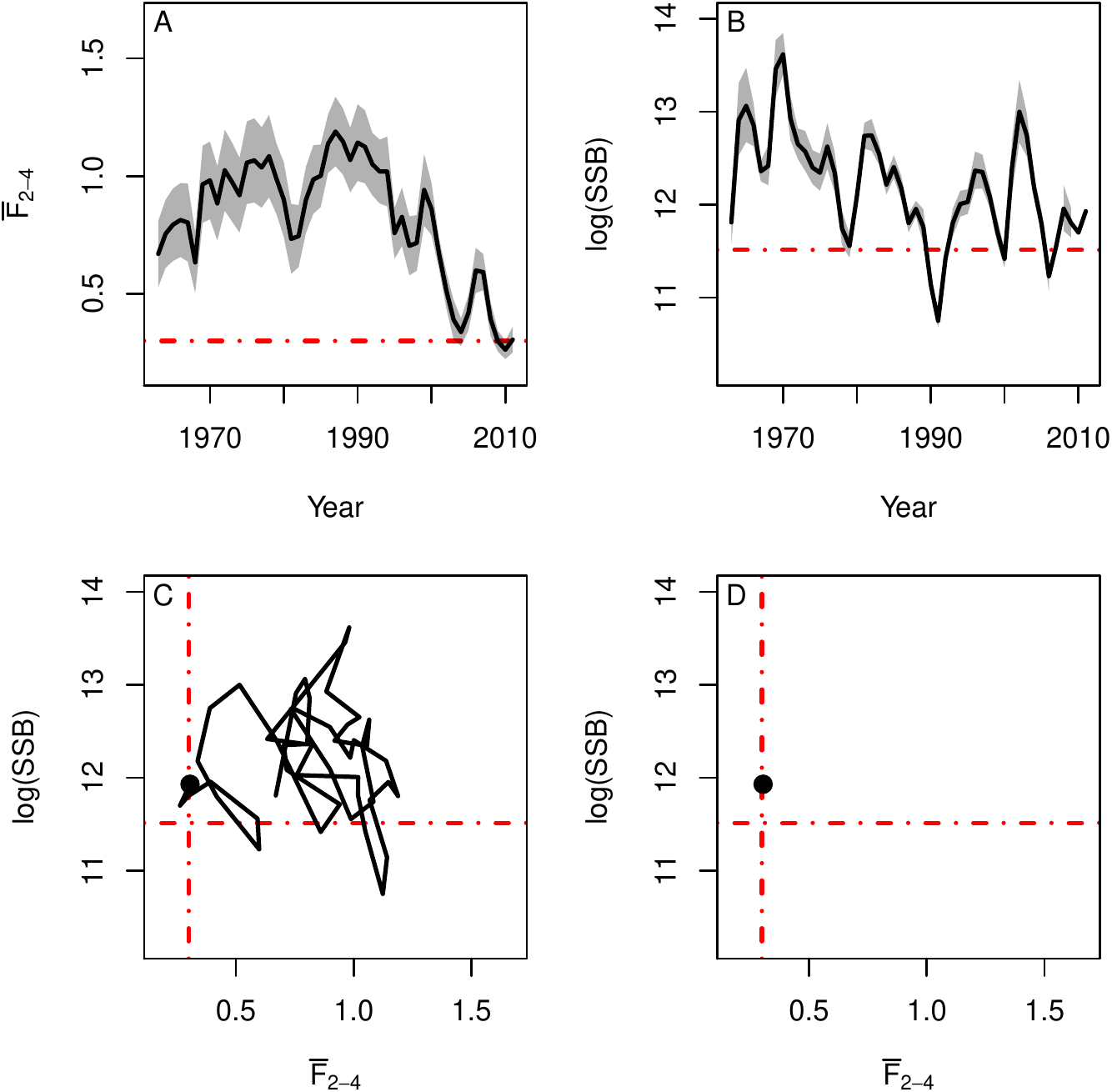}
\caption{Estimated average fishing mortality (A) and log spawning stock
biomass (B) with the Generalized Gamma model for Northern Shelf Haddock
including 95 \% pointwise confidence intervals (grey area); their
estimated trajectory (C); and confidence ellipses in the final year (D)
at 50 \%, 75 \% and 95 \% levels. The red lines indicate the management
plan reference points while the black point is the estimated value in
the final year.}
\end{figure}

\clearpage

\subsubsection{Normal}\label{normal-3}

\begin{figure}[htbp]
\centering
\includegraphics{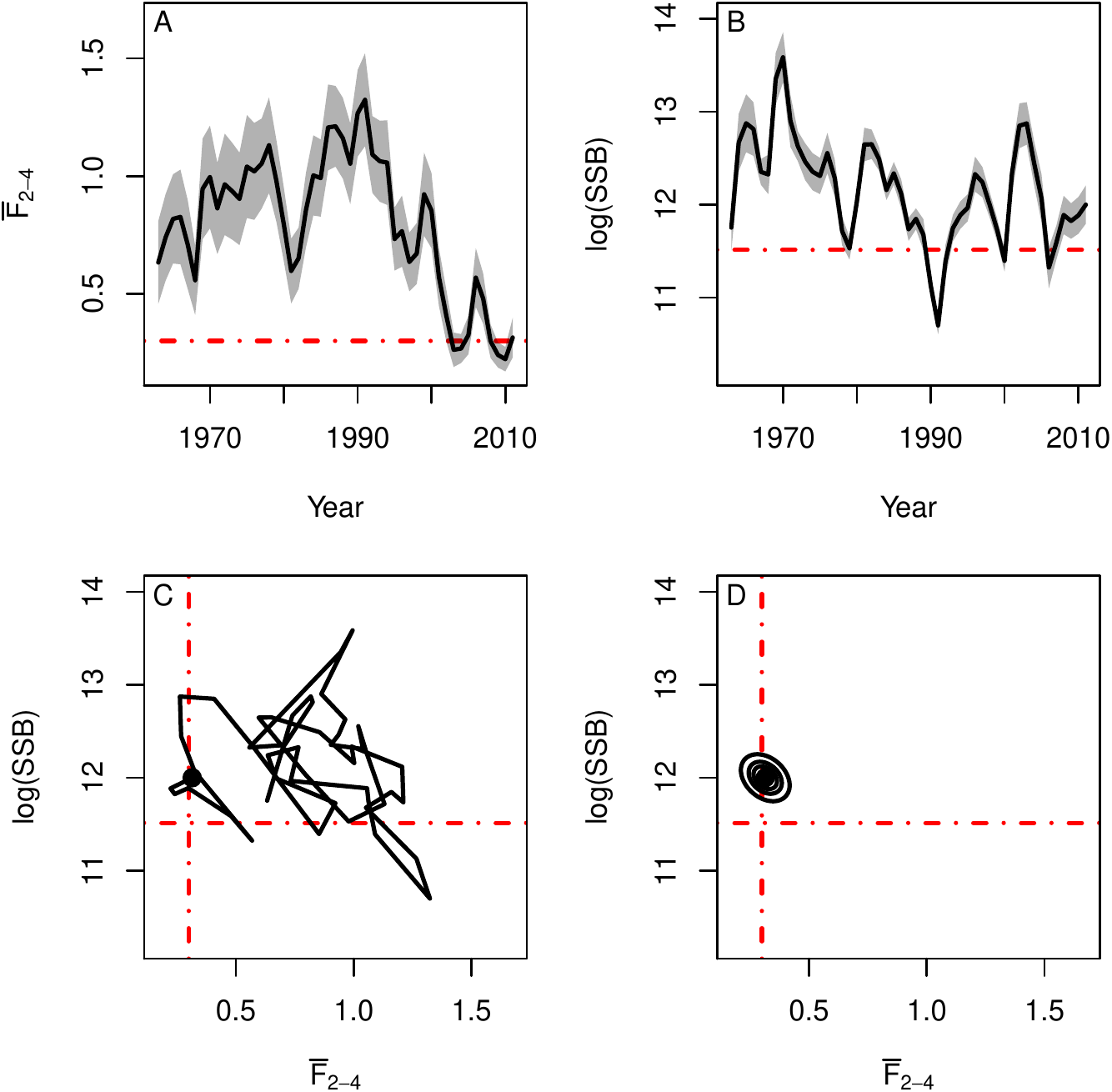}
\caption{Estimated average fishing mortality (A) and log spawning stock
biomass (B) with the Normal model for Northern Shelf Haddock including
95 \% pointwise confidence intervals (grey area); their estimated
trajectory (C); and confidence ellipses in the final year (D) at 50 \%,
75 \% and 95 \% levels. The red lines indicate the management plan
reference points while the black point is the estimated value in the
final year.}
\end{figure}

\clearpage

\subsubsection{Left Truncated Normal}\label{left-truncated-normal-3}

\begin{figure}[htbp]
\centering
\includegraphics{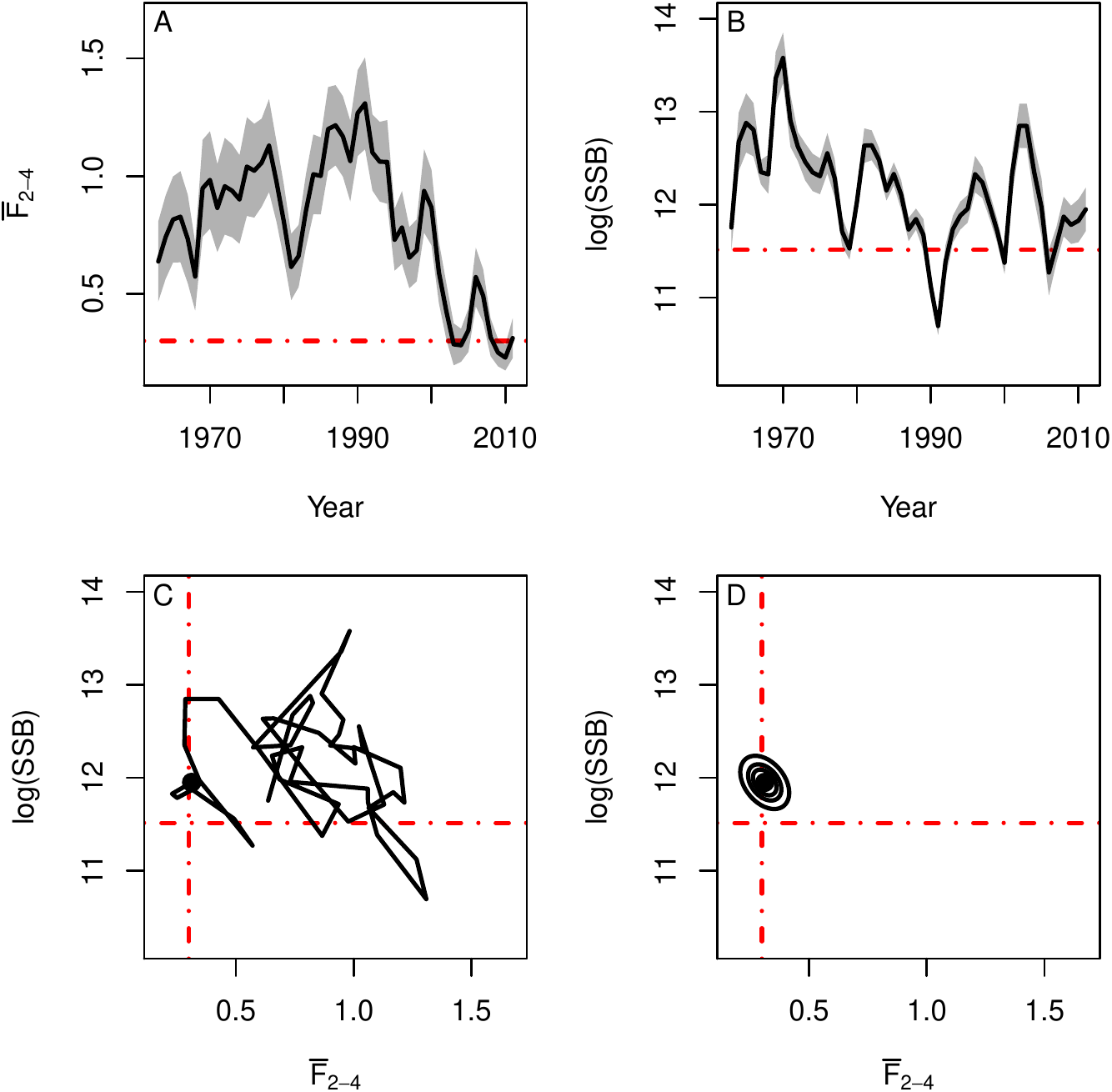}
\caption{Estimated average fishing mortality (A) and log spawning stock
biomass (B) with the Left Truncated Normal model for Northern Shelf
Haddock including 95 \% pointwise confidence intervals (grey area);
their estimated trajectory (C); and confidence ellipses in the final
year (D) at 50 \%, 75 \% and 95 \% levels. The red lines indicate the
management plan reference points while the black point is the estimated
value in the final year.}
\end{figure}

\clearpage

\subsubsection{log-Students t}\label{log-students-t-3}

\begin{figure}[htbp]
\centering
\includegraphics{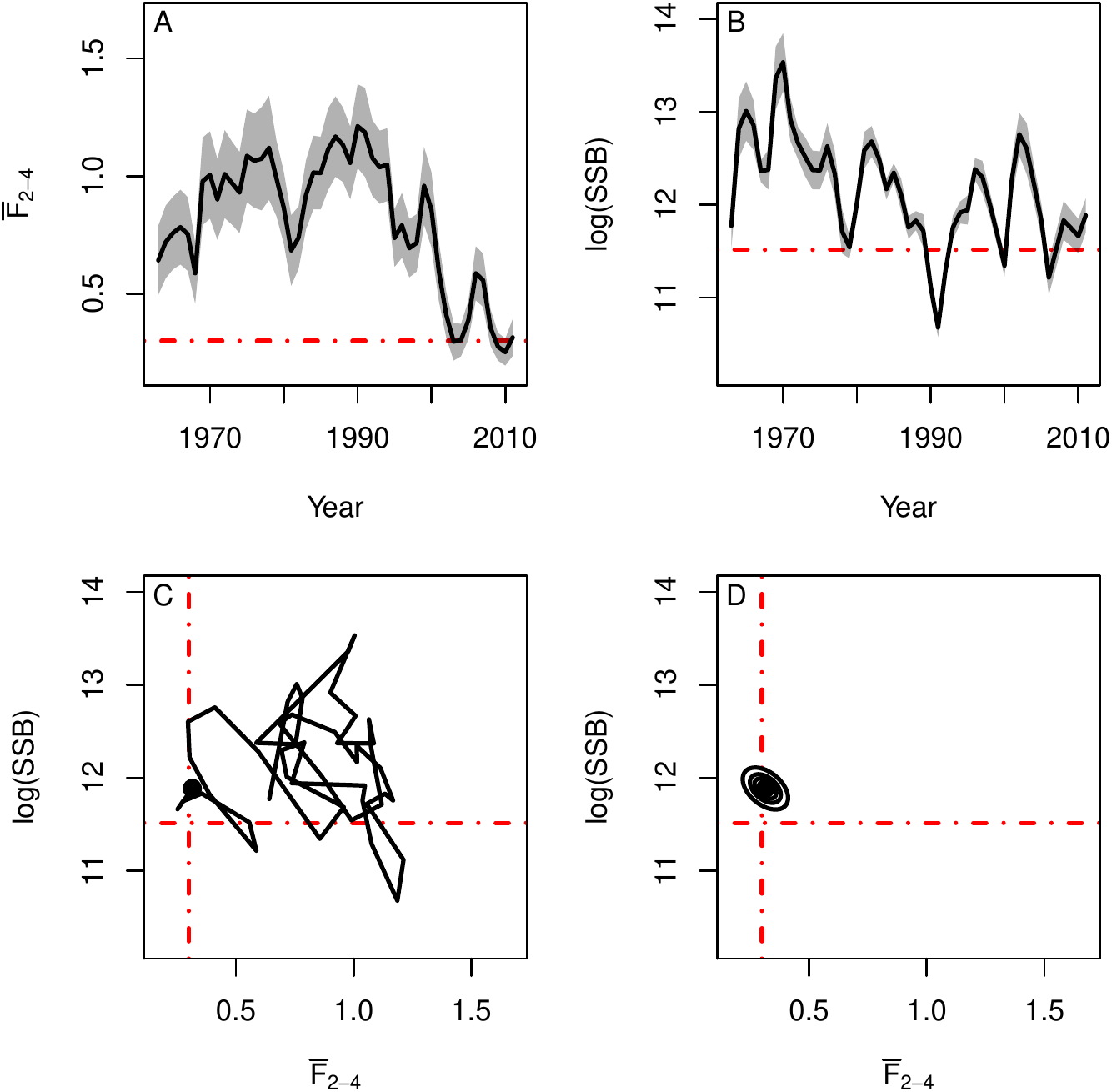}
\caption{Estimated average fishing mortality (A) and log spawning stock
biomass (B) with the log-Students t model for Northern Shelf Haddock
including 95 \% pointwise confidence intervals (grey area); their
estimated trajectory (C); and confidence ellipses in the final year (D)
at 50 \%, 75 \% and 95 \% levels. The red lines indicate the management
plan reference points while the black point is the estimated value in
the final year.}
\end{figure}

\clearpage

\subsubsection{Multivariate log-Normal}\label{multivariate-log-normal-3}

\begin{figure}[htbp]
\centering
\includegraphics{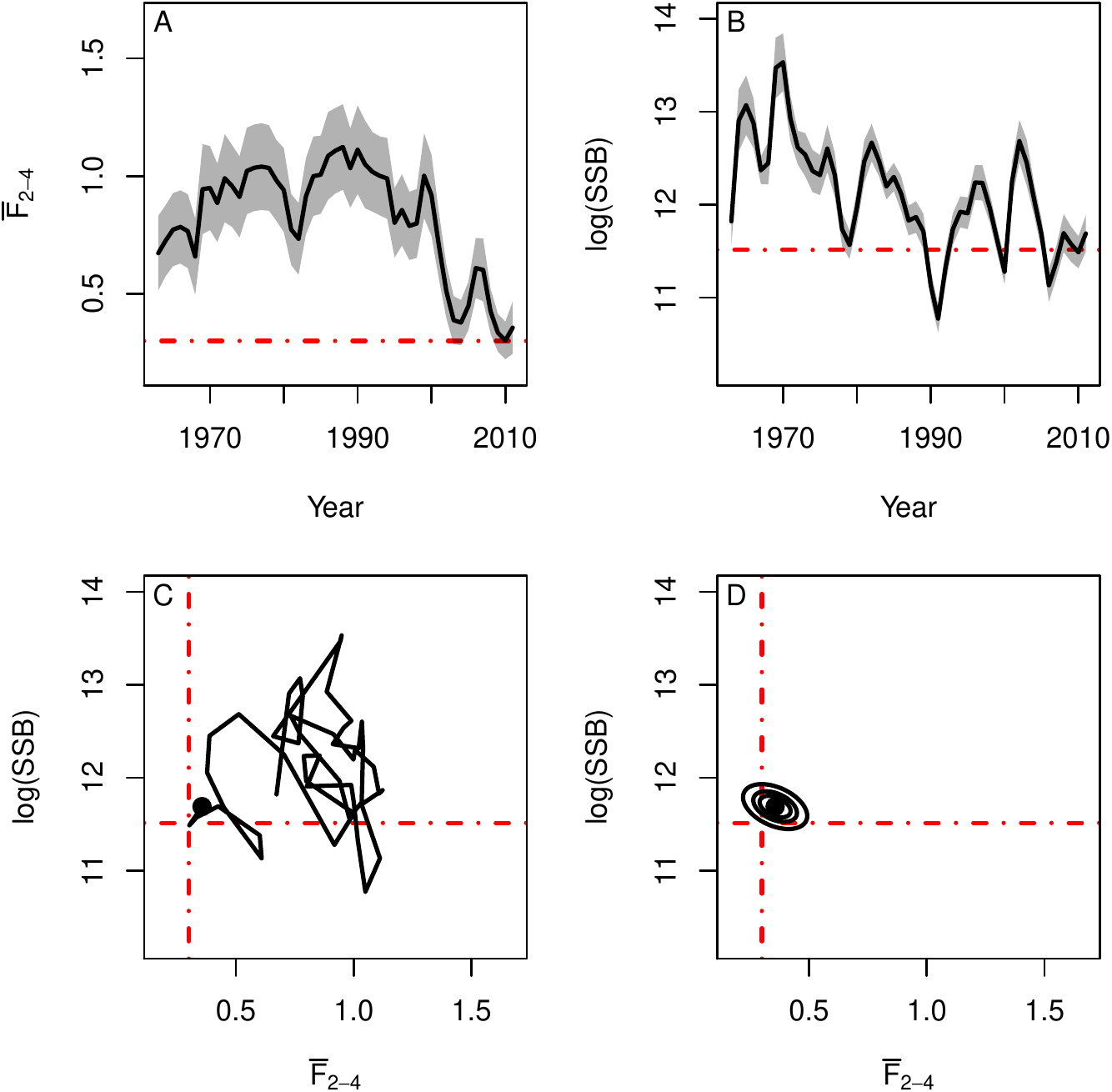}
\caption{Estimated average fishing mortality (A) and log spawning stock
biomass (B) with the Multivariate log-Normal model for Northern Shelf
Haddock including 95 \% pointwise confidence intervals (grey area);
their estimated trajectory (C); and confidence ellipses in the final
year (D) at 50 \%, 75 \% and 95 \% levels. The red lines indicate the
management plan reference points while the black point is the estimated
value in the final year.}
\end{figure}

\clearpage

\subsubsection{Additive Logistic Normal with log-Normal
Numbers}\label{additive-logistic-normal-with-log-normal-numbers-3}

\begin{figure}[htbp]
\centering
\includegraphics{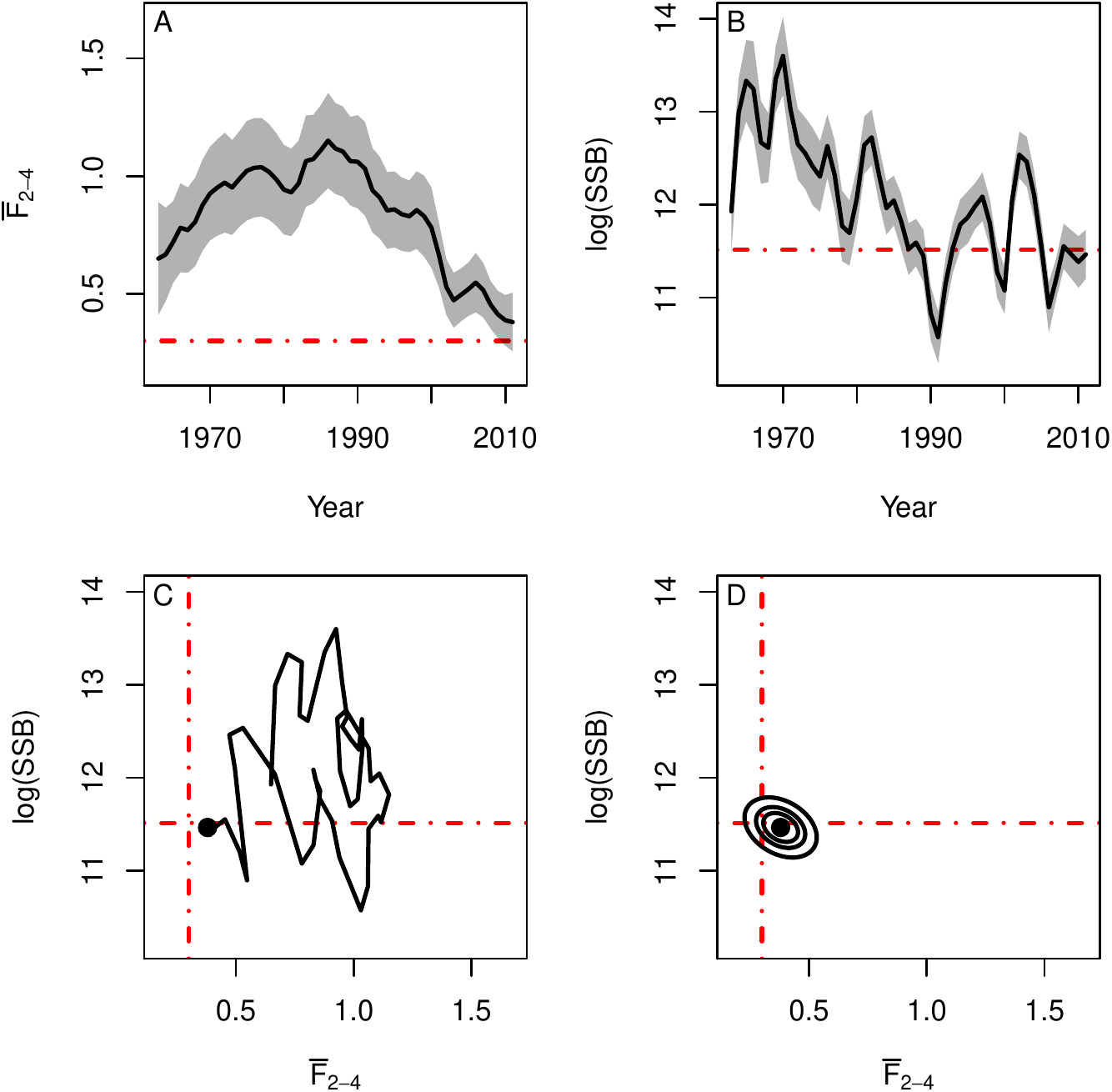}
\caption{Estimated average fishing mortality (A) and log spawning stock
biomass (B) with the Additive Logistic Normal with log-Normal Numbers
model for Northern Shelf Haddock including 95 \% pointwise confidence
intervals (grey area); their estimated trajectory (C); and confidence
ellipses in the final year (D) at 50 \%, 75 \% and 95 \% levels. The red
lines indicate the management plan reference points while the black
point is the estimated value in the final year.}
\end{figure}

\clearpage

\subsubsection{Multiplicative Logistic Normal with log-Normal
Numbers}\label{multiplicative-logistic-normal-with-log-normal-numbers-3}

\begin{figure}[htbp]
\centering
\includegraphics{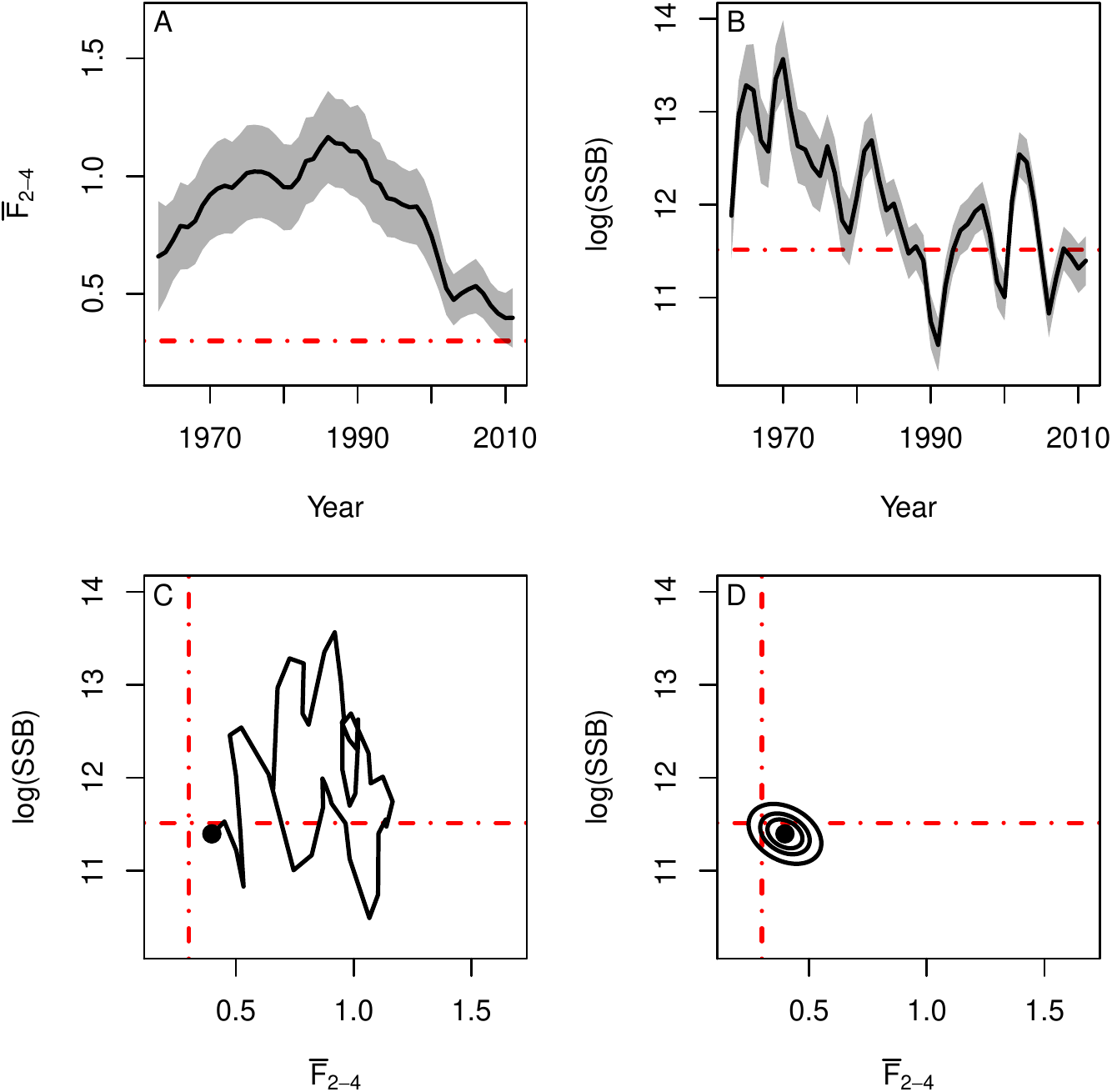}
\caption{Estimated average fishing mortality (A) and log spawning stock
biomass (B) with the Multiplicative Logistic Normal with log-Normal
Numbers model for Northern Shelf Haddock including 95 \% pointwise
confidence intervals (grey area); their estimated trajectory (C); and
confidence ellipses in the final year (D) at 50 \%, 75 \% and 95 \%
levels. The red lines indicate the management plan reference points
while the black point is the estimated value in the final year.}
\end{figure}

\clearpage

\subsubsection{Dirichlet with log-Normal
Numbers}\label{dirichlet-with-log-normal-numbers-3}

\begin{figure}[htbp]
\centering
\includegraphics{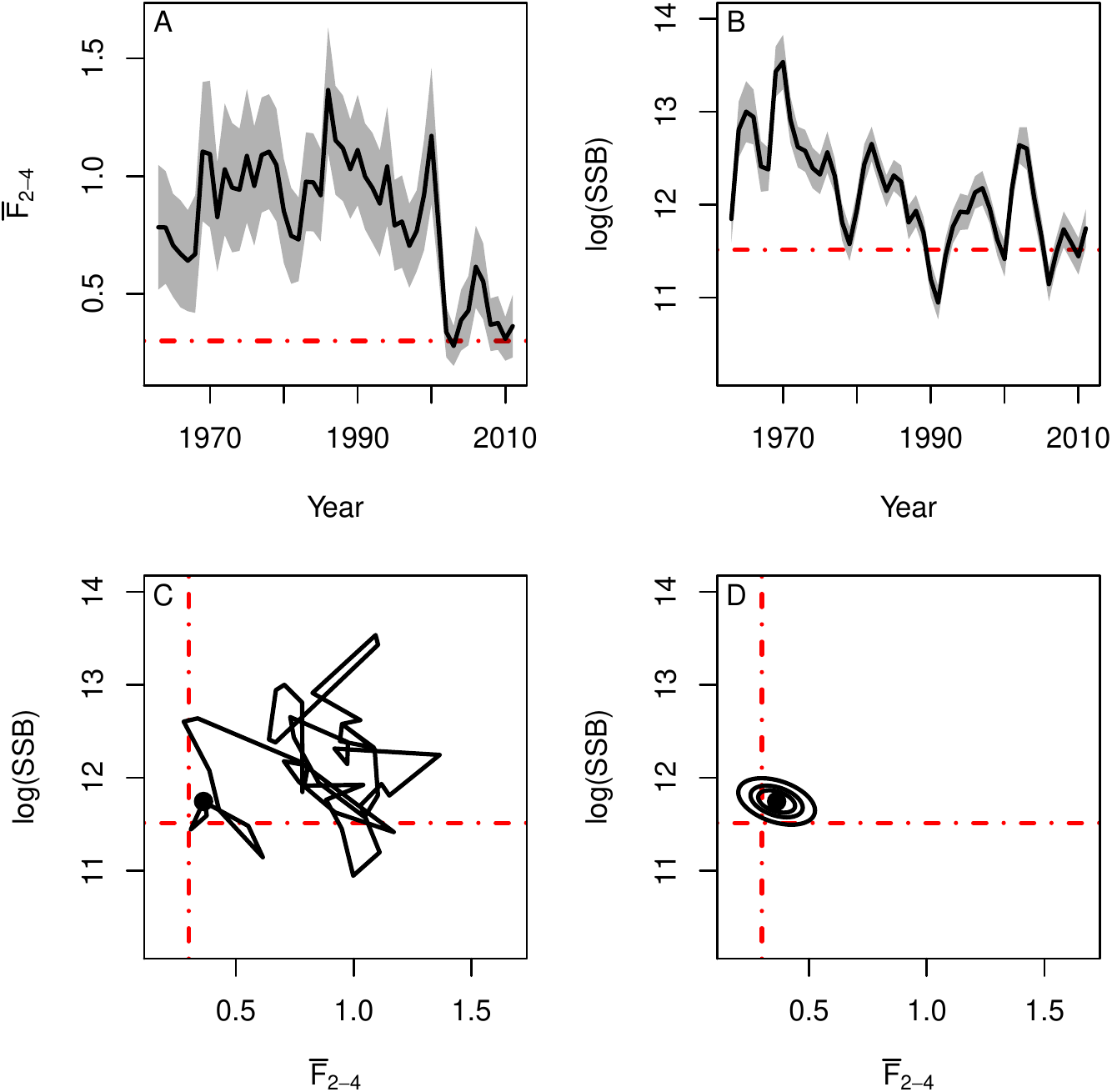}
\caption{Estimated average fishing mortality (A) and log spawning stock
biomass (B) with the Dirichlet with log-Normal Numbers model for
Northern Shelf Haddock including 95 \% pointwise confidence intervals
(grey area); their estimated trajectory (C); and confidence ellipses in
the final year (D) at 50 \%, 75 \% and 95 \% levels. The red lines
indicate the management plan reference points while the black point is
the estimated value in the final year.}
\end{figure}

\clearpage

\subsubsection{Additive Logisitc Normal with log-Normal
Weight}\label{additive-logisitc-normal-with-log-normal-weight-3}

\begin{figure}[htbp]
\centering
\includegraphics{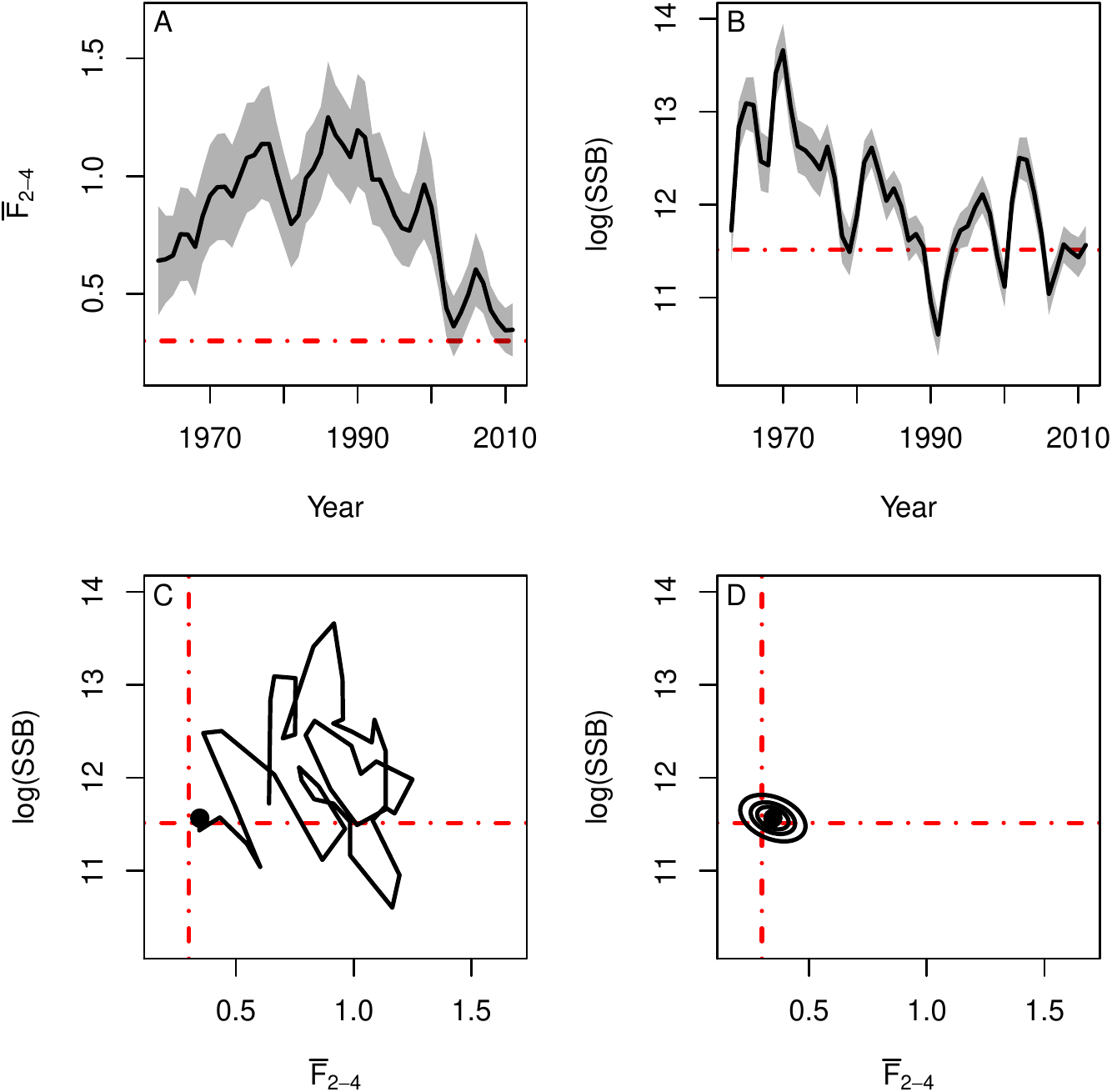}
\caption{Estimated average fishing mortality (A) and log spawning stock
biomass (B) with the Additive Logisitc Normal with log-Normal Weight
model for Northern Shelf Haddock including 95 \% pointwise confidence
intervals (grey area); their estimated trajectory (C); and confidence
ellipses in the final year (D) at 50 \%, 75 \% and 95 \% levels. The red
lines indicate the management plan reference points while the black
point is the estimated value in the final year.}
\end{figure}

\clearpage

\subsubsection{Multiplicative Logistic Normal with log-Normal
Weight}\label{multiplicative-logistic-normal-with-log-normal-weight-3}

\begin{figure}[htbp]
\centering
\includegraphics{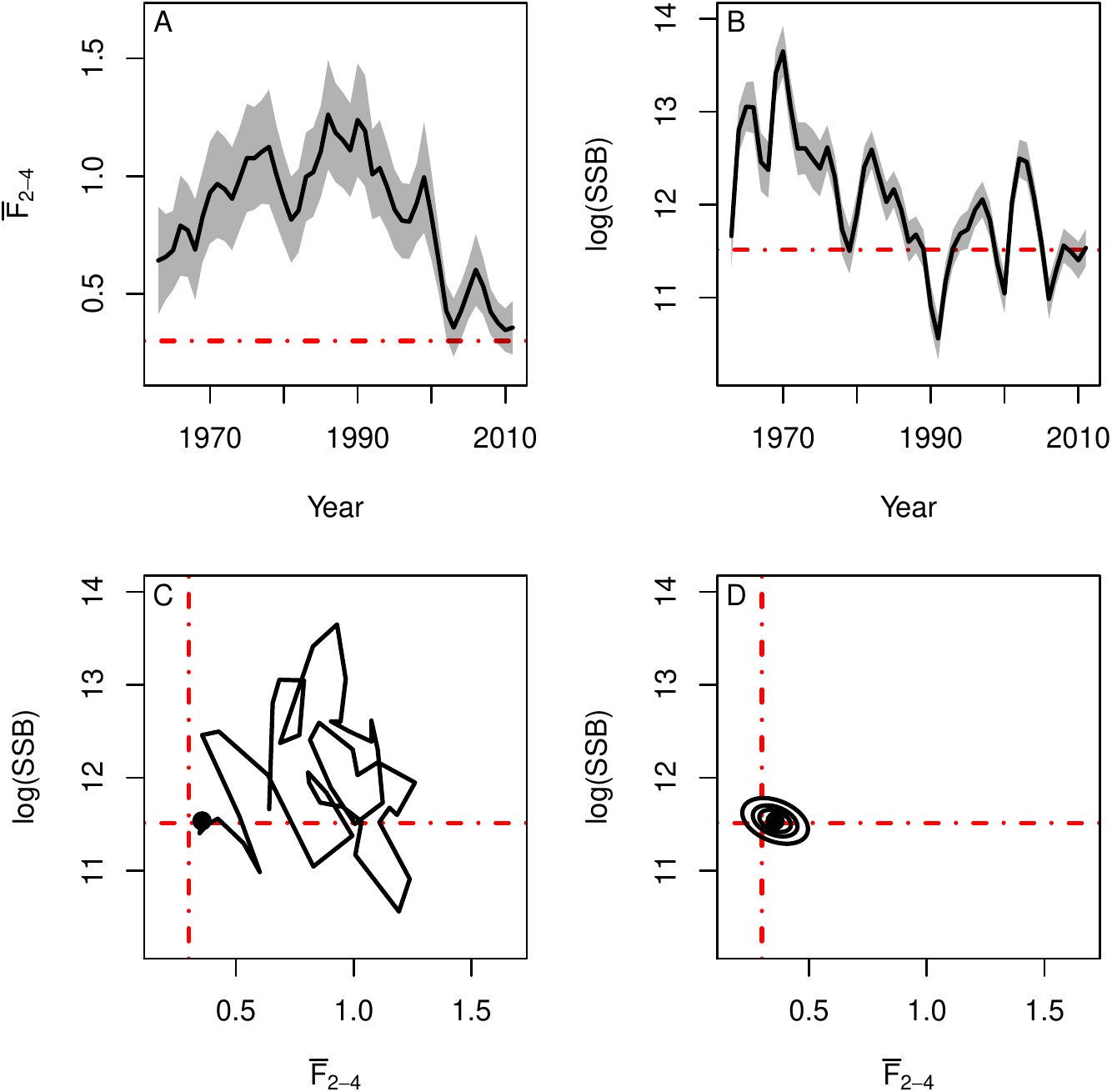}
\caption{Estimated average fishing mortality (A) and log spawning stock
biomass (B) with the Multiplicative Logistic Normal with log-Normal
Weight model for Northern Shelf Haddock including 95 \% pointwise
confidence intervals (grey area); their estimated trajectory (C); and
confidence ellipses in the final year (D) at 50 \%, 75 \% and 95 \%
levels. The red lines indicate the management plan reference points
while the black point is the estimated value in the final year.}
\end{figure}

\clearpage

\subsubsection{Dirichlet with log-Normal
Weight}\label{dirichlet-with-log-normal-weight-3}

\begin{figure}[htbp]
\centering
\includegraphics{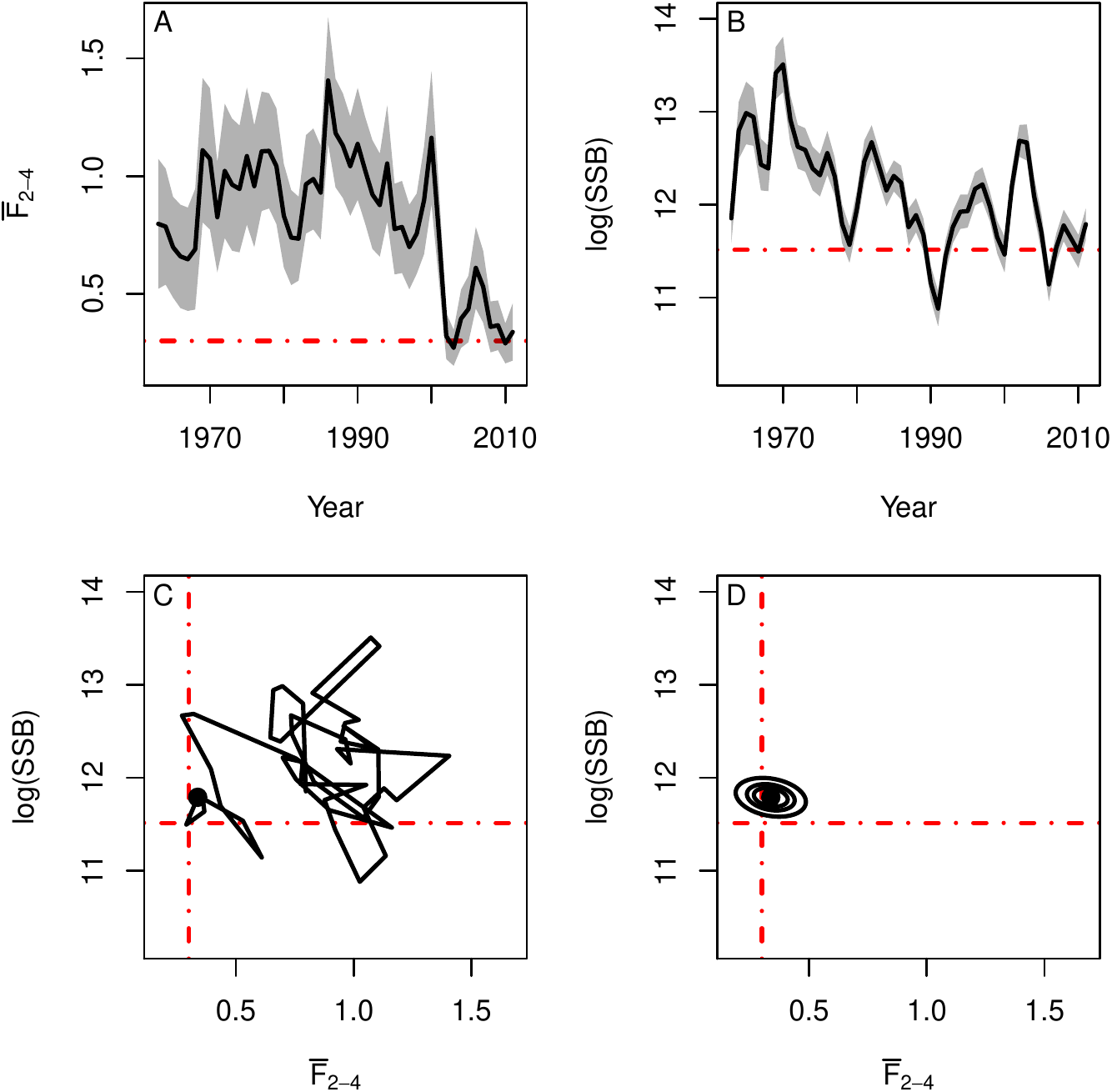}
\caption{Estimated average fishing mortality (A) and log spawning stock
biomass (B) with the Dirichlet with log-Normal Weight model for Northern
Shelf Haddock including 95 \% pointwise confidence intervals (grey
area); their estimated trajectory (C); and confidence ellipses in the
final year (D) at 50 \%, 75 \% and 95 \% levels. The red lines indicate
the management plan reference points while the black point is the
estimated value in the final year.}
\end{figure}

\clearpage
